\documentclass[12pt]{article}
\usepackage{amsfonts}
\usepackage{amssymb}
\usepackage{amsbsy}
\usepackage{graphics}

\input epsf.sty

\newlength{\TYa}
\newlength{\TYb}
\newlength{\TZ}
\newcommand{\BEQ}{\begin{equation}}     % Gleichungen Anfang ..
\newcommand{\BEA}{\begin{eqnarray}}
\newcommand{\EEQ}{\end{equation}}       % .. und Ende
\newcommand{\EEA}{\end{eqnarray}}
\newcommand{\eps}{\varepsilon}          % epsilon
\newcommand{\D}{{\rm d}}                % gerades d fuer Ableitungen
\newcommand{\II}{{\rm i}}               % gerades i fuer komplexe Einheit
\newcommand{\erf}{{\rm erf\,}}          % erf-Funktion
\newcommand{\wit}[1]{\widetilde{#1}}    % weite Schlange
\renewcommand{\vec}[1]{\boldsymbol{#1}} % Vektoren fettgedruckt

\newcommand{\zeile}[1]{\vskip #1 \baselineskip} % N Zeilen ueberschlagen
\newcommand{\vekz}[2]
     {\mbox{${\begin{array}{c} #1  \\ #2 \end{array}}$}}
\newcommand{\appsection}[2]{\setcounter{equation}{0} \section*{Appendix #1. #2}
\renewcommand{\theequation}{#1\arabic{equation}}
              \renewcommand{\thesection}{#1} }
\catcode`\@=11
\def\numberbysection{\@addtoreset{equation}{section}
        \def\theequation{\thesection.\arabic{equation}}}
\numberbysection

\begin{document}

\begin{titlepage}

~~ 

\vskip 1.5 cm
\begin{center}
{\Large \bf Phenomenology of local scale invariance : \\
from conformal invariance to dynamical scaling}
\end{center}

\vskip 2.0 cm
\centerline{  {\bf Malte Henkel} }
\vskip 0.5 cm
\centerline {Laboratoire de Physique des 
Mat\'eriaux,\footnote{Laboratoire associ\'e au CNRS UMR 7556} 
Universit\'e Henri Poincar\'e Nancy I,} 
\centerline{ B.P. 239, 
F -- 54506 Vand{\oe}uvre l\`es Nancy Cedex, France}

~\zeile{2}
\begin{abstract}
\noindent Statistical systems displaying a strongly anisotropic or dynamical
scaling behaviour are characterized by an anisotropy exponent $\theta$ or
a dynamical exponent $z$. For a given value of $\theta$ (or $z$), we construct
local scale transformations, which can be viewed as scale transformations
with a space-time-dependent dilatation factor. Two distinct types of local
scale transformations are found. The first type may describe strongly 
anisotropic scaling of static systems with a given value of $\theta$, whereas 
the second type may describe dynamical scaling with a dynamical exponent $z$. 
Local scale transformations act as a dynamical symmetry group of certain
non-local free-field theories. Known special cases of local scale invariance
are conformal invariance for $\theta=1$ and Schr\"odinger invariance for
$\theta=2$.  

The hypothesis of local scale invariance implies that two-point functions of
quasiprimary operators satisfy certain linear fractional differential
equations, which are constructed from commuting fractional derivatives. The
explicit solution of these yields exact expressions for two-point correlators
at equilibrium and for two-point response functions out of equilibrium. 
A particularly simple and general form is found for the two-time
autoresponse function. These
predictions are explicitly confirmed at the uniaxial Lifshitz points 
in the ANNNI and ANNNS models and in the aging behaviour of simple
ferromagnets such as the kinetic Glauber-Ising model and the
kinetic spherical model with a non-conserved order parameter undergoing
either phase-ordering kinetics or non-equilibrium critical dynamics. 
\end{abstract}
\vfill
Nucl. Phys. {\bf B} (2002), in press
\end{titlepage}

%%%%%%%%%%%%%%%%%%%%%%%%%%%%%%%%%%%%%%%%%%%%%%%%%%%%%%%%%%%%%%%%%%%%%%%%%%%%%%%%
\section{Introduction}
%%%%%%%%%%%%%%%%%%%%%%%%%%%%%%%%%%%%%%%%%%%%%%%%%%%%%%%%%%%%%%%%%%%%%%%%%%%%%%%%

Critical phenomena are known at least since 1822 when Cagnard de la Tour
observed critical opalescence in binary mixtures of alcohol and water. 
The current understanding of (isotropic equilibrium) critical phenomena, 
see e.g. \cite{Fish83,Zinn89,Drou88,Card96}, is based on the covariance of the
$n$-point correlators 
$G_n=G(\vec{r}_1,\ldots,\vec{r}_n)=
\langle \phi_1(\vec{r}_1)\ldots\phi_n(\vec{r}_n)\rangle$ under global
scaling transformations $\vec{r}_i \to b \vec{r}_i$
\BEQ \label{1:Kov}
G(b\vec{r}_1,\ldots,b\vec{r}_n) = b^{-(x_1+\ldots+x_n)} 
G(\vec{r}_1,\ldots,\vec{r}_n) 
\EEQ
precisely {\em at\/} the critical point. 
Here the $\phi_i$ are scaling 
operators\footnote{We follow the terminology of \cite{Card96}: the $\phi_i$
are called scaling {\em operators}, because if the theory is quantized in the
operator formalism, $\phi_i \to \hat{\phi}_i$ becomes a field operator. 
The variables $h_i$ canonically conjugate to $\phi_i$ are called the 
{\em scaling fields}. For the Ising model, the scaling operator 
$\phi_{\sigma}=\sigma$ is the order parameter density and its canonically 
conjugate scaling field $h$ is the magnetic field.}
with scaling dimension $x_i$ and one might consider the formal
covariance of the $\phi_i$
\BEQ
\phi_i(b \vec{r}_i) = b^{-x_i} \phi(\vec{r}_i)
\EEQ
as a compact way to express the covariance (\ref{1:Kov}) of the correlators. 
In isotropic (e.g. rotation-invariant) equilibrium systems, the $\phi_i$
corresponds to the physical order parameter or energy densities and so on. 
Eq.~(\ref{1:Kov}) may be derived from the renormalization group and in turn
implies the phenomenological scaling behaviour of the various observables
of interest. It has been known since a long time that in systems with
sufficiently short-ranged interactions, $G_n$ actually
transform covariantly under the conformal group, that is under space-dependent
or {\em local} scale transformations $b=b(\vec{r})$ such that the angles are
kept unchanged \cite{Poly70}. Since in two dimensions, the Lie algebra of the
conformal group is the infinite-dimensional Virasoro algebra, strong
constraints on the possible $\phi_i$ present in a $2D$ conformally invariant
theory follow \cite{Bela84}. Roughly, for a given unitary $2D$ conformal theory,
the entire set of scaling operators $\phi_i$, the values of their scaling
dimensions $x_i$ and the critical $n$-point correlators $G_n$ can be
found exactly. Furthermore, there is a classification of the modular invariant
partition functions of unitary models {\em at} criticality 
(the ADE classification) which goes a long way towards a classification of the 
universality classes of $2D$ conformally invariant critical points
(for reviews, see \cite{Card96,Fran97,Henk99}). 

Here we are interested in critical systems where the $n$-point
functions satisfy an {\em anisotropic} scale covariance of the form
\BEQ \label{1:Ani}
G(b^{\theta}t_1,b\vec{r}_1,\ldots,b^{\theta}t_n,b\vec{r}_n) = 
b^{-(x_1+\ldots+x_n)} 
G(t_1,\vec{r}_1,\ldots,t_n,\vec{r}_n) 
\EEQ
where we distinguish so-called `spatial' coordinates $\vec{r}_i$ and
`temporal' coordinates $t_i$. The exponent $\theta$ is the 
{\em anisotropy exponent}. By definition, a system whose $n$-point functions
satisfy (\ref{1:Ani}) with $\theta\ne 1$ is a {\em strongly
anisotropic critical system}. Systems of this kind are quite common. 
For example, eq.~(\ref{1:Ani}) is realized in (i) static equilibrium
critical behaviour in anisotropic systems such as dipolar-coupled uniaxial
ferromagnets \cite{Ahar76} and/or at a Lifshitz point in systems with competing 
interactions \cite{Horn75,Selk92,Neub98} or even anisotropic
surface-induced disorder \cite{Turb02},
(ii) anisotropic criticality in steady states of non-equilibrium systems
such as driven diffusive systems \cite{Schm95,Marr99},  
stochastic surface growth \cite{Krug97} or such as directed percolation.
In these cases, $\vec{r}$ and $t$ are merely labels for different directions
in space and the $G_n$ are in the case (i) 
equal to the $n$-point correlators $C_n$ of the physical 
scaling operators. (iii) Further examples are found in quantum critical 
points, see \cite{Hert76,Sach00}.  
Anisotropic scaling also occurs in (iv) critical dynamics of 
statistical systems at equilibrium \cite{Halp77} or (v) 
in non-equilibrium dynamical
scaling phenomena \cite{Bray94,Priv96,Chop98,Marr99,Schu00,Hinr00,Cate00}. 
In the cases (iv) and (v), $t$ represents the physical time and the 
system's behaviour is characterized jointly by the time-dependent 
$n$-point correlators $C_n$ as well as with the $n$-point response functions 
$R_n$. Habitually, $\theta=z$ then is referred to as the 
{\em dynamical exponent}. At equilibrium, the $C_n$ and $R_n$ are related 
by the fluctuation-dissipation theorem \cite{Kreu81,Card96}, but no such 
relation is known to hold for systems far from equilibrium. 

We ask: is it possible to extend the dynamical scaling (\ref{1:Ani}) 
towards space-time-dependent rescaling factors $b=b(t,\vec{r})$ such 
that the $n$-point functions $G_n$ still transform covariantly ? 

It is part of the problem to establish what kind(s) of space-time 
transformations might be sensibly included into the set of generalized
scaling transformations. Also, for non-static systems, covariance under a
larger scaling group may or may not hold simultaneously for correlators 
and response functions. Another aspect of the problem is best illustrated for
the two-point function $G_2=G(t_1,\vec{r}_1,t_2,\vec{r}_2)=G(t,\vec{r})$
where for simplicity we assume for the moment space and time translation
invariance and $t=t_1-t_2$ and $\vec{r}=\vec{r}_1-\vec{r}_2$. From 
(\ref{1:Ani}) one has the scaling form
\BEQ
G_2 = G(t,\vec{r}) = t^{-2x/\theta} {\cal G}(u) \;\; , \;\;
u = |\vec{r}|^{\theta}/t
\EEQ
where -- in contrast to the situation of isotropic equilibrium points with
$\theta=1$ (see below) -- the scaling function ${\cal G}(u)$ is undetermined. 
We look for general arguments which would allow us to determine the form of 
$\cal G$, independently of any specific model. In turn, once we have found some
sufficiently general local scaling transformations, and thus predicted the
form of $G_2$, the explicit comparison with model results, either analytical
or numerical, will provide important tests. Several examples of this kind
will be discussed in this paper. 

Some time ago, Cardy had discussed the presence of local scaling for
critical dynamics \cite{Card85}. Starting from the observation that static
$2D$ critical systems are conformally invariant, he argued that the
response functions should transform covariantly under the set of
transformations $\vec{r}\to b(\vec{r})\vec{r}$ and $t\to b(\vec{r})^{\theta}t$.
Through a conformal transformation, the response function was mapped from
$2D$ infinite space onto the strip geometry and found there through van
Hove theory. Explicit expressions for the scaling function $\cal G$ were
obtained for both non-conserved (then ${\cal G}(u)\sim e^{-u}$, up to
normalization constants) and conserved order parameters \cite{Card85}.
However, these forms have to the best of our knowledge 
so far never been reproduced in any model beyond
simple mean field (i.e. van Hove) theory. That had triggered us to try to study
the construction of groups of local anisotropic scale transformations somewhat 
more systematically, beginning with the simplest case of Schr\"odinger 
invariance which holds for $\theta=2$ \cite{Henk92,Henk94}. 
At the time, it appeared to be suggestive that the exactly known Green's 
function of the $1D$ kinetic Ising
model with Glauber dynamics \cite{Glau63} could be recovered this way. 
How these initial results might be extended beyond the
$\theta=2$ case is the subject of this paper. 

The outline of the paper is as follows: in section 2, we shall review
some basic results of conformal invariance and of the simplest case of 
strongly anisotropic scaling, which 
occurs if $\theta=2$. In this case, there does exist a Lie group of local
scale transformations, which is known as the {\em Schr\"odinger group}
\cite{Nied72,Hage72}. Building on the analogy with this case and conformal
invariance for $\theta=1$, we discuss in section 3 the systematic construction
of infinitesimal local scale transformations which are compatible with 
the anisotropic scaling (\ref{1:Ani}). We shall see that there are two
distinct solutions, one corresponding to strongly anisotropic scaling
at equilibrium and the other corresponding to dynamical scaling. We also show
that the local scale transformations so constructed act as dynamical symmetries
on some linear field equations of fractional order. Furthermore, linear 
fractional differential equations which are satisfied by the two-point scaling 
functions ${\cal G}(u)$ are derived. In section 4, these are solved explicitly 
and the form of ${\cal G}(u)$ is thus determined. In section 5, these explicit
expressions are tested by comparing them with results from several distinct
models with strongly anisotropic scaling, notably Lifshitz points in
systems with competing interactions such as the ANNNI model and for 
some ferromagnetic non-equilibrium spin systems (especially the Glauber-Ising
model in $2D$ and $3D$) undergoing aging after being
quenched from some disordered initial state to a temperature at or below
criticality. We also comment on equilibrium critical dynamics. 
A reader mainly interested in the applications may start reading
this section first and refer back to the earlier ones if necessary. 
Section 6 gives our conclusions. 
Several technical points are discussed in the appendices. 
In appendix A we discuss the construction of commuting fractional derivatives
and prove several simple rules useful for practical calculations. In appendix
B we generalize the generators of the Schr\"odinger Lie algebra to $d>1$
space dimensions and in appendix C we present an alternative route towards
the construction of local scale transformations. Appendix D discusses further 
the solution of fractional-order differential equations through series methods. 

%%%%%%%%%%%%%%%%%%%%%%%%%%%%%%%%%%%%%%%%%%%%%%%%%%%%%%%%%%%%%%%%%%%%%%%%%%%%%%%%
\section{Conformal and Schr\"odinger invariance}
%%%%%%%%%%%%%%%%%%%%%%%%%%%%%%%%%%%%%%%%%%%%%%%%%%%%%%%%%%%%%%%%%%%%%%%%%%%%%%%%

Our objective will be the systematic construction of infinitesimal
local scale transformations with anisotropy exponents $\theta\ne 1$. 
Consider the scaling of a two-point function
\BEQ \label{2:Skal}
G=G(t,\vec{r}) = b^{2x} G(b^{\theta} t, b\vec{r}) = 
t^{-2x/\theta} \Phi(r t^{-1/\theta}) = r^{-2x} \Omega(t r^{-\theta})
\EEQ
where $t=t_1-t_2$, $\vec{r}=\vec{r}_1-\vec{r}_2$, $r=|\vec{r}|$ and $x$ is
a scaling dimension. For convenience, we also assumed spatio-temporal 
translation invariance. The scaling of $G$ is described by the scaling 
functions $\Phi(u)$ or alternatively by 
$\Omega(v)=v^{-2x/\theta}\Phi\left(v^{-1/\theta}\right)$. 
In this section we concentrate on the formal consequences of
the scaling (\ref{2:Skal}) and postpone the question of the physical meaning of
$G$ to a later stage. 

Considering (\ref{2:Skal}) for $\vec{r}=0$, one has 
$G(t,\vec{0})\sim t^{-2x/\theta}$  and if $t=0$, one has 
$G(0,\vec{r})\sim r^{-2x}$. Therefore,
\BEA
\Phi(0) = \Phi_0 &,& \Phi(u) \simeq \Phi_{\infty} u^{-2x} \;\; ~~; \;\;
u\to\infty \nonumber \\
\label{2:SkalFRand}
\Omega(0) = \Omega_0 &,& \Omega(v) \simeq \Omega_{\infty} v^{-2x/\theta}
\;\; ; \;\; v\to\infty 
\EEA
where $\Phi_{0,\infty}$ and $\Omega_{0,\infty}$ are generically non-vanishing
constants. This exhausts the information scale invariance alone can provide. 

\subsection{Conformal transformations} 

Consider static and isotropic systems with short-ranged interactions.
Then the two-point function 
$G(t,\vec{r}) = \langle \phi_1(t_1,\vec{r}_1) \phi_2(t_2,\vec{r}_2)\rangle$
is the correlation function of the physical scaling operators $\phi_{1,2}$. 
If these $\phi_{i}$ are actually quasiprimary \cite{Bela84} scaling operators,
$G$ does transform covariantly under the action of the conformal group. 
To be specific, we restrict ourselves 
to two dimensions (here $t$ and $r$ merely label
the different directions) and introduce the complex
variables
\BEQ 
z = t + \II r \;\; , \;\; \bar{z} = t - \II r
\EEQ
Then the projective conformal transformations are given by
\BEQ \label{2:Moeb}
z\to z' = \frac{\alpha z + \beta}{\gamma z + \delta} \;\; ; \;\;
\alpha\delta-\beta\gamma =1
\EEQ
and similarly for $\bar{z}$. Writing $z'=z'(z)=z+\eps(z)$, the
infinitesimal generators read
\BEQ
\ell_n = - z^{n+1} \partial_z \;\; , \;\; 
\bar{\ell}_n = - \bar{z}^{n+1} \partial_{\bar{z}}
\EEQ
and satisfy the commutation relations
\BEQ \label{2:Vir0} 
\left[ \ell_n , \ell_m \right] = (n-m) \ell_{n+m} \;\; , \;\;
\left[ \ell_n , \bar{\ell}_m \right] = 0 \;\; , \;\;
\left[ \bar{\ell}_n , \bar{\ell}_m \right] = (n-m) \bar{\ell}_{n+m}
\EEQ
In fact, although the generators $\ell_n$ were initially only constructed for
$n=-1,0,1$, the $\ell_n$ can be written down for  
{\em all} $n\in\mathbb{Z}$ and (\ref{2:Vir0}) still holds. 
The existence of this infinite-dimensional Lie
algebra, known as the Virasoro algebra without central charge, is peculiar to
two spatial dimensions. The set (\ref{2:Moeb}) corresponds to the 
finite-dimensional subalgebra $\{\ell_{\pm1,0},\bar{\ell}_{\pm1,0}\}$. 

The simplest possible way scaling operators can transform under the set
(\ref{2:Moeb}) is realized by the {\em quasiprimary} 
operators \cite{Scha75,Bela84}, 
which transform as
\BEQ
\delta \phi_i(z,\bar{z}) = \left( \Delta _i\eps'(z) + \eps(z) \partial_z + 
\overline{\Delta}_i \bar{\eps}'(\bar{z}) + \bar{\eps}(\bar{z}) 
\partial_{\bar{z}}\right) \phi_i(z,\bar{z})
\EEQ
where $\Delta_i$ and $\overline{\Delta}_i$ are called the conformal weights
of the operator $\phi_i$. If $\phi_i$ is a scalar under (space-time) rotations
(we shall always assume this to be the case), 
$\Delta_i = \overline{\Delta}_i=x_i/2$, where $x_i$ is the scaling dimension
of $\phi_i$. If $\eps(z)=\eps\, z^{n+1}$, one then has 
$\delta \phi_i(z,\bar{z})=-\eps(\ell_n + \bar{\ell}_n)\phi_i(z,\bar{z})$
where the generators $\ell_n$, $\bar{\ell}_n$ now read 
\BEQ \label{2:lPhi}
\ell_n = - z^{n+1} \partial_z - \Delta_i (n+1) z^n \;\; , \;\;
\bar{\ell}_n = - \bar{z}^{n+1} \partial_{\bar{z}} - \Delta_i (n+1) \bar{z}^n
\EEQ
and again satisfy (\ref{2:Vir0}). Later on, we shall work with the
generators
\BEQ \label{2:GenXY}
X_n := \ell_n + \bar{\ell}_n \;\; , \;\;
Y_n := \II ( \ell_n - \bar{\ell}_n )
\EEQ
which satisfy the commutation relations
\BEQ \label{2:ConfXYAlg}
\left[ X_n , X_m \right] = (n-m) X_{n+m} \;\; , \;\;
\left[ X_n , Y_m \right] = (n-m) Y_{n+m} \;\; , \;\;
\left[ Y_n , Y_m \right] = -(n-m) X_{n+m}
\EEQ
The covariance of $G$ under finite projective conformal transformations
leads to the projective conformal Ward identities for the $n$-point functions
$G$ of quasiprimary scaling operators (see \cite{Scha75} for a detailed 
discussion on quasiprimary operators)
\BEQ
\ell_n G = \bar{\ell}_n G = 0 \;\; \longleftrightarrow
X_n G = Y_n G = 0
\EEQ
for $n=\pm 1,0$ and the generators as defined in 
eqs.~(\ref{2:lPhi},\ref{2:GenXY}). This gives for the two-point function of
two scalar quasiprimary operators \cite{Poly70}
\BEQ \label{2:ZweiP}
G(t_1,t_2; r_1,r_2) = G_{12}\, \delta_{x_1,x_2} 
\left( ( z_1 - z_2)(\bar{z}_1 - \bar{z}_2) \right)^{-x_1} 
= G_{12}\, \delta_{x_1,x_2} 
\left( (t_1-t_2)^2 + (r_1 - r_2)^2 \right)^{-x_1}
\EEQ
where $G_{12}$ is a normalization constant (usually, one sets $G_{12}=1$). 
Comparison with (\ref{2:Skal}) gives the scaling function
$\Omega(v) \sim (1+v^2)^{-x}$. The constraint $x_1=x_2$
is the only result which goes beyond simple scale and rotation invariance.

The three-point function is \cite{Poly70} 
\BEQ \label{2:DreiP}
G(t_1,t_2,t_3; r_1,r_2,r_3) = G_{123}\, \rho_{12}^{-x_{123}}
\rho_{23}^{-x_{231}} \rho_{31}^{-x_{312}}
\EEQ
where
\BEQ
\rho_{ab}^2 = z_{ab} \bar{z}_{ab} = (t_a-t_b)^2 + (r_a - r_b)^2 \;\; , \;\;
z_{ab} = z_a - z_b
\EEQ
and $x_{abc} := x_a+x_b-x_c$. The constant $G_{123}$ is the operator
product expansion coefficient of the three quasiprimary 
operators $\phi_{1,2,3}$. For scalar quasiprimary operators, the results
(\ref{2:ZweiP},\ref{2:DreiP}) remain also valid in $d>2$ dimensions, since
two or three points can by translations and/or rotations always be brought into
any predetermined plane. 

The conformal invariance of scale- and rotation-invariant systems is well
established. A convenient way to show this proceeds via the derivation of
Ward identities, invoking the (improved) energy-momentum tensor. These Ward
identities hold for systems with local interactions and it can be shown
that any $n$-point function which is translation-, rotation- and scale invariant
is automatically invariant under any projective conformal transformation,
see e.g. \cite{Card96,Drou88,Fran97,Henk99}. 

We have restricted ourselves to quasiprimary operators \cite{Bela84}, 
which is all what we shall need in this paper. 
 
\subsection{Schr\"odinger transformations} 

The Schr\"odinger group in $d+1$ dimensions 
is usually defined \cite{Nied72,Hage72} by the following
set of transformations
\BEQ \label{2:Schr}
\vec{r} \to \vec{r}' = 
\frac{{\cal R}\vec{r} + \vec{v}t + \vec{a}}{\gamma t+\delta} \;\; , \;\;
t \to t' = \frac{\alpha t+\beta}{\gamma t+\delta} \;\; ; \;\;
\alpha\delta - \beta\gamma =1
\EEQ
where $\alpha,\beta,\gamma,\delta,\vec{v},\vec{a}$ are real parameters and
$\cal R$ is a rotation matrix in $d$ spatial dimensions. 
The Schr\"odinger group can be obtained as a semi-direct product of the
Galilei group with the group $Sl(2,\mathbb{R})$ of the real projective
transformations in time. A faithful $d+2$-dimensional matrix representation is
\BEQ
{\cal L}_{g} = \left( \begin{array}{ccc} {\cal R} & \vec{v} & \vec{a} \\
0 & \alpha & \beta \\
0 & \gamma & \delta \end{array} \right) \;\; , \;\;
{\cal L}_{g} {\cal L}_{g'} = {\cal L}_{g g'}
\EEQ
According to Niederer \cite{Nied72}, the group (\ref{2:Schr}) is the largest
group which transforms any solution of the free Schr\"odinger equation
\BEQ \label{2:freiS}
\left( \II \frac{\partial}{\partial t} + \frac{1}{2 m} 
\frac{\partial}{\partial\vec{r}}\cdot\frac{\partial}{\partial\vec{r}}
\right) \psi =0
\EEQ
into another solution of (\ref{2:freiS}) through 
$(t,\vec{r})\mapsto g(t,\vec{r})$, $\psi\mapsto T_g \psi$
\BEQ \label{2:Schrpsi}
\left(T_g \psi\right)(t,\vec{r}) = 
f_g(g^{-1}(t,\vec{r}))\,\psi(g^{-1}(t,\vec{r}))
\EEQ
where \cite{Nied72,Perr77}
\BEQ \label{2:Schrf}
f_{g}(t,\vec{r}) = (\gamma t+\delta)^{-d/2} 
\exp\left[ -\frac{\II m}{2} \frac{\gamma \vec{r}^2+
2{\cal R}\vec{r}\cdot(\gamma\vec{a}-\delta\vec{v})+\gamma\vec{a}^2-
t\delta\vec{v}^2+2\gamma\vec{a}\vec{v}}{\gamma t+\delta}\right]
\EEQ
Independently, it was shown by Hagen \cite{Hage72} that the non-relativistic
free field theory is Schr\"odinger-invariant (see also \cite{Mehe00}). 
Furthermore, according to Barut \cite{Baru73}
the Schr\"odinger group in $d$ space dimension can be obtained by a group
contraction (where the speed of light $c\to\infty$) from the conformal
group in $d+1$ dimensions (this implies a certain rescaling of the mass as
well). Formally, one may go over to the diffusion 
equation by letting $m=(2\II D)^{-1}$, where $D$ is the diffusion constant. 

In order to implement the Galilei invariance of the free Schr\"odinger
equation and of a statistical system described by it, the wave function $\psi$
and the scaling operators $\phi_i$ of such a theory will under a Galilei
transformation pick up a complex phase as described by $f_{g}\ne 0$ and 
characterized by the mass $m$ \cite{Barg54,Levy67}. 
By analogy with conformal invariance \cite{Bela84}, we call those
scaling operators with the simplest possible transformation behaviour
under infinitesimal transformation {\em quasiprimary}, that is
$\delta_X \phi_i = -\eps X_n \phi_i$ and $\delta_Y \phi_i = -\eps Y_m \phi_i$.
In $d=1$ space dimensions, to which we restrict here for simplicity
(then ${\cal R}=1$), we have \cite{Henk94}
\BEA
X_n &=& -t^{n+1}\partial_t -\frac{n+1}{2} t^n r\partial_r -\frac{n(n+1)}{4}
{\cal M} t^{n-1} r^2 - \frac{x}{2}(n+1) t^n \nonumber \\
Y_m &=& -t^{m+1/2}\partial_r -\left( m+\frac{1}{2}\right) {\cal M} t^{m-1/2} r
\label{2:SchrGen} \\
M_n &=& -{\cal M} t^n \nonumber
\EEA
for a quasiprimary operator $\phi$ with scaling dimension $x$ and {\em `mass'}
${\cal M}=\II m$. Here $x$ and $\cal M$ are quantum numbers which can be used
to characterize the scaling operator $\phi$. Extensions to spatial 
dimensions $d>1$ are briefly described in appendix~B. 
Necessarily, any Schr\"odinger-invariant theory 
contains along with $\phi$ also the conjugate scaling
operator $\phi^*$, characterized by the pair $(x,-{\cal M})$. For
$x={\cal M}=0$, we recover the infinitesimal transformations of the Lie
group (\ref{2:Schr}). The commutation relations are
\BEA
\left[ X_n , X_m \right] &=& (n-m) X_{n+m} \;\; , \;\;
\left[ X_n , Y_m \right] \:=\: \left(\frac{n}{2}-m\right) Y_{n+m} \;\; , \;\;
\left[ X_n , M_m \right] \:=\: -m M_{n+m} \nonumber \\
\left[ Y_n , Y_m \right] &=& (n-m) M_{n+m} \;\; , \;\;
\left[ Y_n , M_m \right] \:=\: \left[ M_n, M_m \right] \:=\: 0
\label{2:SchAlg}
\EEA
in $d=1$ space dimensions. The infinitesimal generators of the
finite transformations (\ref{2:Schr}) are given by the set
$\{X_{\pm 1,0}, Y_{\pm 1/2}, M_0\}$. In this case the generator $M_0$ commutes
with the entire algebra. The eigenvalue $-{\cal M}$ of $M_0$ can be used along 
with the eigenvalue $\cal Q$ of
the quadratic Casimir operator \cite{Perr77}
\BEQ
Q := \left( 4 M_0 X_0 - 2 \{ Y_{-\frac{1}{2}}, Y_{\frac{1}{2}} \} \right)^2
-2\left\{ 2 M_0 X_{-1}- Y_{-\frac{1}{2}}^{2},
     2 M_0 X_{1} - Y_{\frac{1}{2}}^2 \right\} 
\EEQ
(where $\{A,B\}:=AB+BA$) 
to characterize the unitary irreducible (projective) representations of the
Lie algebra (\ref{2:SchAlg}) of the Schr\"odinger group \cite{Perr77}. 
One can show that the representations
with ${\cal Q}=0$ realized on scalar functions reproduce the transformation
(\ref{2:Schrpsi},\ref{2:Schrf}) \cite{Perr77}. Frequently, the algebra
with $M_n\ne 0$ is referred to as a centrally extended algebra. However, since
the algebra (\ref{2:SchAlg}) with $M_n=0$ (i.e. ${\cal M}=0$) is not 
semi-simple, its central extension is quite different from those of the 
conformal algebra (\ref{2:Vir0}). Since for the physical applications, we
shall need ${\cal M}\ne 0$ anyway, we shall refer to (\ref{2:SchAlg}) as the
Schr\"odinger Lie algebra {\it tout court} and avoid talking of any `central
extensions' in this context.

As was the case for $2D$ conformal transformations, one may write down
the generators $X_n,M_n$ for any $n\in\mathbb{Z}$ and $Y_m$ for any 
$m\in\mathbb{Z}+\frac{1}{2}$ such that (\ref{2:SchAlg}) 
remains valid \cite{Henk92,Henk94}.

By definition \cite{Henk94}, $n$-point functions $G$ of 
quasiprimary scaling operators $\phi_i$ with respect to
the Schr\"odinger group satisfy
\BEQ \label{2:SchrKov}
X_{n} G = Y_{m} G = 0
\EEQ
with $n=-1,0,1$ and $m=-1/2,+1/2$. 
Consequently, the only non-vanishing two-point function of scalar quasiprimary
scaling operators is, for any spatial dimension $d\geq 1$,
\BEQ \label{2:S2ponto}
\langle \phi_1(t_1,\vec{r}_1)\phi_2^*(t_2,\vec{r}_2)\rangle
= G_{12}\,\delta_{x_1,x_2}\, \delta_{{\cal M}_1,{\cal M}_2} \:
(t_1-t_2)^{-x_1} \exp\left[ - \frac{{\cal M}_1}{2}
\frac{(\vec{r}_1-\vec{r}_2)^2}{t_1-t_2}
\right] \;\; ; \;\; t_1 > t_2
\EEQ
whereas $\langle \phi\phi\rangle=\langle\phi^*\phi^*\rangle=0$ provided
${\cal M}_1\ne 0$ \cite{Henk94}.
Usually, the normalization constant $G_{12}=1$. Comparison with the
form (\ref{2:Skal}) gives the scaling function 
$\Phi(u)\sim e^{-{\cal M}u^2/2}$. 
Similarly, the basic non-vanishing three-point function of 
quasiprimary operators reads \cite{Henk94}
\BEA
\lefteqn{ 
\langle\phi_1(t_1,\vec{r}_1)\phi_2(t_2,\vec{r}_2)
\phi_3^*(t_3,\vec{r}_3)\rangle
= \delta_{{\cal M}_a+{\cal M}_b,{\cal M}_c} \:
t_{13}^{-x_{132}/2} t_{23}^{-x_{231}/2} t_{12}^{-x_{123}/2} } \nonumber\\
&\times& 
\exp\left[-\frac{{\cal M}_1}{2}\frac{\vec{r}_{13}^2}{t_{13}}
-\frac{{\cal M}_2}{2}\frac{\vec{r}_{23}^2}{t_{23}} \right]
{\cal G}_{12,3}\left(\frac{(\vec{r}_{13}t_{23}-\vec{r}_{23}t_{13})^2}
{t_{12}t_{13}t_{23}}
\right) \;\; ; \;\; t_1 > t_3 \;, \; t_2 > t_3 \label{2:S3ponto}
\EEA
with $t_{ab} = t_a-t_b$, $r_{ab}=r_a-r_b$, $x_{abc}=x_a+x_b-x_c$ 
and ${\cal G}_{ab,c}$ is an arbitrary differentiable scaling function. 
A similar expression holds for $\langle\phi\phi^*\phi^*\rangle$, while
$\langle\phi\phi\phi\rangle=\langle\phi^*\phi^*\phi^*\rangle=0$, unless the
`mass' $\cal M$ of the scaling operator $\phi$ vanishes. 

It is instructive to compare the form of the two- and three-point functions
(\ref{2:ZweiP},\ref{2:DreiP}) as obtained from conformal invariance with the
expressions (\ref{2:S2ponto},\ref{2:S3ponto}) following from Schr\"odinger
invariance. As might have been anticipated from comparing the finite
transformations (\ref{2:Moeb}) and (\ref{2:Schr}), the dependence on $z_{ab}$
and $t_{ab}$, respectively, is identical provided the scaling dimensions $x_i$
are replaced by $x_i/\theta$. For the two-point
function, we have in both cases the constraint $x_a=x_b$ and one could
extend the arguments of \cite{Scha75} on $n$-point functions between
derivatives of quasiprimary operators from conformal to Schr\"odinger
invariance. On the other hand, Schr\"odinger invariance
yields the constraints ${\cal M}_a={\cal M}_b$ for the two-point function
$\langle \phi_a\phi_b^*\rangle$ and ${\cal M}_a+{\cal M}_b={\cal M}_c$ for the
three-point function $\langle\phi_a\phi_b\phi_c^*\rangle$. These are examples
of the Bargmann superselection rules \cite{Barg54} and follow already from
Galilei invariance \cite{Levy67,Giul96}. It follows from the Bargmann
superselection rules that no Galilean scaling operator can be hermitian unless
it is massless. The `mass' $\cal M$ therefore plays quite a different role
in (non-relativistic) Galilean theories as 
compared to relativistic ones. It no
longer measures a deviation from criticality, but should rather be considered
as the analogue of a conserved charge. Finally, the explicit from of the
scaling functions in (\ref{2:S2ponto},\ref{2:S3ponto}) depends on the way
the Galilei transformation is realized.\footnote{For example, one may modify
the generators (\ref{2:lPhi}) and (\ref{2:SchrGen}) to emulate the effect
of a discrete lattice with lattice constant $a$. This works for free fields
for both Schr\"odinger \cite{Henk94a} and conformal \cite{Henk98} invariance.
In the context of conformal invariance, the best-known example of the 
dependence of the correlators on the realization are the logarithmic 
conformal field theories, see \cite{Floh01,Rahi01} for recent reviews.}

So far, we have always considered both space and time to be infinite in extent.
In some applications, however, one is interested in situations where the
system is `prepared 'at $t=0$ and is then allowed to `evolve' for positive 
times $t>0$. We must then ask which subset of the Schr\"odinger transformations
will leave the $t=0$ boundary condition invariant as well. Indeed, inspection
of the generators (\ref{2:SchrGen}) shows that the line $t=0$ is only modified
by $X_{-1}$ and that furthermore the subset
$\{X_{0,1},Y_{\pm 1/2},M_0\}$ closes. We may therefore impose the covariance
conditions (\ref{2:SchrKov}) with $n=0,1$ and $m=\pm 1/2$ {\em only} 
\cite{Henk94}. Then the two-point function is
\BEQ \label{2:S2ponto-surf}
\langle \phi_1(t_1,\vec{r}_1)\phi_2^*(t_2,\vec{r}_2)\rangle
= G_{12}\, \delta_{{\cal M}_1,{\cal M}_2} \:
\left(\frac{t_1}{t_2}\right)^{(x_2-x_1)/2}\, (t_1-t_2)^{-x_1} 
\exp\left[ - \frac{{\cal M}_1}{2}\frac{(\vec{r}_1-\vec{r}_2)^2}{t_1-t_2}
\right] \;\; ; \;\; t_1 > t_2 > 0
\EEQ
Compared to eq.~(\ref{2:S2ponto}), there is no more a constraint on the 
exponents $x_{1,2}$, because time translation invariance was no longer assumed.

Although Schr\"odinger invariance of $n$-point functions was imposed at the
beginning of this section {\it ad hoc}, there exist by now quite a few
critical statistical systems with $\theta=2$ where the predictions
(\ref{2:S2ponto},\ref{2:S3ponto},\ref{2:S2ponto-surf}) have been reproduced
\cite{Henk94,Henk94a}. Models where some Green's functions coincide with the 
expressions found from Schr\"odinger invariance include the kinetic Ising 
model with Glauber dynamics \cite{Glau63}, the symmetric exclusion process
\cite{Kand90,Schu94}, the symmetric and asymmetric non-exclusion processes
\cite{Schu94}, a reaction-diffusion model of a single species with reactions
$2A \leftrightarrow 2\emptyset$ \cite{Gryn94} and in the 
axial next-nearest
neighbour spherical model (ANNNS model \cite{Selk92}) {\em at\/} its Lifshitz 
point \cite{Frac93}. In section 5, we shall consider in detail the phase
ordering kinetics of the $2D$ and $3D$ Glauber Ising model and several
variants of the kinetic spherical model with a non-conserved order parameter,
see \cite{Henk01}. Finally, for short-ranged interactions, one may invoke a
Ward identity to prove that invariance under spatio-temporal 
translations, Galilei transformations and dilatations with
$\theta=2$ automatically imply invariance under the `special' Schr\"odinger
transformations \cite{Henk94}. 

%%%%%%%%%%%%%%%%%%%%%%%%%%%%%%%%%%%%%%%%%%%%%%%%%%%%%%%%%%%%%%%%%%%%%%%%%%%%%%%%
\section{Infinitesimal local scale transformations}
%%%%%%%%%%%%%%%%%%%%%%%%%%%%%%%%%%%%%%%%%%%%%%%%%%%%%%%%%%%%%%%%%%%%%%%%%%%%%%%%

We want to construct local space-time transformations which are compatible
with the strongly anisotropic scaling (\ref{1:Ani}) for a given anisotropy
exponent $\theta\ne 1$. The first step in such an undertaking must be
the construction of the analogues of the projective conformal transformations
(\ref{2:Moeb}) and the Schr\"odinger transformations (\ref{2:Schr}). This,
and the derivation and testing of some simple consequences, is the aim of this
paper. A brief summary of some aspects of this construction was already given
in \cite{Henk97,Henk99}. The question whether these `projective' 
transformations can be extended towards some larger algebraic structure 
will be left for future work. 

For simplicity of notation, we shall work in $d=1$ space dimensions throughout.
Extensions to $d>1$ will be obvious. 

\subsection{Axioms of local scale invariance} 

Given the practical success of both conformal ($\theta=1$) and Schr\"odinger 
($\theta=2$) invariance, we shall try to remain as close as possible to these. 
Specifically, our attempted construction is 
based on the following requirements.
They are the defining axioms of our notion of {\em local scale invariance}. 
\begin{enumerate}
\item For both conformal and Schr\"odinger invariance,
 M\"obius transformations
play a prominent role. We shall thus seek space-time transformations such
that the time coordinate undergoes a M\"obius transformation
\BEQ \label{3:Moeb}
t\to t' = \frac{\alpha t + \beta}{\gamma t +\delta} \;\; ; \;\;
\alpha\delta - \beta\gamma =1
\EEQ
If we call the infinitesimal generators of these transformations $X_n$, 
($n=-1,0,1$), we require that even after the transformations on the spatial 
coordinates $\vec{r}$ are included, the commutation relations
\BEQ \label{3:XComm}
\left[ X_n , X_m \right] = (n-m) X_{n+m}
\EEQ
remain valid. Scaling operators which transform 
covariantly under (\ref{3:Moeb}) 
are called {\em quasiprimary}, by analogy with the notion of conformal 
quasiprimary operators \cite{Bela84}. 
\item The generator $X_0$ of scale transformations is
\BEQ \label{3:X0gen}
X_0 = - t \partial_t - \frac{1}{\theta} r \partial_r - \frac{x}{\theta}
\EEQ
where $x$ is the scaling dimension of the quasiprimary operator on which
$X_0$ is supposed to act. 
\item Spatial translation invariance is required. 
\item When acting on a quasiprimary operator 
$\phi$, extra terms coming from the scaling dimension of $\phi$ must be present
in the generators and be compatible with (\ref{3:X0gen}). 
\item By analogy with the `mass' terms contained in the generators 
(\ref{2:SchrGen}) for $\theta=2$, mass terms constructed such as to be 
compatible with $\theta\ne 1,2$ should be expected to be present. 
\item We shall test the notion of local scale invariance by
calculating two-point functions of quasiprimary operators and comparing them 
with explicit model results (see section 5). We require that the generators 
when applied to a quasiprimary two-point function will yield a {\em finite} 
number of independent conditions. 

The simplest way to satisfy this is the requirement that the generators
applied to a two-point function provide a realization of a finite-dimensional
Lie algebra. However, more general ways of finding non-trivial two-point
functions are possible. 
\end{enumerate}
\subsection{Construction of the infinitesimal generators}

The generators $X_n$ which realize (\ref{3:XComm}) will be of the form
\BEQ
X_n = X_n^{(I)} + X_n^{(II)} + X_n^{(III)} 
\EEQ
where $X_n^{(I)}=-t^{n+1}\partial_t$ is the infinitesimal form of 
(\ref{3:Moeb}), $X_n^{(II)}$ contains the action on $r$ and the scaling 
dimensions  while $X_{n}^{(III)}$ will contain the mass terms. From $X_n^{(I)}$,
we already have the commutation relations (\ref{3:XComm}) and
$X_n^{(II,III)}$ will be constructed such as to keep these intact. 

We now find $X_n^{(II)}$. Since for time translations, $X_{-1}=-\partial_t$ and
using (\ref{3:X0gen}), we make the ansatz
\BEQ \label{3:XIIansatz}
X_n = - t^{n+1}\partial_t -  a_n(t,r)\partial_r - b_n(t,r)
\EEQ
and have the initial conditions
\BEQ \label{3:XIIansatz_Ini}
a_{-1} = b_{-1} = 0 \;\; , \;\; a_0(t,r) = \frac{r}{\theta} \;\; , \;\;
b_0(t,r) = \frac{x}{\theta}
\EEQ
To have consistency with (\ref{3:XComm}), we set first $m=-1$, yielding
$[X_n,X_{-1}]=(n+1)X_{n-1}$. This gives the conditions
\BEQ
\frac{\partial a_n}{\partial t} =(n+1) a_{n-1} \;\; , \;\;
\frac{\partial b_n}{\partial t} =(n+1) b_{n-1} 
\EEQ
with the solutions
\BEA
a_n(t,r) &=& \sum_{k=0}^{n} \left(\vekz{n+1}{k+1}\right) A_k(r) t^{n-k} 
\nonumber\\
b_n(t,r) &=& \sum_{k=0}^{n} \left(\vekz{n+1}{k+1}\right) B_k(r) t^{n-k}
\label{3:ab}
\EEA
where $A_n(r)$ and $B_n(r)$ are independent of $t$ and $A_0(r)=r/\theta$ and
$B_0(r)=x/\theta$. Next, we set
$m=0$ in (\ref{3:XComm}), yielding $[X_n,X_0]=n X_n$ and thus obtain the 
conditions
\BEQ \label{3:X0ab}
\left( n+\frac{1}{\theta}\right) a_n = 
t \frac{\partial a_n}{\partial t}
+ \frac{1}{\theta} r \frac{\partial a_n}{\partial r} \;\; , \;\;
n b_n = 
t \frac{\partial b_n}{\partial t}
+ \frac{1}{\theta} r \frac{\partial b_n}{\partial r} 
\EEQ
Inserting (\ref{3:ab}), we easily find
\BEQ \label{3:XIIansatz_AB}
A_k(r) = A_{k0} r^{\theta k +1} \;\; , \;\;
B_k(r) = B_{k0} r^{\theta k}
\EEQ
where $A_{k0}, B_{k0}$ are constants and $A_{00}=1/\theta$, $B_{00}=x/\theta$. 
The values of these constants are found from the condition 
$[X_n,X_1]=(n-1)X_{n+1}$. Using the explicit forms
$a_1(t,r)=2\theta^{-1}tr+A_{10}r^{\theta+1}$ and 
$b_1(t,r)=2x\theta^{-1}t+B_{10}r^{\theta}$, we obtain
\BEA
t\left( t\frac{\partial a_n}{\partial t} +
\frac{2}{\theta}\left( r\frac{\partial a_n}{\partial r} - a_n - t^n r\right) 
+ A_{10} U \left( r \frac{\partial a_n}{\partial r} -(\theta+1)a_n \right) 
\right)  &=& (n-1) a_{n+1}
\nonumber \\
t\left( t\frac{\partial b_n}{\partial t} + 
\frac{2}{\theta}\left( r\frac{\partial b_n}{\partial r} - xt^n\right)
+ A_{10} U r\frac{\partial b_n}{\partial r}  - 
B_{10} \theta U  r^{-1} a_n \right)  &=& 
(n-1)b_{n+1}
\label{3:X1ab}
\EEA
where $U=r^{\theta}/t$. Using (\ref{3:X0ab}), the first of these becomes
\BEQ
t\left( \frac{1}{\theta} r\frac{\partial a_n}{\partial r} 
+ \left( n-\frac{1}{\theta}\right) a_n - \frac{2}{\theta} t^n r 
+A_{10} U \left( r\frac{\partial a_n}{\partial r} -(\theta+1) a_n \right)
\right) = (n-1) a_{n+1}
\EEQ 
Insertion of the explicit form of the $a_n$ known from 
(\ref{3:ab},\ref{3:XIIansatz_AB}) leads to (terms with $k=0,1$ cancel)
\BEA 
& &\sum_{k=2}^{n}\left[ 
\left( \left(\vekz{n+1}{k+1}\right) (n+k) - 
\left(\vekz{n+2}{k+1}\right)(n-1) \right) A_{k0} 
+\theta \left(\vekz{n+1}{k}\right)(k-2) A_{k-1,0}A_{10}\right] U^k
 \nonumber \\
&& +\left[ (n-1)\left( \theta A_{n0}A_{10} - A_{n+1,0}\right)
\right] U^{n+1} =0
\label{3:BedingungAn}
\EEA
which must be valid for all values of $U$. This leads to the conditions
\BEQ \label{3:BedAn1}
\left[ \left( \left(\vekz{n+1}{k+1}\right) (n+k) - 
\left(\vekz{n+2}{k+1}\right)(n-1) \right) A_{k0} 
+\theta \left(\vekz{n+1}{k}\right)(k-2) A_{k-1,0}A_{10}\right] = 0 
\EEQ
\BEQ \label{3:BedAn2}
A_{n+1,0} = \theta A_{n0} A_{10} \;\; ; \;\; \forall n\geq 2
\EEQ
{}From (\ref{3:BedAn2}), we have $A_{n0}=\theta^{n-2} A_{20} A_{10}^{n-2}$ for
all $n\geq 2$. Inserting this into (\ref{3:BedAn1}), this is automatically
satisfied for all $k\geq 2$ because of the identity 
\BEQ \label{3:BinomId}
\left(\vekz{n+1}{k+1}\right)(n+k) - \left(\vekz{n+2}{k+1}\right)(n-1) +
\left(\vekz{n+1}{k}\right)(k-2) = 0 
\EEQ
In particular, $A_{20}$ remains arbitrary. Finally, using the identity
\BEQ
\sum_{k=2}^{n} \left(\vekz{n+1}{k+1}\right) x^k = \frac{(1+x)^{n+1}-1}{x}
-(n+1) - \frac{1}{2}n(n+1)x
\EEQ
and using the initial conditions for $a_n$, we obtain the closed form
\BEA
\lefteqn{ 
a_n(t,r) = \sum_{k=0}^{n} \left(\vekz{n+1}{k+1}\right) A_{k0} r^{\theta k+1}
t^{n-k} } \label{3:anAusd} \\
&=& \left( \frac{n+1}{\theta}t^n r + 
\frac{1}{2}n(n+1)A_{10} t^{n-1} r^{\theta+1} \right)
\left( 1 - \frac{A_{20}}{\theta A_{10}^2}\right) 
+ \frac{A_{20}}{(\theta A_{10})^3} t^{n+1}r^{1-\theta} 
\left[ \left(1+\theta A_{10}r^{\theta}/t\right)^{n+1} -1\right] 
\nonumber 
\EEA
which depends on the three free parameters $A_{10}, A_{20}$ and $\theta$.

Next, using (\ref{3:X0ab}), the second of the relations
(\ref{3:X1ab}) becomes
\BEQ
t\left( \frac{r}{\theta} \frac{\partial b_n}{\partial r} + n b_n -
\frac{2x}{\theta} t^n + A_{10} U r \frac{\partial b_n}{\partial r} 
- B_{10} U r^{-1} a_n \right) = (n-1) b_{n+1}
\EEQ
which we now analyse. Inserting the form (\ref{3:ab}) for $a_n$ and $b_n$,
we obtain the condition
\BEA
& &  
\sum_{k=1}^{n} \left[ \left(\vekz{n+1}{k+1}\right)(n+k) B_{k0} + \theta
\left(\vekz{n+1}{k}\right) \left( (k-1)B_{k-1,0}A_{10} - A_{k-1,0}B_{10}\right)
\right. 
\nonumber \\
& & \left.-\left(\vekz{n+2}{k+1}\right)(n-1) B_{k0} \right] U^k \; 
+ \left( (n+1)n -(n-1)(n-2) -2\right) \frac{x}{\theta} \\
&+& \left[\theta\left( n B_{n0}A_{10} - A_{n0} B_{10}\right) - 
(n-1) B_{n+1,0} \right] U^{n+1}  = 0 \nonumber 
\EEA
which must be valid for all values of $U$. Now the term of order $O(U^0)$
vanishes and the term of order $O(U^{n+1})$ yields the recurrence
\BEA
B_{n+1,0} &=& \frac{\theta}{n-1}\left( n B_{n0}A_{10} - A_{n0} B_{10}\right)
\nonumber \\
&=& \frac{\theta}{n-1} \left( n B_{n0}A_{10} - 
\theta^{n-2} A_{10}^{n-2} A_{20} B_{10} \right)
\EEA
If we insert this into the above condition for terms of order $O(U^k)$, 
$k=1,\ldots,n$ and use again the identity (\ref{3:BinomId}), we see that all
these terms vanish. The final solution for the coefficients $B_{n0}$ is
\BEQ
B_{n0} = (n-1) \left(\theta A_{10}\right)^{n-2} B_{20} 
-(n-2) \theta^{n-2} A_{10}^{n-3} A_{20} B_{10}
\EEQ
and where $B_{10}$ and $B_{20}$ remain free parameters. Using the identities
\BEA
\sum_{k=2}^{n} \left(\vekz{n+1}{k+1}\right) (k-1) x^k 
&=& \frac{((n-1)x-2)(1+x)^n}{x} + \frac{2}{x} +(n+1) 
\nonumber \\
\sum_{k=2}^{n} \left(\vekz{n+1}{k+1}\right) (k-2) x^k 
&=& \frac{((n-2)x-3)(1+x)^n}{x} + \frac{3}{x} +2(n+1) +\frac{1}{2}n(n+1)x
\EEA
and the initial conditions for the $b_n$, 
the closed form for $b_{n}(t,r)$ reads
\BEA
\lefteqn{ 
b_{n}(t,r) = \frac{n+1}{\theta} x t^n 
+\frac{n(n+1)}{2} t^{n-1}r^{\theta} B_{10} 
\left( 1 - \frac{A_{20}}{\theta A_{10}^2}\right)} 
\nonumber \\
&+& t^n \frac{A_{10}B_{20} -2A_{20}B_{10}}{\theta^2 A_{10}^3} 
\left[ (n+1)+(n-1)\left(1+\theta A_{10} r^{\theta}/t\right)^n\right]
\nonumber \\
&+& t^{n+1} r^{-\theta} \frac{2A_{10}B_{20}-3A_{20}B_{10}}{\theta^3 A_{10}^4}
\left[ 1-\left(1+\theta A_{10} r^{\theta}/t\right)^n\right]
\label{3:bnAusd} \\
&+& n t^n \frac{A_{20}B_{10}}{\theta^2 A_{10}^3} 
\left(1+\theta A_{10} r^{\theta}/t\right)^n
\nonumber 
\EEA
and depends on the free parameters $A_{10}, A_{20}, B_{10}, B_{20}$ and 
$\theta$.   

The results obtained so far give the most general form for the generators
$X_n$ satisfying $[X_n,X_m]=(n-m)X_{n+m}$ with $m=-1,0,1$ and $n\in\mathbb{Z}$
and with $X_0$ given by (\ref{3:X0gen}). 
If we were merely interested in the subalgebra spanned
by $\{X_{-1},X_0,X_1\}$, we could simply set $A_{20}=B_{20}=0$, since those
parameters do not enter anyway in these three generators. 

We now inquire the additional conditions needed for 
$[X_n,X_m]=(n-m)X_{n+m}$ to hold with $n,m\in\mathbb{Z}$. Given the
complexity of the expressions (\ref{3:anAusd},\ref{3:bnAusd}) for 
$a_n$ and $b_n$, respectively, it is helpful to start with an example. A
straightforward calculation shows that
\BEQ
\left[ X_3, X_2 \right] = X_5 + \theta^2 r^{5\theta} 
\left( \theta A_{10}^2 - A_{20} \right) \left\{
A_{10} A_{20}  r \partial_r + 
\left( 4 A_{10} B_{20} - 3 A_{20} B_{10} \right) \right\}
\EEQ
The extra terms on the right must vanish. This leads to the distinction
of four cases which are collected in the following table
\begin{center}
\begin{tabular}{l|cccc}  
   & $A_{10}$ & $A_{20}$          & $B_{10}$ & $B_{20}$ \\ \hline
1. & $\ne 0$  & $\theta A_{10}^2$ & $\ne 0$  & $\ne 0$  \\
2. & $0$      & $0$               & $\ne 0$  & $\ne 0$  \\
3. & $\ne 0$  & $0$               & $\ne 0$  & $0$      \\
4. & $0$      & $\ne 0$           & $0$      & $\ne 0$  \\
\end{tabular}
\end{center}
and we now have to see to what extent these necessary conditions are also
sufficient. Indeed, straightforward but tedious explicit computation of the
commutator $[X_n,X_m]$ shows that it is equal to $(n-m)X_{n+m}$ in all four
cases. While for case 1 the expressions for $a_n$ and $b_n$ are still lengthy,
they simplify for the three other cases
\BEQ
a_n(t,r) = \left\{ 
\begin{array}{ll}
\theta^{-1} (n+1) t^n r & \mbox{\rm ~;~~ case 2} \\
\theta^{-1} (n+1) t^n r + \frac{1}{2}n(n+1) t^{n-1} r^{\theta+1} A_{10} 
& \mbox{\rm ~;~~ case 3} \\
\theta^{-1} (n+1) t^n r + \frac{1}{6}(n^3-n) t^{n-2} r^{2\theta+1} A_{20} 
& \mbox{\rm ~;~~ case 4} 
\end{array} \right.
\EEQ
\BEQ
b_n(t,r) = \frac{(n+1)}{\theta} x t^n + \left\{ 
\begin{array}{ll}
\frac{n(n+1)}{2} t^{n-1}r^{\theta} B_{10} 
+ \frac{n^3-n}{6} t^{n-2} r^{2\theta} B_{20}  
& \mbox{\rm ~;~~ case 2} \\
\frac{n(n+1)}{2} t^{n-1}r^{\theta} B_{10} 
& \mbox{\rm ~;~~ case 3} \\
\frac{n^3-n}{6} t^{n-2} r^{2\theta} B_{20}  
& \mbox{\rm ~;~~ case 4} 
\end{array} \right.
\EEQ

\noindent We summarize our result as follows. 

\noindent {\bf Proposition 1:} {\it The generators}
\BEQ \label{3:XIIgen}
X_n = - t^{n+1} \partial_t - a_n(t,r) \partial_r - b_n(t,r)
\EEQ
{\it where}
\BEA 
a_n(t,r)   
&=& \left( \frac{n+1}{\theta}t^n r + 
\frac{1}{2}n(n+1)A_{10} t^{n-1} r^{\theta+1} \right)
\left( 1 - \frac{A_{20}}{\theta A_{10}^2}\right) 
\nonumber \\
& & + \frac{A_{20}}{(\theta A_{10})^3} t^{n+1}r^{1-\theta} 
\left[ \left(1+\theta A_{10}r^{\theta}/t\right)^{n+1} -1\right] 
\EEA
{\it and} 
\BEA
b_{n}(t,r) &=& \frac{n+1}{\theta} x t^n 
+\frac{n(n+1)}{2} t^{n-1}r^{\theta} B_{10} 
\left( 1 - \frac{A_{20}}{\theta A_{10}^2}\right) 
+ n t^n \frac{A_{20}B_{10}}{\theta^2 A_{10}^3} 
\left(1+\theta A_{10} r^{\theta}/t\right)^n
\nonumber \\
& &+ t^n \frac{A_{10}B_{20} -2A_{20}B_{10}}{\theta^2 A_{10}^3} 
\left[ (n+1)+(n-1)\left(1+\theta A_{10} r^{\theta}/t\right)^n\right]
\nonumber \\
& &+ t^{n+1} r^{-\theta} \frac{2A_{10}B_{20}-3A_{20}B_{10}}{\theta^3 A_{10}^4}
\left[ 1-\left(1+\theta A_{10} r^{\theta}/t\right)^n\right]
\EEA
{\it and where one of the following conditions}
\BEA
(1.)   & & A_{10} \ne 0 \;\; , \;\; A_{20} = \theta A_{10}^2 \;\; , \;\;
B_{10}\ne 0 \;\; , \;\; B_{20}\ne 0
\nonumber \\ 
(2.)  & & A_{10} = A_{20} = 0 \;\; , \;\;  
B_{10}\ne 0 \;\; , \;\; B_{20}\ne 0
\nonumber \\ 
(3.) & & A_{10} \ne 0 \;\; , \;\; A_{20} = 0 \;\; , \;\; 
B_{10} \ne 0 \;\; , \;\; B_{20} = 0 
\label{3:CondiAB} \\
(4.) & & A_{10}= 0 \;\; , \;\; A_{20} \ne 0 \;\; , \;\; 
B_{10} = 0 \;\; , \;\; B_{20} \ne  0
\nonumber 
\EEA
{\it holds, are the most general linear (affine) first-order operators in 
$\partial_t$ and $\partial_r$ consistent with the axioms 1 and 2 and 
which satisfy the commutation relations 
$[X_{n},X_{m}]=(n-m)X_{n+m}$ for all $n,m\in\mathbb{Z}$. If only the 
subalgebra $\{X_{\pm 1,0}\}$ is considered, $A_{10}$ and $B_{10}$ remain 
arbitrary and the generators $X_{\pm 1,0}$ are those of case 3.}

Before we construct the mass terms, we consider space translations, generated
by $-\partial_r$. There will be a second set of generators
\BEQ
Y_m = Y_{m}^{(II)} + Y_{m}^{(III)}
\EEQ
where $Y_{m}^{(II)}$ contains the action on $r$ and $Y_{m}^{(III)}$ contains
the mass terms. Given the form (\ref{2:SchrGen}) of the generators of the
Schr\"odinger algebra, we do not expect any terms proportional to
$\partial_t$ to be present in the $Y_m$. Indeed, if we tried to include
terms of this form, it is easy to see that one were back to the case
$\theta=1$, that is conformal invariance. 

The following notation will be useful. Let
\BEQ
\theta = 2/N
\EEQ
which defines $N$. We write (up to mass terms to be included later), 
\BEQ \label{3:YIIgen}
Y_m = Y_{k-N/2} = - \frac{2}{N(k+1)}\left( 
\frac{\partial a_k(t,r)}{\partial r}\partial_r 
+\frac{\partial b_k(t,r)}{\partial r} \right) 
\EEQ
where $m=-\frac{N}{2} + k$ and $k$ is an integer. 
Here, $a_n$ and $b_n$ are those of the proposition 1. 
In particular, $Y_{-N/2}=-\partial_r$ and $Y_{n-N/2}$ is obtained from 
$[X_n,Y_{-N/2}]=\frac{1}{2}N(n+1) Y_{-N/2+n}$. 
If $A_{10}$, $A_{20}$, $B_{10}$, $B_{20}$
would all vanish, we have indeed $[X_n,Y_m]=(\frac{1}{2}Nn-m)Y_{n+m}$
and we now look for the conditions on the parameters which will retain
this commutator for all values of $n$ and $m$. 
 
For the general situation given by (\ref{3:CondiAB}), direct calculations show
that
\BEQ
\left[ X_{-1}, Y_m \right] = \left( -\frac{N}{2} - m \right) Y_{m-1} \;\; , \;\;
\left[ X_{0}, Y_m \right] = - m  Y_{m}
\EEQ
throughout, but the commutator with $X_1$ is 
more complicated. We shall consider the four cases one by one. 

{\bf 1.} For case 1, we consider
\BEQ
K_{1k} := \left[ X_1, Y_{k-N/2} \right] - \left( N-k\right) Y_{k+1-N/2}
\EEQ
Since this is still very complex, we expand in $A_{10}$ and find
\BEA
K_{1k} &=& -\frac{2\theta(\theta-2)}{3}k B_{20} t^{k-1} r^{2\theta-1} 
- \theta^2 k A_{10} B_{10} t^{k-1} r^{2\theta -1}
\nonumber \\
& & + \frac{\theta^2(7-5\theta)}{6} k(k-1) t^{k-2} r^{3\theta-1} A_{10} B_{10}
+ O\left( A_{10}^2\right)
\EEA 
Therefore, if $A_{10}, B_{10}, B_{20}$ are all independent, it follows that
$B_{10}=B_{20}=0$ (the other possibility $A_{10}=B_{20}=0$ reduces to a
special case of either case 2 or 3 and will be treated below). In this
case, we expand further and find
\BEQ
K_{1k} = - \frac{\theta(4\theta+1)(\theta-1)}{6} k A_{10}^2 t^{k-1} r^{2\theta}
\partial_r + O\left( A_{10}^3\right)
\EEQ
Therefore, $A_{10}=0$ and the case 1 has become trivial, unless $\theta=1$. 
On the other hand, for $\theta=1$ there is a non-trivial solution of the
$K_{1k}=0$, namely $B_{20}=\frac{3}{2} A_{10} B_{10}$. It is now 
straightforward to check that the algebra of the generators $X_n, Y_m$ indeed
closes for all values of $n$ and $m$. 

{\bf 2.} For case 2, we have
\BEQ
K_{1k} = -\frac{2\theta(\theta-2)}{3}k B_{20} t^{k-1} r^{2\theta-1} 
\EEQ
which implies either $B_{20}=0$ for generic $\theta$ or else $\theta=2$ and
$B_{20}$ arbitrary. The remaining commutators $[X_n, Y_m]$  
are equal to $(nN/2-m)Y_{n+m}$ for both possibilities.  

{\bf 3.} For case 3, we have
\BEQ
K_{1k} = -\frac{\theta(\theta+1)}{2}k A_{10}^2 t^{k-1} r^{2\theta} \partial_r 
- \theta^2 k t^{k-1} r^{2\theta-1} A_{10} B_{10} 
\EEQ
which implies $A_{10}=0$. We therefore recover the case 2. 

{\bf 4.} Finally, for the case 4, we have 
\BEQ
K_{1k} = -\frac{(2\theta+1)(\theta-2)}{3}k A_{20} t^{k-1} r^{2\theta}\partial_r 
-\frac{2\theta(\theta-2)}{3} k B_{20} t^{k-1} r^{2\theta-1}
\EEQ
and therefore for generic $\theta$, we must have $A_{20}=B_{20}=0$ which is
trivial or else we must have $\theta=2$. In that last case, we consider
\BEQ
\left[ X_2, Y_{k-N/2} \right] = \left( \frac{3}{2}N -k\right) Y_{k+2-N/2}
-\frac{5}{3}k(k-1) A_{20}^2 t^{k-2} r^8 \partial_r 
-\frac{8}{3}k(k-1) A_{20} B_{20} t^{k-2} r^7 
\EEQ
and the extra terms on the right only vanish if $A_{20}=0$. This reproduces
a special situation of case 2.

In conclusion, the unwanted extra terms in $[X_n, Y_m]$ are eliminated in 
three cases, namely
\BEA
\mbox{\rm (i)}   & & \mbox{\rm $N$ generic} \;\; , \;\; B_{10}\ne 0 \;\; , \;\;
B_{20} =  0 \;\; , \;\; A_{10} = 0 \;\; , \;\; A_{20} = 0\nonumber \\
\mbox{\rm (ii)}  & & \mbox{\rm $N=1$~~~~~} \,\; , \;\; B_{10}\ne 0 \;\; , \;\;
B_{20}\ne 0 \;\; , \;\; A_{10} = 0 \;\; , \;\; A_{20} = 0 
\label{3:YBedAB} \\
\mbox{\rm (iii)} & & \mbox{\rm $N=2$~~~~~} \,\; , \;\; B_{10}\ne 0 \;\; , \;\;
B_{20}= \frac{3}{2} A_{10}B_{10} \;\; , \;\; A_{10} \ne 0 \;\; , \;\; 
A_{20}= A_{10}^2\nonumber 
\EEA
For the three cases (\ref{3:YBedAB}) we list the explicit form of the 
generators $X_n$ with $n\in\mathbb{Z}$ and $Y_m$ with $m=k-N/2$ and
$k\in\mathbb{Z}$ in table~\ref{tab0}. In all three cases, 
the generators depend on two free parameters. 

%%~~~~~~~~~~~~~~~~~~~~~~~~~~~~~~~~~~~~~~~~~~~~~~~~~~~~~~~~~~~~~~~~~~~~~~~~~~~~~~
\begin{table}
\caption{Generators $X_n$ and $Y_{k-N/2}$ without mass terms and 
with $n,k\in\mathbb{Z}$ according to the conditions (i), (ii), (iii) of 
{\protect eq.~(\ref{3:YBedAB})}.\label{tab0}}
\begin{center}
\begin{tabular}{|l|lcl|} \hline
(i) & $X_n$ & = & $-t^{n+1}\partial_t - \frac{n+1}{2}Nt^{n} r\partial_r 
-\frac{(n+1)x}{2}Nt^n - \frac{n(n+1)}{2}B_{10} t^{n-1} r^{2/N}$  \\[\TYa]
    & $Y_{k-N/2}$ &  = & $-t^k\partial_r 
      - \frac{2}{N^2}k B_{10} t^{k-1}r^{-1+2/N}$  \\[\TYb]
(ii) & $X_n$ & = & $-t^{n+1}\partial_t - \frac{1}{2}(n+1)t^{n} r\partial_r 
-\frac{1}{2}(n+1)xt^n - \frac{n(n+1)}{2}B_{10} t^{n-1} r^2 
-\frac{(n^2-1)n}{6} B_{20} t^{n-2} r^{4}$ \\[\TYa]
    & $Y_{k-1/2}$ &  = & $-t^k\partial_r - 2k B_{10} t^{k-1}r 
      - \frac{4}{3} k(k-1) B_{20} t^{k-2} r^3$ \\[\TYb]
(iii) & $X_n$ & = & $-t^{n+1}\partial_t 
        - A_{10}^{-1}[(t+A_{10}r)^{n+1}-t^{n+1}]\partial_r
        - (n+1)x t^n -\frac{n+1}{2}\frac{B_{10}}{A_{10}}[(t+A_{10}r)^n-t^n]$ 
        \\[\TYa]
    & $Y_{k-1}$ & = & $-(t+A_{10}r)^k \partial_r  
                  -\frac{k}{2} B_{10} (t+A_{10}r)^{k-1}$ \\ \hline
\end{tabular} \end{center}
\end{table}
%%~~~~~~~~~~~~~~~~~~~~~~~~~~~~~~~~~~~~~~~~~~~~~~~~~~~~~~~~~~~~~~~~~~~~~~~~~~~~~~

We still have to consider the commutators $[Y_{m},Y_{\ell}]$. 
Indeed, in the first case (\ref{3:YBedAB}), the commutator 
between the $Y_m$ is non-vanishing
\BEQ
\left[ Y_m , Y_{\ell} \right] = (m-\ell) (N-2) \frac{2 B_{10}}{N^3}  
t^{m+\ell+N-1} r^{2/N-2}
\EEQ
Unless $N=2/(2+n)=1,\frac{2}{3},\frac{1}{2},\frac{2}{5},\ldots$ or $N=2$, 
that is $\theta=1,2,3,4,\ldots$, 
there will be an infinite series of further generators. 
In the second case (\ref{3:YBedAB}), there are three series of new generators 
$Z_n^{(i)}$, $i=0,1,2$, see below. 
Finally, in the third case (\ref{3:YBedAB}), 
the commutator $[Y_m,Y_{\ell}]=A_{10}(m-\ell)Y_{m+\ell}$, see below. 
Our results so far can be summarized as follows.

\noindent {\bf Proposition 2:} 
{\it The generators $X_n$ defined in eq.~(\ref{3:XIIgen})
with $n\in\mathbb{Z}$ and the generators $Y_m$ defined in
eq.~(\ref{3:YIIgen}) with $m=-N/2+k$ and $k\in\mathbb{Z}$ and where
$a_n$ and $b_n$ are as in proposition 1 satisfy the commutation relations
\BEQ \label{3:XYComm_II}
\left[ X_n , X_{n'} \right] = (n-n') X_{n+n'} \;\; , \;\;
\left[ X_n , Y_m \right] = \left( n \frac{N}{2} - m \right) Y_{n+m}
\EEQ
in one of the following three cases:\\
(i) $B_{10}$ arbitrary, $A_{10}=A_{20}=B_{20}=0$ and $N$ arbitrary.  \\
(ii) $B_{10}$ and $B_{20}$ arbitrary, $A_{10}=A_{20}=0$ and $N=1$. 
In this case, there is a closed Lie algebra spanned
by the set $\{X_n, Y_m, Z_n^{(2)}, Z_m^{(1)}, Z_n^{(0)}\}$ of generators where
$n\in\mathbb{Z}$ and $m\in\mathbb{Z}+\frac{1}{2}$ and}
\BEQ
Z_{n}^{(2)} := -\, n t^{n-1} r^2 \;\; , \;\;
Z_{m}^{(1)} := -2\, t^{m-1/2} r \;\; , \;\;
Z_{n}^{(0)} := -2\, t^n
\EEQ
{\it with the following non-vanishing commutators, 
in addition to (\ref{3:XYComm_II})}
\BEA
& & 
\left[ Y_m, Y_{m'} \right] = (m-m') \left( 4 B_{20} Z_{m+m'}^{(2)} 
+B_{10} Z_{m+m'}^{(0)}\right) \;\; , \;\;
\nonumber \\
& & 
\left[ X_n, Z_{n'}^{(2)}\right] = -n'\, Z_{n+n'}^{(2)} \;\; , \;\;
\left[ Y_m, Z_{n}^{(2)}\right] = -n Z_{n+m}^{(1)} \;\; , \;\;
\nonumber \\
& &
\left[ X_n, Z_{m}^{(1)}\right] = -\left(\frac{n}{2}+m\right) Z_{n+m}^{(1)}
\;\; , \;\;
\left[ Y_m, Z_{m'}^{(1)}\right] = - Z_{m+m'}^{(0)} \;\; , \;\;
\left[ X_n, Z_{n'}^{(0)}\right] = -n'\, Z_{n+n'}^{(0)}
\EEA
{\it where $n,n'\in\mathbb{Z}$ and $m,m'\in\mathbb{Z}+\frac{1}{2}$. The
Lie algebra structure is determined by the parameter $B_{10}/B_{20}$. \\
(iii) $A_{10}$ and $B_{10}$ arbitrary, $A_{20}=A_{10}^2$, 
$B_{20}=\frac{3}{2}A_{10} B_{10}$ and $N=2$. 
Then for all $n,m\in\mathbb{Z}$ one has}
\BEQ
\left[ X_n , X_m \right] = (n-m) X_{n+m} \;\; , \;\;
\left[ X_n , Y_m \right] = (n-m) Y_{n+m} \;\; , \;\;
\left[ Y_n , Y_m \right] = A_{10} (n-m) Y_{n+m}
\EEQ
The verification of the commutators is straightforward. 

For case (i), if $N\in\mathbb{N}$ and $B_{10}=0$, there is a maximal 
finite-dimensional subalgebra, namely 
$\{X_{\pm1,0}, Y_{-N/2},Y_{-N/2+1},\ldots,Y_{+N/2}\}$. For case (ii), the
maximal finite-dimensional subalgebra is span\-ned by 
$\{X_{\pm1,0}, Y_{\pm1/2}, 4 B_{20}Z_{0}^{(2)}+B_{10} Z_{0}^{(0)}\}$.
The Schr\"odinger algebra eq.~(\ref{2:SchAlg}) is recovered for
$N=1$, $B_{10}={\cal M}/2$ and $B_{20}=0$. The inequivalent realizations of the
Schr\"odinger algebra are classified in \cite{Lahn98} and two distinct 
realizations were found. The first one of that list \cite{Lahn98} is the one
discussed here and the second realization is excluded 
by our axiom 1. For $N=2$ in case (i), the conformal
generators will be fully recovered once the mass terms have been included. 
Finally, case (iii) is isomorphic to the conformal algebra (\ref{2:Vir0}) 
through the correspondence $X_n=\ell_n+\bar{\ell}_n$, 
$Y_m=A_{10} \bar{\ell}_n$.

We now construct the mass terms contained in $X_{n}^{(III)}$ and $Y_m^{(III)}$.
For us, a mass term is a contribution to the generators which generically is
not proportional to a term of either zeroth or first order in $\partial_t$ or
$\partial_r$. The preceeding discussion has shown that 
the terms merely built from first order derivatives 
$\partial_t,\partial_r$ or without derivatives at all have already been 
found. The simple example outlined in appendix~C rather illustrates the need 
for `derivatives' $\partial_r^a$ of arbitrary order $a$. For our limited 
purpose, namely the construction of generators which satisfy 
eqs.~(\ref{3:XYComm_II}), we require the operational rules
\BEQ \label{3:FrakRegel}
\partial_r^{a+b} = \partial_r^a \partial_r^b \;\; , \;\;  
[ \partial_r^a, r ] = a \partial_r^{a-1}
\EEQ
together with the scaling 
$\partial_r^a f(\lambda r) = \lambda^a \partial_{\lambda r}^a 
f(\lambda r)$ and that for $a=n\in\mathbb{N}$, we recover the usual derivative.
However, the commutativity of fractional derivatives is not at all trivial and
several of the existing definitions, such as the Riemann-Liouville or the
Gr\"unwald-Letnikov fractional derivatives, are not commutative 
\cite{Samk93,Mill93,Podl99,Hilf00}. On the other hand, the Gelfand-Shilov 
\cite{Gelf64,Podl99} or Weyl \cite{Mill93} fractional derivatives or a recent
definition in the complex plane based on the Fourier transform \cite{Zava98} 
do commute. To make this paper self-contained, we shall present in appendix~A a
definition which gives a precise meaning to the symbol $\partial_r^a$ and allows
the construction of $X_n$ and $Y_m$ to proceed. In the sequel, the identities 
(\ref{A:Lem11},\ref{A:Lem12},\ref{A:Lem13},\ref{A:Lem15},\ref{A:Lem31})
will be used frequently. 

For generic $N$, setting $x=0$ and $B_{10}=0$ for the moment, we make the ansatz
\BEQ
X_n = -t^{n+1}\partial_t - \frac{N}{2} (n+1) t^n r \partial_r
- A_n(t,r) \partial_t^{a(n)} - B_n(t,r) \partial_r^{b(n)} 
- \partial_r^{c(n)} C_n(t,r) 
\EEQ
where the functions $A_n, B_n, C_n$ and the constants $a(n), b(n), c(n)$
have to be determined. From the condition $[X_n,X_0]=nX_n$, we find the
equations
\BEA
\left( t \partial_t + \frac{N}{2} r \partial_r \right) A_n(t,r)
- a(n) A_n(t,r) &=& n A_n (t,r) 
\nonumber \\
\left( t \partial_t + \frac{N}{2} r \partial_r \right) B_n(t,r)
- \frac{N}{2} b(n) B_n(t,r) &=& n B_n (t,r) 
\nonumber \\
\left( t \partial_t + \frac{N}{2} r \partial_r \right) C_n(t,r)
-  \frac{N}{2} c(n) C_n(t,r) &=& n C_n (t,r)
\EEA
with the solutions, where $u=r^{2/N} t^{-1}$
\BEQ
A_n(t,r) = t^{n+a(n)} {\cal A}_n(u) \;\; , \;\;
B_n(t,r) = t^{n+N b(n)/2} {\cal B}_n(u) \;\; , \;\;
C_n(t,r) = t^{n+N c(n)/2} {\cal C}_n(u)
\EEQ
Next, we require that $[X_n, X_{-1}]=(n+1)X_{n-1}$ and find
\BEA
t^{n-1+a(n)} \left( \left(n+a(n)\right){\cal A}_n - \frac{r^{2/N}}{t} 
{{\cal A}_n}' \right) &=& (n+1) t^{n-1+a(n-1)} {\cal A}_{n-1}
\nonumber \\
t^{n-1+N b(n)/2} \left( \left(n+\frac{N}{2}b(n)\right){\cal B}_n 
- \frac{r^{2/N}}{t} {{\cal B}_n}' \right) &=& 
(n+1) t^{n-1+N b(n-1)/2} {\cal B}_{n-1}
\nonumber \\
t^{n-1+Nc(n)/2} \left( \left(n+\frac{N}{2}c(n)\right){\cal C}_n 
- \frac{r^{2/N}}{t} 
{{\cal C}_n}' \right) &=& (n+1) t^{n-1+N c(n-1)/2} {\cal C}_{n-1}
\EEA
where the prime denotes the derivative with respect to $u$. Since this
must be valid for all values of $t$ and $r$ (or $t$ and $u$), we find
\BEQ
a(n) = a(n-1) = a \;\; , \;\;
b(n) = b(n-1) = \frac{2b}{N}  \;\; , \;\;
c(n) = c(n-1) = \frac{2c}{N}  
\EEQ
where $a,b,c$ are $n$-independent constants, and
\BEA
\left(n+a\right) {\cal A}_n(u) - u {{\cal A}_n}'(u) 
&=& (n+1) {\cal A}_{n-1}(u)
\nonumber \\
\left(n+b\right) {\cal B}_n(u) - u {{\cal B}_n}'(u) 
&=& (n+1) {\cal B}_{n-1}(u)
\nonumber \\
\left(n+c\right) {\cal C}_n(u) - u {{\cal C}_n}'(u) 
&=& (n+1) {\cal C}_{n-1}(u)
\EEA
In addition, we have the initial conditions
\BEQ
{\cal A}_{-1}(u) = {\cal A}_{0}(u) = 
{\cal B}_{-1}(u) = {\cal B}_{0}(u) = 
{\cal C}_{-1}(u) = {\cal C}_{0}(u) = 0
\EEQ
The solution of this is, e.g. for ${\cal A}_n$, 
\BEQ \label{3:LoesungA}
{\cal A}_n(u) = \sum_{k=1}^{n} \alpha_{k} \left(\vekz{n+1}{k+1}\right)
u^{k+a}
\EEQ
where the $\alpha_k$ are free parameters. Similar expressions hold for
${\cal B}_n$ and ${\cal C}_n$, where $a$ is replaced by $b$ and $c$,
respectively, and free parameters $\beta_k,\gamma_k$ are introduced. 

In the sequel, we shall concentrate on quasiprimary operators which
are assumed to transform covariantly under the action of $X_{\pm 1,0}$ only. 
We repeat the explicit expression for the generator $X_1$ of `special'
transformations, in the simplest case,
\BEQ
X_1 = -t^2\partial_t - N t r \partial_r - \alpha r^{2(1+a)/N} \partial_t^a
- \beta r^{2(1+b)/N} \partial_r^{2b/N} - \gamma \partial_r^{2c/N} r^{2(1+c)/N} 
\EEQ
and where $\alpha,\beta,\gamma$ are free parameters. 

For the physical applications, it is now important to check the consistency
with the invariance under spatial translations, generated by 
$Y_{-N/2}=-\partial_r$. In particular, from eq. (\ref{3:XYComm_II}), we should
have $[X_1, Y_{-N/2}]=N Y_{-N/2+1}$. From this, we easily find
\BEQ
Y_{-N/2+1} = -t\partial_r 
-\frac{2\alpha}{N^2}(1+a) r^{2(1+a)/N-1}\partial_t^a
-\frac{2\beta}{N^2}(1+b) r^{2(1+b)/N-1}\partial_r^{2b/N}
-\frac{2\gamma}{N^2}(1+c) \partial_r^{2c/N} r^{2(1+c)/N-1}
\EEQ
Acting again on this with $Y_{-N/2}$, we have the commutator
\BEA
\lefteqn{
\left[ Y_{-N/2+1},Y_{-N/2}\right] = 
-\frac{4\alpha}{N^3} (1+a)(1+a-N/2) 
r^{2(1+a)/N-2}\partial_t^a } 
\nonumber \\
& &-\frac{4\beta}{N^3} (1+b)(1+b-N/2) 
r^{2(1+b)/N-2}\partial_r^{2b/N}
-\frac{4\gamma}{N^3} (1+c)(1+c-N/2) 
\partial_r^{2c/N} r^{2(1+c)/N-2}
\nonumber 
\EEA
and a sequence of further generators may be constructed through the repeated
action of $Y_{-N/2}$. The number of these generators will be finite only if the
conditions
\BEQ
\frac{2}{N} (1+a) = k_1 \in\mathbb{N} \;\; , \;\;
\frac{2}{N} (1+b) = k_2 \in\mathbb{N} \;\; , \;\;
\frac{2}{N} (1+c) = k_3 \in\mathbb{N} 
\EEQ
are satisfied. That means that the realizations under construction will
be characterized by the value of $N$ and the three positive integers
$k_1,k_2,k_3$. We shall call the $k_i$ the {\em degrees} of the realization. 

A further consistency check, for $N$ integer, is provided by the condition
$[X_1,Y_{N/2}]=0$ or equivalently, for the $N+1$-th iterated commutator
\BEQ
\left[ X_1 \left[ \cdots \left[ X_1, Y_{-N/2} \right] \cdots \right] \right] = 0
\EEQ
Direct, but tedious calculations show that this is satisfied if either
(i) $\alpha\ne0$ and $\beta=\gamma=0$ or alternatively (ii) $\alpha=0$ and 
$\beta,\gamma\ne0$. We call the first case {\em Typ I} and the second case 
{\em Typ II}. 

If we consider the generators of Typ I, we see that for $N=1$ and $k_1=2$, 
we recover the generators (\ref{2:SchrGen}) of the Schr\"odinger algebra,
with $\alpha={\cal M}/2$. This explains the origin of the name `mass term' for
the contributions to $X_n,Y_m$ parametrized by $\alpha,\beta,\gamma$. 
Furthermore, for $N=2$ and $k_1=2$, let 
\BEQ 
z=t+\sqrt{\alpha}r \;\; , \;\; \bar{z}=t-\sqrt{\alpha}r
\EEQ
with $\alpha=-1/c^2$ where $c$ is the `speed of light' (or `speed of sound'). 
Therefore  $X_{n}=\ell_n+\bar{\ell}_n$ and $Y_{n}=\II(\ell_n-\bar{\ell}_n)$ 
where the conformal generators $\ell_n,\bar{\ell}_n$ are given in 
(\ref{2:lPhi}). Usually one sets $c=1$ and the presence of a dimensionful 
constant is then no longer visible. The Schr\"odinger algebra generators 
(\ref{2:SchrGen}) are also recovered for Typ II with $k_2=k_3=2$ and 
$\beta+\gamma={\cal M}/2$. These special cases already suggest that the choice 
$k_i=2$ may be particularly relevant for physical applications. We find a third
example with a finite-dimensional closed Lie algebra in the presence of mass
terms for Typ II with $N=2$ and degree $k_2=k_3=2$. Then the commutators read 
\BEQ
\left[ X_n , X_m \right] = (n-m) X_{n+m} \;\; , \;\;
\left[ X_n , Y_m \right] = (n-m) Y_{n+m} \;\; , \;\;
\left[ Y_n , Y_m \right] = (\beta+\gamma) (n-m) Y_{n+m}
\EEQ
and we recover case (iii) of the proposition 2, after identifying
$A_{10}=\beta+\gamma$ and $B_{10}=2\gamma$.

{}From now on and for the rest of this paper, we shall always take $B_{10}=0$. 
\subsection{Dynamical symmetry} 

For $k_1=k_2=k_3=2$, our realization acts as a dynamical symmetry
on certain linear (inte\-gro-)dif\-fe\-ren\-tial equations with constant 
coefficients as we now show. In $d$ spatial dimensions, 
generalizing the above constructions 
along the same lines as in appendix B, we consider the generalized 
Schr\"odinger operator
\BEQ
\mathcal{S} = -\alpha \partial_t^N + \left(\frac{N}{2}\right)^2 
\vec{\partial}_{r}\cdot\vec{\partial}_{r}
\EEQ
and the generators $X_{-1}=-\partial_t$, $Y_{-N/2}^{(i)}=-\partial_{r_i}$ 
together with the Typ I generators with $k_1=2$
\BEA
X_0 &=& -t\partial_t - \frac{N}{2} \vec{r}\cdot\vec{\partial}_{r} 
- \frac{N}{2} x 
\nonumber \\
X_1 &=& -t^2 \partial_t - N t \vec{r}\cdot\vec{\partial}_r - N x t 
- \alpha \vec{r}^2 \partial_t^{N-1}
\label{3:TypI} \\
Y_{-N/2+1}^{(i)} &=& -t\partial_{r_i} - \frac{2\alpha}{N} r_i \partial_t^{N-1}
\nonumber 
\EEA
with $i=1,\ldots,d$ and we have
\BEQ \label{3:GCasimirGal}
\left[ \mathcal{S}, X_{-1} \right] = 
\left[ \mathcal{S}, Y_{-N/2}^{(i)} \right ] = 
\left[ \mathcal{S}, Y_{-N/2+1}^{(i)} \right ] = 0
\EEQ
which shows that $\mathcal{S}$ is a Casimir 
operator of the `Galilei'-type sub-algebra generated from 
\linebreak $\{X_{-1}, Y_{-N/2}^{(i)}, Y_{-N/2+1}^{(i)}\}$ as 
given in (\ref{3:TypI}). Furthermore,
\BEQ \label{3:DynaSymmI}
\left[ \mathcal{S}, X_0 \right] = - N \mathcal{S} \;\; , \;\;
\left[ \mathcal{S}, X_1 \right] = - 2 N t \mathcal{S} 
+ \alpha N^2 \left( x - \frac{d}{2} + \frac{N-1}{N}\right)
\EEQ
In addition, since $[X_1, Y_{-N/2+k}^{(i)}] = (N-k) Y_{-N/2+k+1}^{(i)}$, 
it follows immediately that $[\mathcal{S},Y_{-N/2+k}^{(i)}]=0$ for all 
$k\ne N$, since from the Jacobi identities
\BEQ
\left[ \mathcal{S}, Y_{-N/2+k+1}^{(i)} \right] = (N-k)^{-1} \left( 
\left[ X_1, \left[ \mathcal{S}, Y_{-N/2+k}^{(i)} \right] \right] - 
\left[ Y_{-N/2+k}^{(i)}, \left[ \mathcal{S}, X_{1} \right] \right] \right) = 0 
\EEQ
on the solutions of $\mathcal{S}\psi=0$ and $[\mathcal{S},X_1]\psi=0$ by 
induction over $k$. Additional generators created from the commutators
$[Y_m, Y_{m'}]$ are treated similarly. Therefore, we have shown:

\noindent {\bf Proposition 3:} {\it The realization 
(\ref{3:TypI}) of Typ I generated from 
$\{X_{-1}, X_{1}, Y_{-N/2}^{(i)}\}$, $i=1,\ldots,d$ sends any 
solution $\psi(t,\vec{r})$ with scaling dimension} 
\BEQ \label{3:xI}
x = \frac{d}{2} - \frac{N-1}{N}
\EEQ
{\it of the differential equation}
\BEQ \label{3:WelleI}
\mathcal{S} \psi(t,\vec{r}) = 
\left( -\alpha \partial_t^N + \left(\frac{N}{2}\right)^2 
\vec{\partial}_{r}\cdot\vec{\partial}_{r} \right) \psi(t,\vec{r}) = 0
\EEQ
{\it into another solution of the same equation.} If we construct a 
free-field theory such that (\ref{3:WelleI}) is the equation of motion,
then $x$ as given in (\ref{3:xI}) is the scaling dimension of that free
field $\psi$. That theory is non-local when $N$ is not a positive integer. 

Similarly, for Typ II with $k_2=k_3=2$ and $d=1$ for simplicity (we shall refer
to this case in the sequel as {\em Typ IIa}), we consider
\BEQ
\mathcal{S} = -(\beta+\gamma) \partial_t + 
\frac{1}{\theta^2} \partial_r^{\theta}
\EEQ
and the generators of (\ref{3:TypI}) are replaced by ($k_2=k_3=2$)
\BEA
X_0 &=& -t\partial_t - \frac{1}{\theta} {r}{\partial}_{r} 
- \frac{x}{\theta} 
\nonumber \\
X_1 &=& -t^2\partial_t -\frac{2}{\theta} t {r}{\partial}_r-\frac{2x}{\theta}t
- (\beta+\gamma) {r}^2 \partial_r^{2-\theta}  
- \gamma 2(2-\theta) r\partial_r^{1-\theta} 
- \gamma (2-\theta)(1-\theta) \partial_r^{-\theta} ~~~~~~~
\label{3:TypIIa} \\
Y_{-N/2+1} &=& -t\partial_{r} - (\beta+\gamma)\theta r \partial_r^{2-\theta}
-\gamma\theta(2-\theta) \partial_r^{1-\theta} 
\nonumber 
\EEA
Again, (\ref{3:GCasimirGal}) holds so that $\mathcal{S}$ is a Casimir 
operator of the `Galilei' sub-algebra generated from
\linebreak $\{X_{-1}, Y_{-N/2}, Y_{-N/2+1}\}$. In
addition, instead of (\ref{3:DynaSymmI}) we have
\BEQ
\left[ \mathcal{S} , X_0 \right] = - \mathcal{S} \;\; , \;\; 
\left[ \mathcal{S}, X_1 \right] = -2 t\mathcal{S} +\frac{\beta+\gamma}{\theta} 
\left( 2 x  - \theta +1 - \frac{2\gamma}{\beta+\gamma}\left( 2 - \theta\right)
\right)
\EEQ
Therefore, we have the following dynamical symmetry.

\noindent {\bf Proposition 4:} {\it The realization 
(\ref{3:TypIIa}) of Typ IIa generated from $\{X_{-1},X_{1},Y_{-N/2}\}$ 
sends any solution $\psi(t,r)$ with scaling dimension} 
\BEQ
x = \frac{\theta-1}{2} + \frac{\gamma}{\beta+\gamma} (2-\theta)
\EEQ
{\it of the differential equation}
\BEQ \label{3:WelleII}
\mathcal{S} \psi(t,r) = 
\left( -(\beta+\gamma) \partial_t + \frac{1}{\theta^2} \partial_r^{\theta}
\right) \psi(t,r) = 0 
\EEQ
{\it into another solution of the same equation.} This means that the
ratio $\beta/\gamma$ is a universal number and will be independent of the
irrelevant details (in the renormalization group sense) 
of a given model. As before, $x$ is the
scaling dimension of the free field whose equation of motion is given by
(\ref{3:WelleII}). 

While these dynamical 
symmetries were found for $k_1=k_2=k_3=2$, there is one more possibility for
Typ II with $k_2=k_3=3$ (to be called Typ IIb in the sequel). Consider
\BEQ
\mathcal{S} = - 3(3-\theta)\gamma \partial_t 
+ \frac{1}{\theta^2} \partial_r^{\theta}
\EEQ
and the generators now read ($k_2=k_3=3$)
\BEA
X_0 &=& -t\partial_t - \frac{1}{\theta} {r}{\partial}_{r} 
- \frac{x}{\theta} 
\nonumber \\
X_1 &=& -t^2 \partial_t - \frac{2}{\theta} t {r}{\partial}_r 
-\frac{2x}{\theta}t
- \beta {r}^3 \partial_r^{3-\theta}  
- \gamma \partial_r^{3-\theta} r^3
\label{3:TypIIb} \\
Y_{-N/2+1} &=& -t\partial_{r} - 
\frac{3}{2}\beta\theta r^2 \partial_r^{3-\theta}
-\frac{3}{2}\gamma\theta\partial_r^{3-\theta} r^2
\nonumber 
\EEA
If we take $\beta+\gamma=0$, we recover indeed eqs. (\ref{3:GCasimirGal}) so 
that $\mathcal{S}$ is again Casimir operator of the `Galilei' sub-algebra and
\BEQ
\left[ \mathcal{S}, X_0 \right] = - \mathcal{S} \;\; , \;\;
\left[ \mathcal{S}, X_1 \right] = - 2t \mathcal{S} + \frac{6(3-\theta)}{\theta}
\gamma \left( x - \frac{1}{2}\right)
\EEQ
and we have the following statement.

\noindent {\bf Proposition 5:} {\it The realization 
(\ref{3:TypIIb}) of Typ IIb with $k_2=k_3=3$ 
and $\beta=-\gamma$ sends any solution $\psi(t,r)$ with scaling dimension 
$x=1/2$ of the differential equation}
\BEQ
\mathcal{S} \psi(t,r) = \left( - 3(3-\theta)\gamma \partial_t 
+ \frac{1}{\theta^2} \partial_r^{\theta}\right) \psi(t,r) = 0
\EEQ
{\it to another solution of the same equation.} It follows that for Typ II 
there are two distinct ways of realizing a dynamical symmetry, if 
$\theta\ne 2$. 

All possibilities to obtain linear wave equations with
constant coefficients from Casimir operators of the above simple
form are now exhausted. We illustrate this for Typ I. 
A convenient generalized Schr\"odinger operator
$\mathcal{S}$ should satisfy $[\mathcal{S},X_{-1}]=[\mathcal{S},Y_{-N/2}]=0$ 
and this implies $\mathcal{S}=\mathcal{S}(\partial_t,\partial_r)$. 
Using the identity (\ref{A:Lem32}) and writing $\mathcal{S}=\mathcal{S}(u,v)$,
one has for $k_1=k$
\BEQ
\left[ \mathcal{S}, Y_{-N/2+1} \right] = 
- \frac{\partial \mathcal{S}}{\partial u}\partial_r 
-\frac{\alpha k}{N} \sum_{\ell=1}^{k-1} 
\left(\vekz{k-1}{\ell}\right) r^{k-1-\ell}
\frac{\partial^{\ell} \mathcal{S}}{\partial v^{\ell}} \partial_{t}^{kN/2-1}
\EEQ
which for $k>2$ contains several terms with powers of $r$ which cannot be made
to disappear and therefore $[\mathcal{S},Y_{-N/2+1}]=0$ is impossible. 
Similarly, for Typ II only for the case $k_2=k_3=3$ a compensation of one more
term is feasible. For the other cases, Casimir operators can of course be
constructed in a straightforward way, but we shall not perform this here.
 
For $N=1$ (or $\theta=2$), Typ I ($k_1=2$), Typ IIa ($k_2=k_3=2$) and
Typ IIb ($k_2=k_3=3$) coincide and we recover as expected the
known dynamical symmetry of the free Schr\"odinger equation 
\cite{Nied72,Hage72,Baru73}. For $N=2$ and $k_1=2$, Typ I gives the 
Klein-Gordon equation with its dynamical conformal symmetry. Finally, for
$N=2$, $k_2=k_3=2$ and $\beta+\gamma=A_{10}$, Typ IIa is identical to the
generators (\ref{3:XIIgen},\ref{3:YIIgen}) with the identification
$B_{10}=2\gamma$. 

{}From the different forms of the wave equations 
(\ref{3:WelleI},\ref{3:WelleII}) we see that Typ I and Typ II describe
physically distinct systems. The propagator resulting from Typ I, which in
energy-momentum space is of the form $(\omega^N + p^2)^{-1}$, is typical
for equilibrium systems with so strongly anisotropic interactions, that the
quadratic term $\sim \omega^2$ which is usally present is cancelled and the
next-to-leading term becomes important. This occurs in fact at the Lifshitz
point of spin systems with competing interactions, as we shall see in a
later section. On the other hand, the propagator from Typ II, of the form
$(\omega+ p^{\theta})^{-1}$ in energy-momentum space is reminiscent of a
Langevin equation describing the time evolution of a non-equilibrium system 
(furthermore, if we had tried to describe such a real time evolution in terms
of the propagators found for Typ I, we would have encountered immediate
problems with causality.)
Indeed, we shall see that aspects of aging phenomena in simple ferromagnets
can be understood this way. 

\subsection{The two-point function}

The main objective of this paper is the calculation of two-point functions
\BEQ
G = G(t_1,t_2;r_1,r_2) = \langle \phi_1(t_1, r_1) \phi_2(t_2, r_2) \rangle
\EEQ
of scaling operators $\phi_i$ from its covariance properties under local
scale transformations. We shall assume
that spatio-temporal translation invariance holds and therefore
\BEQ
G = G (t,r) \;\; , \;\; t = t_1 - t_2 \;\; , \;\; r = r_1 - r_2
\EEQ
Since for scaling operators $\phi$ invariant under spatial rotations, the
two points can always be brought to lie on a given line, the case $d=1$ is
enough to find the functional form of the scaling function present in $G$. 
To do so, we have to express the action of the generators $X_{0,1}$ and
$Y_{m}$ on $G$. Each scaling operator $\phi_i$ is characterized by either
the pair ($x_i,\alpha_i$) for Typ I or the triplet ($x_i,\beta_i,\gamma_i$)
for Typ II.\footnote{The indices $\alpha_i,\beta_i,\gamma_i$ refer here to the
two scaling operators and have nothing to do with the indices used in
eq. (\ref{3:LoesungA}).} By definition, two-point functions formed from
{\em quasiprimary} scaling operators satisfy the covariance conditions
\BEQ \label{3:allKova}
X_n G = Y_m G = 0
\EEQ
Since all generators can be obtained from commutators of the three
generators $X_{\pm 1}, Y_{-N/2}$, explicit consideration of a subset is
sufficient. For the three cases considered above, the results are as follows.

{\bf 1.} For Typ I, the single condition
\BEQ \label{3:TypIM}
\alpha_2 = (-1)^{-N} \alpha_1
\EEQ
is sufficient to guarantee that (\ref{3:allKova}) is satisfied, together with
the covariance under all commutators which can be constructed from the
$X_n, Y_m$. We merely need to satisfy explicitly the following conditions
\BEA
X_0 G(t,r) &=& \left( -t\partial_t - \frac{N}{2} r\partial_r - {N} x\right)
G(t,r)  = 0 \nonumber \\
X_1 G(t,r) &=& \left( -t^2\partial_t - N t r \partial_r - 2 N x_1 t 
-\alpha_1 r^2 \partial_t^{N-1}\right) G(t,r)  \nonumber \\ 
& & +2t_2 X_0 G(t,r) + N r_2 Y_{-N/2+1} G(t,r) = 0 \label{3:TypINull}\\
Y_{-N/2+1} G(t,r) &=& \left( - t \partial_r - 
\frac{2\alpha_1}{N} r \partial_t^{N-1} \right) G(t,r) = 0
\nonumber 
\EEA
where $2x=x_1+x_2$. This makes it clear that time and space translation
invariance are implemented. If we multiply the first of eqs.~(\ref{3:TypINull})
by $-t$ and add it to the second one and then multiply the third of
eqs.~(\ref{3:TypINull}) by $-2/Nr$ and also add it, the condition $X_1G(t,r)=0$
simplifies to
\BEQ
\frac{1}{2} N t \left( x_1 - x_2 \right) G(t,r) = 0
\EEQ
which implies the constraint 
\BEQ \label{3:TypIx}
x_1 = x_2
\EEQ
The two remaining eqs.~(\ref{3:TypINull}) may be solved by the ansatz
\BEQ \label{3:TypIA}
G(t,r) = \delta_{x_1,x_2} r^{-2x_1} \Omega\left( t r^{-2/N} \right)
\EEQ
which leads to an equation for the scaling function $\Omega(v)$,
where $v=t r^{-2/N}$
\BEQ \label{3:TypIDGL}
\left( \alpha_1 \partial_v^{N-1} - v^2 \partial_v - N x_1 \right) 
\Omega(v) = 0
\EEQ
and the boundary conditions (see section 2)
\BEQ \label{3:TypIR}
\Omega(0) = \Omega_0 \;\; , \;\; \Omega(v) \simeq \Omega_{\infty} v^{-Nx_1}
\;\; \; \;\; \mbox{\rm for $v\to\infty$}
\EEQ
where $\Omega_{0,\infty}$ are constants. 
Eqs. (\ref{3:TypIA},\ref{3:TypIDGL},\ref{3:TypIR}) together
with the constraints (\ref{3:TypIM},\ref{3:TypIx}) 
determine the two-point function and its scaling function $\Omega(v)$ and 
constitute the main result of this section for the realizations of Typ I. 

{\bf 2.} Similarly, for Typ IIa ($k_2=k_3=2$), the conditions
\BEQ \label{3:TypIIaM}
\beta_2 = (-1)^{-1+\theta} \beta_1 \;\; , \;\;
\gamma_2 = (-1)^{-1+\theta} \gamma_1
\EEQ
are enough to guarantee that the quasiprimarity conditions (\ref{3:allKova}) 
hold and
\BEA
X_0 G(t,r) &=& \left( -t\partial_t - \frac{1}{\theta} r\partial_r 
- \frac{2x}{\theta} \right) G(t,r) = 0  
\nonumber \\
X_1 G(t,r) &=& \left( -t^2\partial_t -\frac{2}{\theta} t r \partial_r 
- \frac{2x_1}{\theta} t - (\beta_1+\gamma_1) r^2 \partial_r^{2-\theta} 
-2(2-\theta)\gamma_1 r \partial_r^{1-\theta} \right) G(t,r) \nonumber \\
& & +2t_2 X_0 G(t,r) + \frac{2}{\theta} r_2 Y_{-N/2+1} G(t,r) = 0
\label{3:TypIIaNull} \\
Y_{-N/2+1} &=& \left( - t \partial_r - 
\theta(\beta_1+\gamma_1)r\partial_r^{2-\theta}
-2\theta(2-\theta)\gamma_1 \partial_r^{1-\theta} \right) G(t,r) = 0
\nonumber
\EEA
In the same way as before, the condition $X_1G(t,r)=0$ can be simplified into
\BEQ
\frac{1}{\theta} t \left( x_1 - x_2 \right) G(t,r) = 0 
\EEQ
which implies again the constraint (\ref{3:TypIx}).  
The two remaining eqs.~(\ref{3:TypIIaNull}) can be solved by the ansatz
\BEQ \label{3:TypIIaA}
G(t,r) = \delta_{x_1,x_2}t^{-2x_1/\theta} \Phi\left(r t^{-1/\theta}\right)
\EEQ
and lead to the following equation for the scaling function $\Phi(u)$,
where $u=r t^{-1/\theta}$
\BEQ \label{3:TypIIaDGL}
\left( \partial_u +\theta(\beta_1+\gamma_1)u \partial_u^{2-\theta} 
+2\theta(2-\theta) \gamma_1 \partial_u^{1-\theta} \right) \Phi(u) = 0
\EEQ
with the boundary conditions
\BEQ \label{3:TypIIaR}
\Phi(0) = \Phi_0 \;\; , \;\; \Phi(u) \simeq \Phi_{\infty} u^{-2x_1} 
\;\; \mbox{\rm for $u\to\infty$}
\EEQ
where $\Phi_0,\Phi_{\infty}$ are constants. Eqs. 
(\ref{3:TypIIaA},\ref{3:TypIIaDGL},\ref{3:TypIIaR})
together with the constraints (\ref{3:TypIIaM},\ref{3:TypIx}) 
determine the two-point function and its scaling function $\Phi(u)$ and 
constitute the main result of this section for the realizations of Typ IIa.

{\bf 3.} For Typ IIb ($k_2=k_3=3$), there are two possibilities. First, we 
might take $\gamma_2=(-1)^{-1+\theta}\gamma_1$, which would be the same
as in eq.~(\ref{3:TypIIaM}). It turns out, however, that this leads to the
same equation for the scaling function as for Typ IIa, upon identification of
parameters. We therefore examine the second possibility 
\BEQ \label{3:TypIIbM}
\gamma_2 = (-1)^{-2+\theta} \gamma_1
\EEQ
Then, in contrast to the previous cases, we need the additional generator
\BEQ
M := \frac{1}{3\theta} \left[ Y_{-N/2+1}, Y_{-N/2} \right] 
= -(\beta+\gamma) r \partial_r^{3-\theta} 
  -(3-\theta)\gamma \partial_{r}^{2-\theta}
\EEQ
We let $\beta_{1,2} = -\gamma_{1,2}$ and find 
\BEA
X_0 G(t,r) &=& \left( -t\partial_t - \frac{1}{\theta} r\partial_r 
- \frac{2x}{\theta} \right) G(t,r) = 0 
\nonumber \\
X_1 G(t,r) &=& \left( -t^2\partial_t -\frac{2}{\theta} t r \partial_r 
- \frac{2x_1}{\theta} t + \gamma_1 r^3 \partial_{r}^{3-\theta} 
-\gamma_1 \partial_r^{3-\theta} r^3 
- 2(3-\theta)(2-\theta)(1-\theta)\gamma_1 \partial_r^{-\theta} \right) 
G(t,r) \nonumber \\
& & +\left( 2t_2 X_0 + \frac{2}{\theta} r_2 Y_{-N/2+1} -3r_2^2 M\right) 
G(t,r) = 0  
\label{3:TypIIbNull} \\
Y_{-N/2+1} &=& \left( - t \partial_r  
- 3\theta(3-\theta)\gamma_1 r\partial_r^{2-\theta} \right) G(t,r)
+ 3\theta r_2 M G(t,r) = 0 
\nonumber 
\EEA
In addition to the usual conditions (\ref{3:allKova}) for quasiprimarity, we 
also need explicitly that $M G(t,r)=0$. As we did before, the condition
$X_1G(t,r)=0$ can be simplified and we find
\BEA
\left( t \partial_t + \frac{1}{\theta} r \partial_r + \frac{x_1+x_2}{\theta}
\right) G(t,r) &=& 0 \nonumber \\
\left( \gamma_1 \theta(2-\theta)(3-\theta) \left( 3 r \partial_r + 2(1-\theta)
\right) \partial_r^{-\theta} + (x_1 - x_2) t \right) G(t,r) &=& 0 
\label{3:TypIIaGfunktion}\\
\left( t \partial_r + \gamma_1 3 \theta(3-\theta) r \partial_{r}^{2-\theta}
\right) G(t,r) &=& 0 \nonumber 
\EEA 
To analyse these further, let $H(t,r):=\partial_{r}^{-\theta} G(t,r)$. Then,
the last two of the above equations take the form, after having also acted with 
$\partial_r$ on the second equation (\ref{3:TypIIaGfunktion}) 
\BEA
\left( \gamma_1 3\theta(3-\theta)(2-\theta) \left( r\partial_r^2 +\frac{1}{3}
(5-2\theta)\partial_r \right) + (x_1-x_2)t \partial_r^{1+\theta} 
\right) H(t,r) &=& 0 \nonumber \\
\left( t \partial_r^{1+\theta} + \gamma_1 3\theta(3-\theta) r \partial_r^2
\right) H(t,r) &=& 0 
\EEA
The only apparent way to make these two equations compabtible is to make 
one of them trivial or else to make them
coincide. The first one is possible if $\theta=2$ and $x_1=x_2$ and the
second possibility occurs if
\BEQ \label{3:TypIIbx}
\theta = \frac{5}{2} \;\; , \;\; x_2 = x_1 + 1/2
\EEQ
and these conditions make Typ IIb a very restricted one. In order to have
$x_1\ne x_2$ in either Schr\"odinger or conformal invariance, time translation
invariance must be broken \cite{Henk94} but remarkably enough for Typ IIb, 
in spite of the presence of time translation invariance, the scaling dimensions
of the two quasiprimary operators are {\em not} the same. 
If the above conditions (\ref{3:TypIIbx}) hold, one can make the ansatz
\BEQ \label{3:TypIIbA}
G(t,r) = \delta_{x_1+1/2,x_2} t^{-(x_1+x_2)/\theta} 
\Phi\left(r t^{-1/\theta}\right)
\EEQ
where $\Phi(u)$ satisfies the equation
\BEQ \label{3:TypIIbDGL}
\left( \partial_u + \gamma_1 3\theta(3-\theta) u \partial_u^{2-\theta} \right)
\Phi(u) = 0 
\EEQ
with the usual boundary conditions. In addition, we still have to take
into account the condition 
$M G(t,r)=-(3-\theta)2\gamma_1 t^{-(x_1+x_2+2-\theta)/\theta}
\partial_u^{2-\theta} \Phi(u)=0$. Therefore $\partial_u^{2-\theta}\Phi(u)=0$
and from eq. (\ref{3:TypIIbDGL}) it follows $\partial_u\Phi(u)=0$. 
The fact that $\Phi(u)={\rm const.}$, together with the
restrictiveness of this case suggests that Typ I and Typ IIa, where
$k_i=2$, might be more relevant for physical applications. 

\begin{table}
\caption{Some properties of three basic types of local scale invariance.
The upper part collects operator properties and the lower part specific
properties of the two-point function. 
Here $\mathcal{S}$ is the generalized Schr\"odinger operator and $x$
is the scaling dimension of the field $\psi$ wich solves $\mathcal{S}\psi=0$,
in $d=1$ spatial dimensions. The scaling function for the two-point function 
satisfies the differential equation
$\mathcal{D}\Omega(v)=0$ or $\mathcal{D}\Phi(u)=0$, respectively. In the
boundary conditions, the first line refers to $u=0$ or $v=0$,
and the second line to $u\to\infty$ or $v\to\infty$, respectively.\label{tab1}}

%%~~~~~~~~~~~~~~~~~~~~~~~~~~~~~~~~~~~~~~~~~~~~~~~~~~~~~~~~~~~~~~~~~~~~~~~~~~~~~~
\begin{center}
\begin{tabular}{|l|l|l|l|} \hline
       & \multicolumn{1}{c|}{Typ I} & \multicolumn{1}{c}{Typ IIa} 
 &\multicolumn{1}{c|}{Typ IIb} \\ \hline
degree & $k_1=2$          & $k_2=k_3=2$ & $k_2=k_3=3$ \\
masses & $\beta=\gamma=0$ & $\alpha=0$  & $\alpha=\beta+\gamma=0$ \\
$\mathcal{S}$ & $-\alpha\partial_t^N+\frac{1}{4}N^2\partial_r^2$ &
$-(\beta+\gamma)\partial_t+\theta^{-2}\partial_r^{\theta}$ &
$-3(3-\theta)\gamma\partial_t+\theta^{-2}\partial_r^{\theta}$ \\
$x$ & 
$\frac{1}{2}-{(N-1)}/{N}$ & 
$\frac{1}{2}(\theta-1)+{(2-\theta)}/{(1+\beta/\gamma)}$ &
$\frac{1}{2}$ \\ \hline \hline
constraints & $x_2 = x_1$ & $x_2 = x_1$ & 
$x_2=x_1+\frac{1}{2}$, $\theta=\frac{5}{2}$ \\
  & $\alpha_2 = (-1)^{-N} \alpha_1$ & $\beta_2=-(-1)^{\theta}\beta_1$, 
    $\gamma_2=-(-1)^{\theta}\gamma_1$ & $\gamma_2=(-1)^{\theta}\gamma_1$
\\ \hline
scaling & $G=r^{-2x_1}\Omega(v)$ & \multicolumn{2}{c|}{ 
$G=t^{-(x_1+x_2)/\theta}\Phi(u)$} \\
        & $v=t r^{-2/N}$ & \multicolumn{2}{c|}{$u=rt^{-1/\theta}$} \\\hline
$\mathcal{D}$ & 
$\alpha_1\partial_v^{N-1}-v^2\partial_v-Nx_1v$ &
$\partial_u+\theta(\beta_1+\gamma_1)u\partial_u^{2-\theta}
+2\theta(2-\theta)\gamma_1\partial_u^{1-\theta}$&
$\partial_u + 3\theta(3-\theta)\gamma_1 
u\partial_u^{2-\theta}$
\\ \hline
auxiliary & & &  \\
condition & & & $\partial_u^{2-\theta} \Phi(u)=0$ \\ \hline
boundary & $\Omega(0)=\Omega_0$ &
\multicolumn{2}{c|}{$\Phi(0)=\Phi_0$} \\
conditions & $\Omega(v)\sim \Omega_{\infty} v^{-Nx_1}$ &
\multicolumn{2}{c|}{$\Phi(u)\sim \Phi_{\infty} u^{-2x_1}$} \\ \hline
\end{tabular}
\end{center}
\end{table}
%%~~~~~~~~~~~~~~~~~~~~~~~~~~~~~~~~~~~~~~~~~~~~~~~~~~~~~~~~~~~~~~~~~~~~~~~~~~~~~~

In summary, starting from certain axioms of local scale invariance, 
we have constructed infinitesimal local scaling transformations. We have
shown that these act as dynamical symmetries of certain linear equations of
motion of free fields. We have also derived the equations which the scaling 
functions of quasiprimary two-point functions must satisfy. 
For easy reference, the main results are collected in table~\ref{tab1}. 

Generalizing from Schr\"odinger invariance, it may be useful to 
introduce a conjugate scaling operator $\phi^*$ characterized by
\BEQ
\phi^*\;: \left\{ \begin{array}{ll} 
(x,(-1)^N\alpha) & \mbox{\rm , for Typ I} \\
(x,-(-1)^{-\theta} \beta, -(-1)^{-\theta} \gamma ) & \mbox{\rm , for Typ IIa}
\end{array} \right.
\EEQ
The two-point functions then become
\BEQ
G(t,r) = \langle \phi_1(t_1,r_1) \phi_2^*(t_1,r_1) \rangle
= \left\{ \begin{array}{ll}
\delta_{x_1,x_2} \delta_{\alpha_1,\alpha_2} t^{-2x_1} 
\Omega\left(t r^{-2/N}\right) 
& \mbox{\rm , for Typ I} \\
\delta_{x_1,x_2} \delta_{\beta_1,\beta_2}\delta_{\gamma_1,\gamma_2}
t^{-2x_1/\theta} \Phi\left(r t^{-1/\theta}\right) 
& \mbox{\rm , for Typ IIa} 
\end{array} \right.
\EEQ
Only for Typ I with $N$ even, the distinction between $\phi$ and $\phi^*$ is
unneccessary. For Typ II, we shall in section 5 identify $\phi^*$ with
the response operator $\wit{\phi}$ in the context of dynamical scaling. 

In general, the infinitesimal generators $X_n, Y_m$ do not close into a 
finite-dimensional Lie algebra. However, if we apply them to certain states
(which we characterized for two-body operators by the conditions 
(\ref{3:TypIM}) and (\ref{3:TypIIaM}) for Typ I and IIa, respectively) 
all generators built from $[Y_{m},Y_{m'}]$ vanish on these states. One might
call such a structure a {\em weak Lie algebra}. For the cases considered
here, it is generated {\em from} the minimal set $\{X_{-1}, X_{1}, Y_{-N/2}\}$.

Technically, we have been led to consider derivatives $\partial^a$ 
of arbitrary real order $a$. We emphasize that all results in this section 
only use the abstract properties of commutativity and scaling of these 
fractional derivatives, as stated in (\ref{3:FrakRegel}) and formulated 
precisely in appendix A, see  
eqs.~(\ref{A:Lem11},\ref{A:Lem12},\ref{A:Lem13}). They are 
therefore {\em independent} of the precise form (\ref{A:Ablei}) which 
is used in appendix A to prove them and any other fractional derivative which
satisfies these three properties could have been used instead. 
On the other hand, the differential equations 
(\ref{3:TypIDGL},\ref{3:TypIIaDGL}) for the two-point 
functions can only be solved explicitly if a particular choice for the action 
of $\partial^a$ on functions is made.

%%%%%%%%%%%%%%%%%%%%%%%%%%%%%%%%%%%%%%%%%%%%%%%%%%%%%%%%%%%%%%%%%%%%%%%%%%%%%%%%
\section{Determination of the scaling functions}
%%%%%%%%%%%%%%%%%%%%%%%%%%%%%%%%%%%%%%%%%%%%%%%%%%%%%%%%%%%%%%%%%%%%%%%%%%%%%%%%

We now derive the solutions of the differential equations
(\ref{3:TypIDGL},\ref{3:TypIIaDGL}), together with the boundary conditions
(\ref{3:TypIR},\ref{3:TypIIaR}), see also table~\ref{tab1}. 
{}From now on, we make use of the specific definition eq.~(\ref{A:Ablei}) for
the fractional derivatives. 
The comparison with concrete model results will be presented in the 
next section. 

\subsection{Scaling function for Typ II}

Having seen that for the realizations of Typ II 
considered in section 3 only the one of degree 
$k_2=k_3=2$ will lead to a non-trivial scaling function, we shall concentrate
on this case, called Typ IIa in table~\ref{tab1}. 
We now discuss the solutions of eq.~(\ref{3:TypIIaDGL}), 
which we repeat here
\BEQ \label{4:DGLII}
\left( \partial_u +\theta(\beta_1+\gamma_1)u \partial_u^{2-\theta} 
+2\theta(2-\theta) \gamma_1 \partial_u^{1-\theta} \right) \Phi(u) = 0
\EEQ

Before we discuss the general solution of eq.~(\ref{4:DGLII}), it may
be useful to consider the special case $\theta=1$ first. Then the scaling
function satisfies
\BEQ \left( \partial_u +(\beta_1+\gamma_1)u \partial_u +2\gamma_1 \right)
\Phi(u) = 0 
\EEQ
The solution is promptly found
\BEQ
\Phi(u) = \left\{ \begin{array}{ll}
\Phi_0 \left( 1+(\beta_1+\gamma_1) u \right)^{-2/(1+\beta_1/\gamma_1)} &
\mbox{\rm ~~; ~~ $\gamma_1 \ne -\beta_1$} \\
\Phi_0 \exp\left( -2\gamma_1 u \right) & 
\mbox{\rm ~~; ~~ $\gamma_1 = -\beta_1$} \end{array}\right.
\EEQ
where $\Phi_0$ is a constant. Although $\theta=1$, this situation does
{\em not\/} correspond to the case of conformal invariance, as a comparison
with the conformal two-point function $\Phi_{\rm conf}(u) \sim (1+u^2)^{-x}$
from eq.~(\ref{2:ZweiP}) shows. Furthermore, although eq.~(\ref{3:TypIIaDGL})
does not contain the scaling dimensions explicitly, 
the ratio $\beta_1/\gamma_1\ne -1$ is indeed a universal number and related 
to the scaling dimension $x_1$ via
\BEQ \label{4:xbetagamma}
x_1 = \frac{1}{1+\beta_1/\gamma_1}
\EEQ
as the consideration of the boundary condition eq.~(\ref{2:SkalFRand}) shows. 
Then we finally have
\BEQ \label{4:theta1Loes}
\Phi(u) =  \left\{\begin{array}{ll} 
\Phi_0 \left( 1 + \gamma_1 x_1^{-1} u \right)^{-2x_1} & 
\mbox{\rm ~~; ~~ $\gamma_1 \ne -\beta_1$} \\
\Phi_0 \exp\left( -2\gamma_1 u \right) & 
\mbox{\rm ~~; ~~ $\gamma_1 = -\beta_1$} \end{array}\right.
\EEQ 
We remark that the case $\beta_1+\gamma_1=0$ is one of the rare
instances where Cardy's \cite{Card85} prediction of a simple exponential 
scaling function is indeed satisfied. The special case $\theta=1$ may be
of relevance in spin systems with their own dynamics
(which is not generated from a heat bath through the master equation). This
situation occurs for example in the easy-plane Heisenberg ferromagnet,
where the dynamical exponent $z=1.00(4)$ in $2D$ \cite{Ever96}. 
The dynamical exponent $z=1$ also occurs 
in a recently introduced model of a fluctuating interface \cite{Gier02} and
related to conformal invariance. 
Another example with a dynamical exponent $z\simeq 1$ is provided by the 
phase-ordering kinetics of binary alloys in a gravitational 
field \cite{Datt93}. 

We now turn towards the general case. In order to find the solution of
eq.~(\ref{3:TypIIaDGL}) by standard series expansion methods, 
see \cite{Mill93,Podl99}, we set
\BEQ
\theta = \frac{p}{q}
\EEQ
where $p,q$ are coprime integers and make the ansatz (with $c_0\ne 0$)
\BEQ
\Phi(u) = \sum_{n=0}^{\infty} c_n u^{n/q+s}
\EEQ
where the $c_n$ and $s$ are to be determined. 
Although this procedure certainly will only give a solution for rational values
of $\theta$, we can via analytic continuation formulate an educated guess
for the scaling function for arbitrary values of $\theta$. Furthermore,
since we are only interested in the situations where the scaling variable
$u$ is positive, we can discard the singular terms which might be generated
by applying the definition (\ref{A:Ablei}) of the fractional derivative
$\partial_u^a$. At the end, we must consider the right-hand limit $u\to 0+$
and check that indeed $\lim_{u\to 0+} \Phi(u)=\Phi_0$ exists.\footnote{This 
formal calculation works here in the same way as for ordinary derivatives since 
the highest derivative in (\ref{4:DGLII}) is indeed of integer order.} 

Insertion of the above ansatz into the differential equation leads to 
\BEA
\lefteqn{
u^{s-1} \left( \sum_{n=0}^{\infty} \left( \frac{n}{q}+s\right) c_n u^{n/q}
\right.} \nonumber \\ 
&+&\left.
\sum_{n=p}^{\infty} c_{n-p} \frac{\theta\Gamma\left(\frac{n-p}{q}+s+1\right)}
{\Gamma\left(\frac{n}{q}+s\right)}
\left[ (\beta_1+\gamma_1)\left(\frac{n}{q}+s-1\right)+2(2-\theta)\gamma_1
\right] u^{n/q} \right) = 0
\EEA
and must be valid for all positive values of $u$. 
This leads to the conditions $s=0$, 
$c_n=0$ for $n=1,2,\ldots,p-1$ and, if we set $\Phi_{\ell} := c_{p\ell}$,
to the recurrence
\BEQ
\Phi_{\ell} = -\frac{\gamma_1\Gamma(\theta(\ell-1)+1)}{\ell\Gamma(\theta\ell)}
\left[ A\ell + B\right] \Phi_{\ell-1}
\EEQ
where we have set
\BEQ
A := \theta \left( 1 + \beta_1/\gamma_1\right) \;\; , \;\;
B := 3 - 2\theta - \beta_1/\gamma_1
\EEQ
To solve this, we set
\BEQ
\Phi_{\ell} = P(\ell) \ell^{-1} \sigma_{\ell} \;\; ; \;\; \ell\geq 1
\EEQ
where the function $P(n)$ is defined by $P(1) := 1$ and for $n\geq 2$ by
\BEQ
P(n) := \prod_{k=1}^{n-1} \frac{\Gamma(\theta k)}{\Gamma(\theta k +\theta)} 
= \frac{\Gamma(\theta)}{\Gamma(n\theta)} 
\EEQ
We first discuss the case $\beta_1+\gamma_1\ne 0$. Then $A\ne 0$ and we find 
the following recurrence for the $\sigma_{\ell}$
\BEQ
\sigma_{\ell+1} = -\gamma_1 \theta A \left(\ell +1 + B/A\right)\sigma_{\ell}
\;\; ; \;\; \ell\geq 1
\EEQ
which is solved straightforwardly and leads to
\BEQ
\Phi_{\ell} = (-\gamma_1\theta A)^{\ell} \frac{\Gamma(\ell+1+B/A)}{
\Gamma(\ell\theta+1)\Gamma(1+B/A)} \Phi_{0}
\EEQ
Similarly, for the other case $\beta_1+\gamma_1$ we have $A=0$ and we find in 
an analogous way
\BEQ
\Phi_{\ell} = (-\gamma_1 \theta B)^{\ell} \frac{1}{\Gamma(\ell\theta+1)}
\Phi_{0}
\EEQ
Therefore the sought-after series solution finally is
\BEQ \label{4:Phiu}
\Phi(u) = \sum_{\ell=0}^{\infty} \Phi_{\ell} u^{\theta\ell}
= \left\{ \begin{array}{ll} 
\Phi_0 
\mathfrak{E}_{\theta,\Lambda}(-\theta^2(\beta_1+\gamma_1) u^{\theta}) &
\mbox{\rm ~~; ~~ $\gamma_1 \ne -\beta_1$} \\
\Phi_0 E_{\theta,1}(-2\theta(2-\theta)\gamma_1 u^{\theta}) &
\mbox{\rm ~~; ~~ $\gamma_1 = -\beta_1$} 
\end{array} \right.
\EEQ
where 
\BEQ \label{4:Lambda}
\Lambda = \Lambda(\theta,\beta_1/\gamma_1) 
:= \frac{(\theta-1)(1+\beta_1/\gamma_1)+2(2-\theta)}
{\theta(1+\beta_1/\gamma_1)}
\EEQ 
and the function $\mathfrak{E}_{a,b}(z)$ and the Mittag-Leffler function 
$E_{a,b}(z)$ are defined as
\BEQ \label{4:EEFunk}
\mathfrak{E}_{a,b}(z) := \frac{1}{\Gamma(b)} \sum_{k=0}^{\infty} 
\frac{\Gamma(k+b)}{\Gamma(ak+1)} z^k
\;\; , \;\;
E_{a,b}(z) = \sum_{k=0}^{\infty} \frac{1}{\Gamma(ak+b)} z^k
\EEQ
From this is it obvious that the scaling function $\Phi(u)$ as given in
(\ref{4:Phiu}) does satisfy 
the boundary condition $\Phi(0)=\Phi_0={\rm const}$ and has an infinite radius
of convergence for $\theta>1$ if $\beta_1+\gamma_1\ne 0$ and for $\theta>0$
if $\beta_1+\gamma_1=0$. Using the identities
\BEQ
\mathfrak{E}_{1,b}(z) = (1-z)^{-b} \;\; , \;\;
\mathfrak{E}_{2,b}(z) = {_1F_1}\left(b;\frac{1}{2};\frac{z}{4}\right)\;\; ,\;\;
\mathfrak{E}_{2,1/2}(z) = e^{z/4} \;\; , \;\;
\EEQ
\BEQ
E_{1,1}(z) = e^z \;\; , \;\;
E_{1/2,1}(-z) = e^{z^2} \left( 1 - \erf(z) \right)
\EEQ
it is easily checked that the known solutions 
(\ref{4:theta1Loes}) and (\ref{2:S2ponto}) are recovered, for $\theta=1$ and
$\theta=2$, respectively.

%%------------------------------------------------------------------------------
\begin{figure}[t]
\centerline{\epsfxsize=3.4in\ \epsfbox{
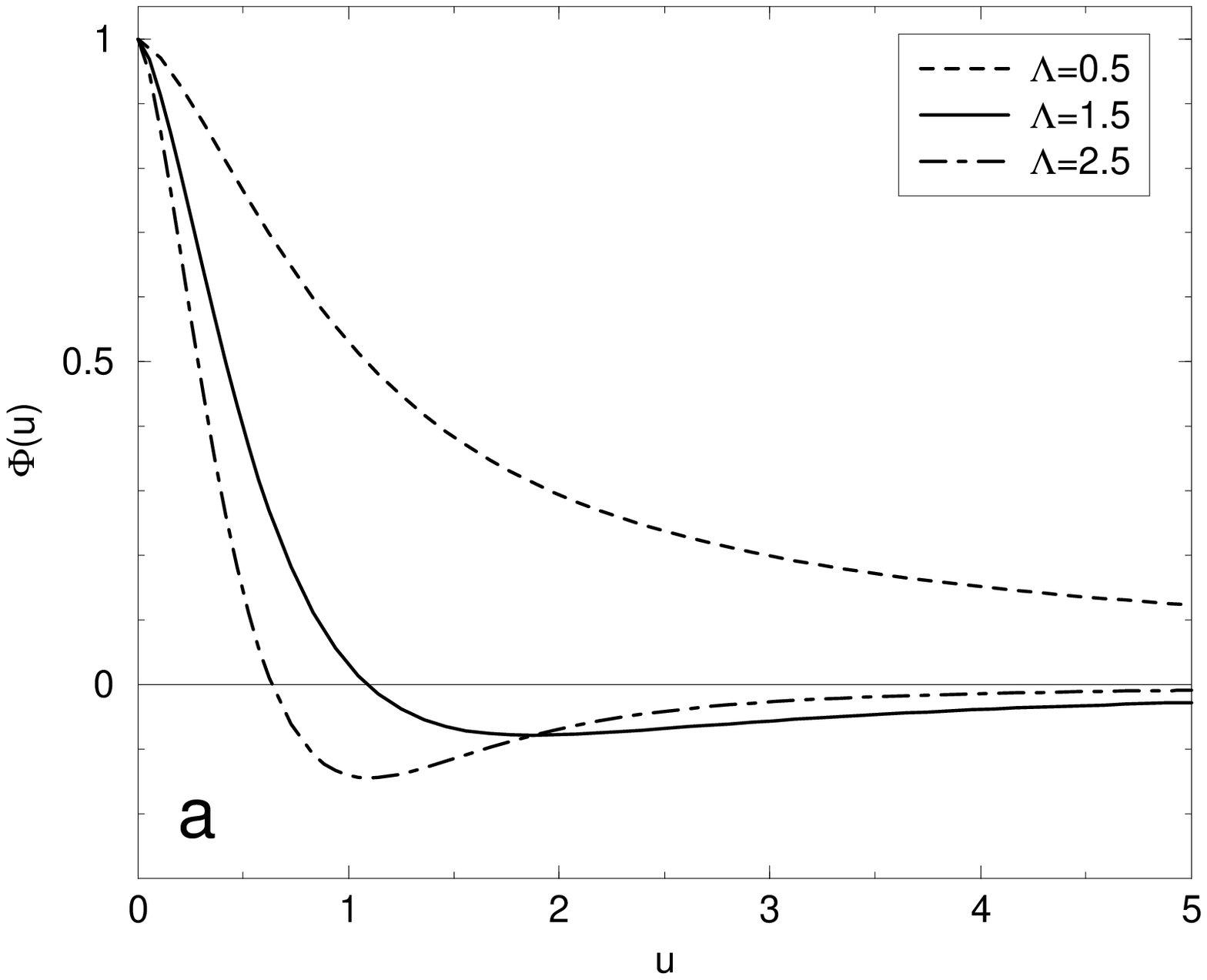}
\epsfxsize=3.4in\ \epsfbox{
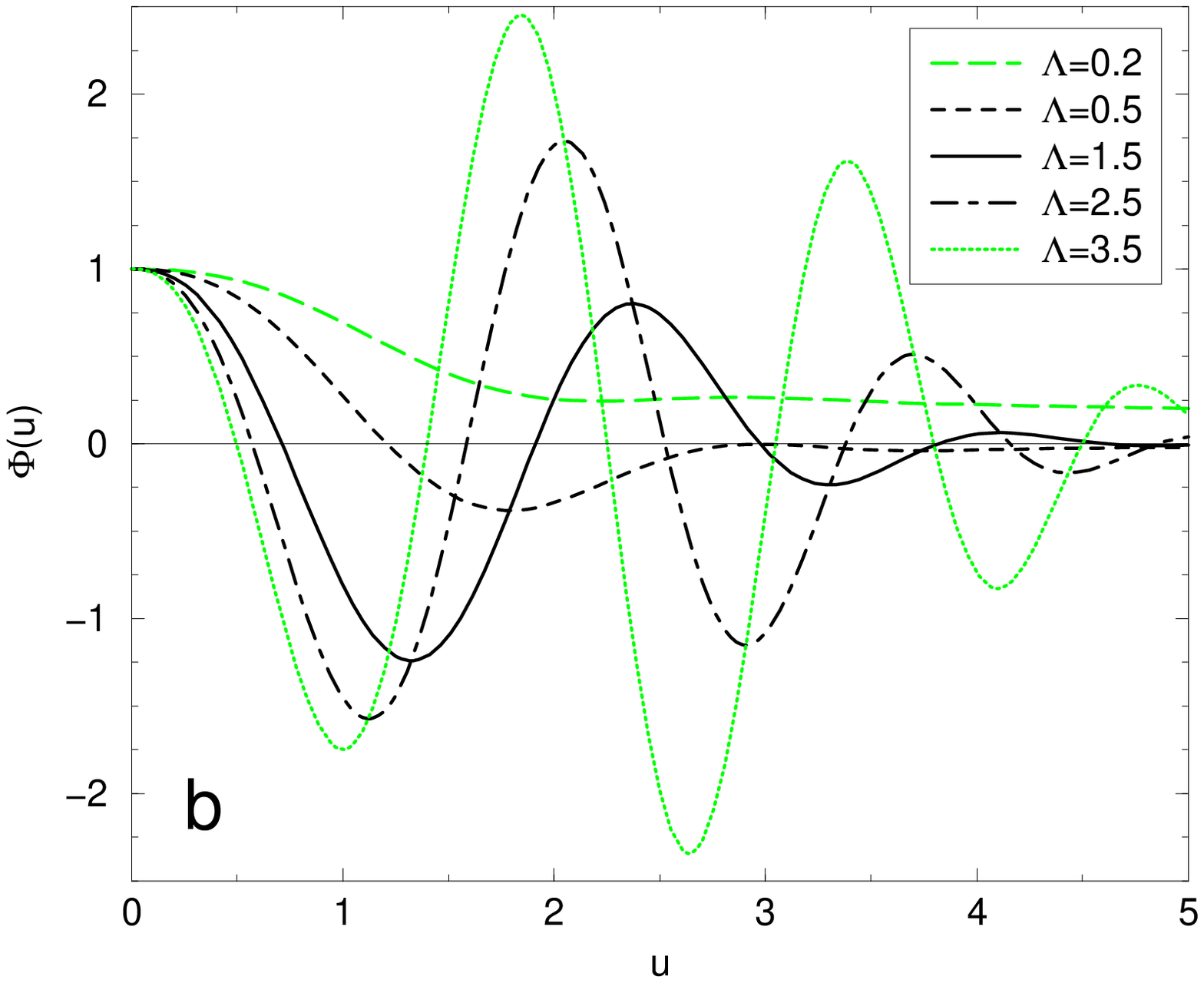}
}
\caption{Scaling functions $\Phi(u)$ for Typ II
as given by {\protect eq.~(\ref{4:Phiu})} with $\beta_1+\gamma_1=1$ for 
(a) $\theta=1.5$ and (b) $\theta=2.5$, for several values of $\Lambda$ as 
indicated. 
\label{Abb_43}}
\end{figure}
%%------------------------------------------------------------------------------
%%------------------------------------------------------------------------------
\begin{figure}[t]
\centerline{
\epsfxsize=3.4in\ \epsfbox{
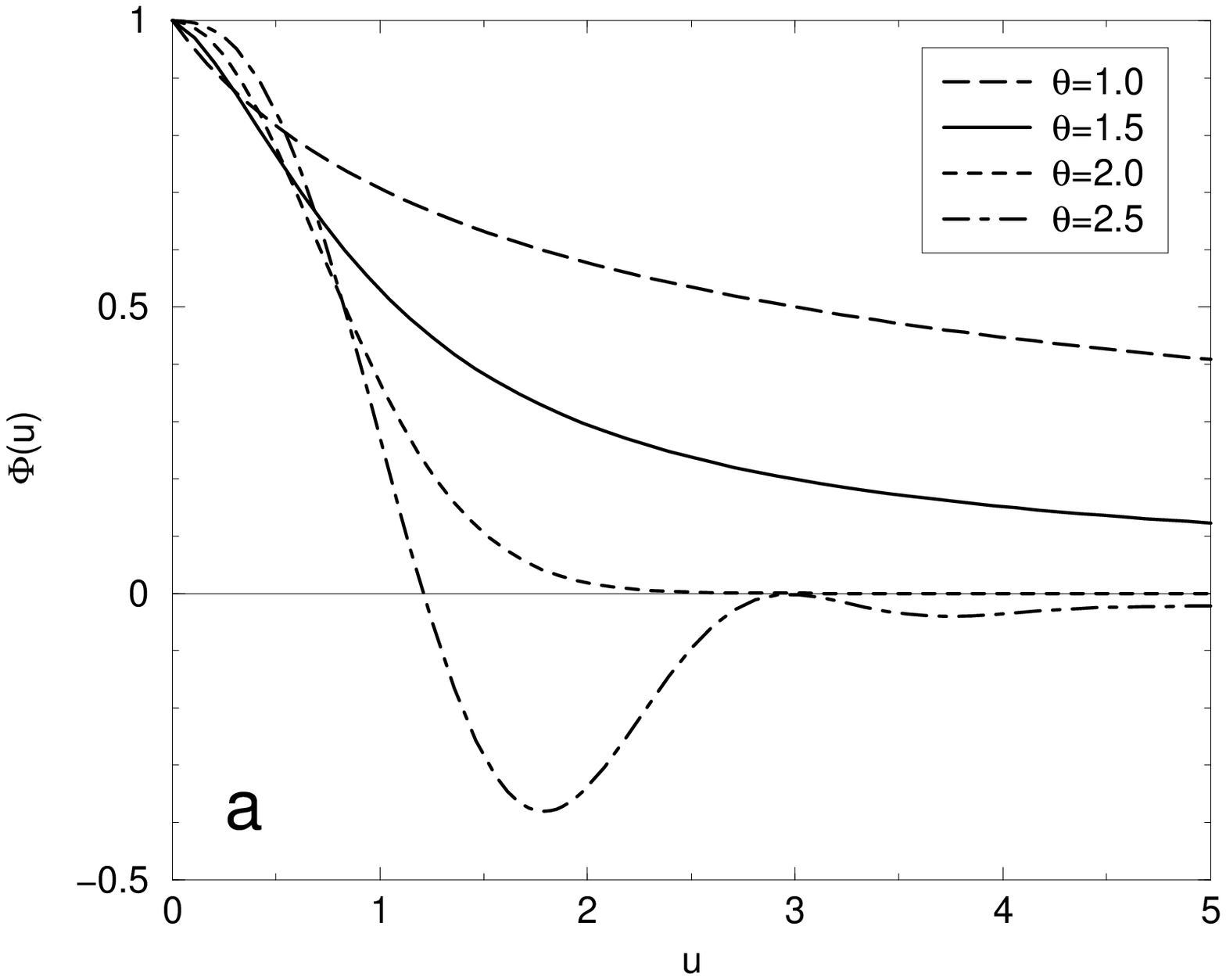}
\epsfxsize=3.4in\ \epsfbox{
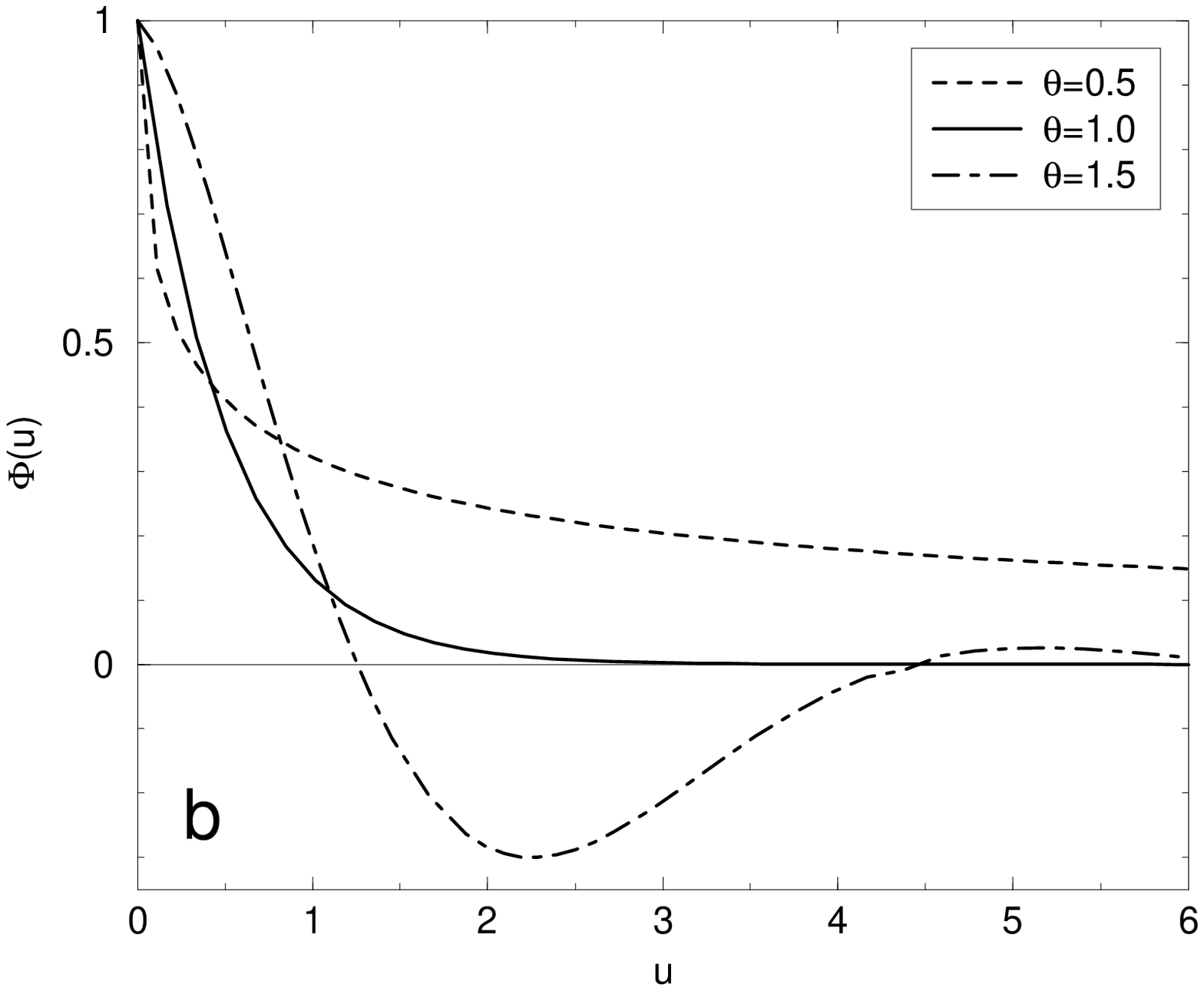}
}
\caption{Scaling functions $\Phi(u)$ for Typ II
as given by {\protect eq.~(\ref{4:Phiu})}.   
(a) Several values of $\theta$ as indicated, $\Lambda=1/2$ and 
$\beta_1+\gamma_1=1$. (b) Several values of $\theta$ and $\gamma_1=-\beta_1=1$.
\label{Abb_44}}
\end{figure}
%%------------------------------------------------------------------------------
In figures~\ref{Abb_43} and \ref{Abb_44} the behaviour of the scaling function 
$\Phi(u)$ is illustrated for several values of $\theta$ and of $\Lambda$.
The case $\beta_1+\gamma_1=1$ is governed by the properties of the function
$\mathfrak{E}_{a,b}(z)$. In figure~\ref{Abb_43} for
two fixed values of $\theta$ the effect of varying $\Lambda$ is displayed. 
Since the universal ratio $\beta_1/\gamma_1$ is arbitrary, the parameter
$\Lambda$ as defined in (\ref{4:Lambda}) can take any positive value. Only for
$\theta=2$ we have $\Lambda=1/2$ fixed. 
The dependence of the scaling function $\Phi(u)$ for $\beta_1+\gamma_1\ne 0$ 
on $\theta$ is shown in figure~\ref{Abb_44}a, where we fix $\Lambda=1/2$.
Finally, the scaling function found in the peculiar case $\beta_1+\gamma_1=0$ 
is illustrated in figure~\ref{Abb_44}b. This case is governed by the 
Mittag-Leffler function $E_{\theta,1}(z)$ whose properties are reviewed in some
detail in \cite{Podl99}.  

In the following cases, the asymptotic behaviour of the scaling function is
algebraic, as follows from the identities, see \cite{Wrig35}
\BEA
E_{a,b}(-x) &\simeq& \frac{1}{\Gamma(b-a)} {x}^{-1} + O\left( x^{-2}\right) 
\;\; , \;\; 0 < a < 2 \nonumber \\
\mathfrak{E}_{a,b}(-x) &\simeq& \frac{1}{b\Gamma(1-a)} x^{-1} 
+ \frac{\Gamma(1-b)}{\Gamma(1-ab)}  x^{-b} +
O\left( x^{-2},x^{-1-a}\right) \;\; , \;\; 1 < a < 3
\EEA
In these cases, the physically required boundary condition is satisfied and
one may deduce the relation between $\Lambda$ and the scaling dimension
$x_1$, thereby generalizing (\ref{4:xbetagamma}). 

We also see from figure~\ref{Abb_44} that on a qualitative level, the cases
$\beta_1+\gamma_1\ne 0$ and $\beta_1+\gamma_1=0$ are broadly similar. For
sufficiently small values of $\theta$, the scaling function $\Phi(u)$ 
descreases monotonically towards zero when $u$ increases. With increasing
$\theta$, the scaling function decays faster for large $u$ and at a certain
value of $\theta$ (at $\theta=2$ or $\theta=1$, respectively), 
the asymptotic amplitude vanishes and the decay becomes
exponential. If we now increase $\theta$ slightly, the scaling function
$\Phi(u)$ starts to oscillate. Oscillations also arise if for $\theta$ fixed
the parameter $\Lambda$ is made sufficiently large, as can be seen from
figure~\ref{Abb_43}. 

\subsection{Scaling function for Typ I}

We write eq. (\ref{3:TypIDGL}) in the form
\BEQ \label{4:DGLI}
\left( \alpha_1 \partial_v^{N-1} - v^2 \partial_v - v \zeta \right)
\Omega(v) = 0 \qquad , \qquad
\zeta = N x_1 = \frac{2x_1}{\theta} 
\EEQ
It is useful to begin with the special case when $N$ is an 
integer \cite{Henk97}. 
Then the anisotropy exponent 
$\theta=2/N=2,1,\frac{2}{3},\frac{1}{2},\frac{2}{5},\frac{1}{3},\ldots$
and one merely has a finite number of generators $Y_m$, $m=-N/2,\ldots,N/2$. 
For $N=1$ and $N=2$ we recover the scaling functions found 
from Schr\"odinger and conformal invariance (see section 2) and now concentrate 
on the new situations $N\geq 3$.  
%%~~~~~~~~~~~~~~~~~~~~~~~~~~~~~~~~~~~~~~~~~~~~~~~~~~~~~~~~~~~~~~~~~~~~~~~~~~~~~~
\begin{table}[t]
\caption{Some solutions $\Omega(v)$ of {\protect eq.~(\ref{4:DGLI})} for $N=4$ 
with $\Omega(0)=1$ and their leading asymptotics for $\beta'=0$ as
$v\to\infty$. Here ${\rm I}_{\nu}, {\rm K}_{\nu}$
are modified Bessel functions, ${\bf L}_{\nu}$ is a modified Struve
function, $\erf$ is the error function (see \cite{Abra65}) and $\beta,\beta'$ 
are constants. The abbreviation
$y := v^2/(2\sqrt{\alpha_1})$ is used throughout. \label{tab2}}
\begin{tabular}{|c|cl|c|} \hline
        &             & & asymptotics \\
$\zeta$ & $\Omega(v)$ & & (for $\beta'=0$)\\ \hline
$1$ & $\left[\frac{\Gamma(3/4)^2}{\pi^2} \sqrt{y}\,
{\rm K}_{1/4}^2\left(\frac{y}{2}\right) 
+\beta v {\rm I}_{1/4}\left(\frac{y}{2}\right) 
{\rm K}_{1/4}\left(\frac{y}{2}\right)\right]$
& $+\beta' v {\rm I}_{1/4}^2\left(\frac{y}{2}\right)$
& $4\sqrt{\alpha_1}\,\beta\, v^{-1}$ \\[\TZ]
$2$ & $\left[ 
-\beta\sqrt{\pi}\Gamma\left(\frac{1}{4}\right)
\left(\frac{y}{2}\right)^{1/4}\left[ 
{\bf L}_{-1/4}\left(y\right)
-{\rm I}_{1/4}\left(y\right) \right]
\right.$ & & $4\sqrt{\alpha_1}\,\beta\, v^{-2}$ \\
& $\left.+\Gamma\left(\frac{3}{4}\right) 
\left(\frac{y}{2}\right)^{1/4}\left[ 
{\rm I}_{-1/4}\left(y\right)
-{\rm I}_{1/4}\left(y\right) \right] \right]$
& $+\beta' {v}^{1/2} {\rm I}_{1/4}\left(y\right)$
&  \\[\TZ]
$3$ & $\left[ e^{-y}+\beta \left(\frac{\pi^2\alpha_1}{64}\right)^{1/4}\left[ 
e^{y} \erf\left(\sqrt{y}\right) -\II e^{-y}\erf\left(\II\sqrt{y}\right) 
-2\sinh y\right]\right]$ & 
$+\beta'\,\sinh y$ & $\alpha_1\, \beta\, v^{-3}$ \\[\TZ] 
$4$ &$\left[ 1+ \sqrt{\pi}\Gamma\left(\frac{3}{4}\right)
\left(\frac{y}{2}\right)^{3/4}  \left[ 
{\bf L}_{1/4}\left(y\right)
-{\rm I}_{-1/4}\left(y\right) \right]\right.$ 
& & $-{2\alpha_1}\, v^{-4}$ \\ 
& $\left.+\beta \left(\frac{y}{2}\right)^{3/4}
\left[ 
{\rm I}_{-1/4}\left(y\right)
-{\rm I}_{1/4}\left(y\right) \right] \right]$
& $+\beta' {v}^{3/2} {\rm I}_{1/4}\left(y\right)$
&  \\
\hline
\end{tabular}
\end{table}
%%~~~~~~~~~~~~~~~~~~~~~~~~~~~~~~~~~~~~~~~~~~~~~~~~~~~~~~~~~~~~~~~~~~~~~~~~~~~~~~
For $N=4$, some explicit solutions for a few integer values of $\zeta$ are
given in table~\ref{tab2}. Given the boundary condition $\Omega(0)=1$, these
still depend on two free parameters $\beta,\beta'$. If $\beta'\ne0$, these
solutions diverge exponentially fast as $v\to\infty$ but if we take $\beta'=0$,
we find $\Omega(v)\sim v^{-\zeta}$ in agreement with the required boundary 
condition, see table~\ref{tab1}. 

Using these examples as a guide, we now study the more general case with 
integer $N$ and $\zeta$ arbitrary. The general solution of 
eq.~(\ref{3:TypIDGL}) for integer $N\geq 2$ is readily found 
\BEQ \label{4:LoesFHyp}
\Omega(v) = \sum_{p=0}^{N-2} b_p v^p {\cal F}_p \qquad ; \qquad 
{\cal F}_p = {_{2}F_{N-1}} \left( \frac{\zeta+p}{N}, 1 ; 1+\frac{p}{N}, 
1+\frac{p-1}{N}, \ldots, \frac{p+2}{N} ; \frac{ v^N }{N^{N-2} \alpha_1} \right) 
\EEQ
where ${_{2}F_{N-1}}$ is a generalized hypergeometric function and the
$b_p$ are free parameters. To be physically acceptable, the boundary condition
(\ref{3:TypIR}) must be satisfied. The leading asymptotic behaviour of the
${\cal F}_p$ for $v\to\infty$ can be found from the general theorems of
Wright \cite{Wrig35} (see \cite{Frac93} for a brief summary) 
and the asymptotics of $\Omega(v)$ is given by
\BEA
\lefteqn{ \Omega(v) \simeq \sqrt{\frac{4\pi^2N}{N-2}} 
\left(\frac{v^{1/(N-2)}}{(\alpha_1 N)^{1/N}}\right)^{\zeta+1-N} 
\exp\left( \frac{N-2}{N\alpha_1^{1/(N-2)}} v^{N/(N-2)}\right) 
} \nonumber \\
&\times& 
\sum_{p=0}^{N-2} b_p\, \frac{\Gamma(p+1)}{\Gamma((p+1)/N)\Gamma((p+\zeta)/N)}
\left(\frac{\alpha_1}{N^2}\right)^{p/N} 
\left( 1 + O\left( v^{-N/(N-2)}\right)\right)
\EEA 
which grows exponentially as $v\to\infty$ if $N>2$. Clearly, this leading term
must vanish, which imposes the following condition on the $b_p$
\BEQ \label{4:bpZwang}
\sum_{p=0}^{N-2} b_p\, \frac{\Gamma(p+1)}{\Gamma((p+1)/N)\Gamma((p+\zeta)/N)}
\left(\frac{\alpha_1}{N^2}\right)^{p/N} =0 
\EEQ 
Remarkably, this condition is already sufficient to cancel not only the
leading exponential term but in fact the {\em entire} series of exponentially
growing terms. Eliminating $b_{N-2}$, the final solution for 
$N$ integer becomes
\BEA
\Omega(v) &=& \sum_{p=0}^{N-3} b_p \Omega_p(v) \nonumber \\
\Omega_p(v) &=& v^p {\cal F}_p - 
\frac{\Gamma(p+1)}{\Gamma(\frac{p+1}{N})\Gamma(\frac{p+\zeta}{N})} 
\frac{\Gamma(\frac{N-1}{N})\Gamma(1+\frac{\zeta-2}{N})}{\Gamma(N-1)}
\left(\frac{\alpha_1}{N^2}\right)^{(p+2-N)/N} 
v^{N-2} \, {\cal F}_{N-2} \label{4:OmeFinal}
\EEA
Up to normalization, the form of $\Omega(v)$ depends on $\zeta$ and on $N-3$
free parameters $b_p$, while $\alpha_1$ merely sets a scale. 
The independent solutions $\Omega_p$ ($p=0,1,\ldots,N-3$) satisfy the 
boundary conditions
\BEQ \label{4:RandOme}
\Omega_p(v) \simeq \left\{ \begin{array}{ll}
v^p & \; ; \; \mbox{\rm $v\rightarrow 0$} \\
\Omega_{p,\infty} v^{-\zeta} & \; ; \; \mbox{\rm $v\rightarrow\infty$} 
\end{array} \right.
\EEQ
where explicitly
\BEQ \label{4:Omeinf}
\Omega_{p,\infty} = -\left( \frac{\alpha_1}{N^2}\right)^{(\zeta+p)/N}
\frac{\Gamma(\frac{1-\zeta}{N})}{\Gamma(1-\zeta)}
\frac{\Gamma(p+1)}{\Gamma(\frac{p+1}{N})}
\frac{\pi \sin\left(\frac{\pi}{N}(p+2)\right)}{\Gamma(\frac{p+\zeta}{N})
\sin\left(\frac{\pi}{N}(p+\zeta)\right)
\sin\left(\frac{\pi}{N}(\zeta-2)\right) }
\EEQ
Therefore, we have not only eliminated the entire exponentially growing series,
but furthermore, the $\Omega_p$ satisfy exactly the physically required
boundary condition (see table~\ref{tab1}) for $v\to\infty$ \cite{Henk97}. 
Indeed, for $N=3$ this
cancellation of the exponential terms is a known property of the 
Kummer function ${}_{1}F_1$ and we have for $N=3$ the scaling function
\BEA
\Omega_{N=3}(v) &=& b_0 \left[ {}_1F_1\left(\frac{\zeta}{3},\frac{2}{3};
\frac{v^3}{3\alpha_1}\right) - 
\frac{\Gamma((\zeta+1)/3)\Gamma(2/3)}{\Gamma(\zeta/3)\Gamma(4/3)} 
\frac{v}{(3\alpha_1)^{1/3}} 
{}_1F_1\left(\frac{\zeta}{3},\frac{4}{3};
\frac{v^3}{3\alpha_1}\right) \right] 
\nonumber \\
&=& b_0 \frac{\sqrt{3}}{2\pi} \Gamma\left(\frac{\zeta+1}{3}\right)
\Gamma\left(\frac{2}{3}\right) U\left(\frac{\zeta}{3},\frac{2}{3};
\frac{v^3}{3\alpha_1}\right)
\EEA
where $U$ is the Tricomi function and its known asymptotic behaviour 
$U(a,b;z)\simeq z^{-a}$ as $z\to\infty$ \cite[eq. (13.1.8)]{Abra65} 
reproduces the boundary condition eq.~(\ref{4:RandOme}). 
We have not found an analogous statement for $N\geq 4$ in the literature.
Wright's formulas \cite{Wrig35} simply list the dominant and the 
subdominant parts of the $v\to\infty$ asymptotic expansion but without
any statement when the dominant part may cancel.  
Rather than giving a formal and lenghty proof of the cancellation of the
entire asymptotic exponential series, we merely argue in favour
of its plausibility through a few tests. In table~\ref{tab2} we list a few
closed solutions for $N=4$ with satisfy the boundary condition $\Omega(0)=1$. 
By varying the parameters $\beta$ and $\beta'$ one obtains the three independent
solutions of the third-order differential equation (\ref{4:DGLI}). Furthermore,
from the explicit form of the solutions we see that only the contribution
parametrized by $\beta'$ diverges exponentially as $v\to\infty$. There is a
second solution which decays as $v^{-\zeta}$ for $v\to\infty$ and the third
solution vanishes exponentially fast in the $v\to\infty$ limit. 
From the last two solutions we can therefore construct scaling functions 
with the physically expected asymptotic
behaviour. In addition we illustrate the convergence of $v^{\zeta}\Omega_p(v)$ 
towards $\Omega_{p,\infty}$, as given in (\ref{4:Omeinf}), by plotting
$v^{\zeta}\Omega_p(v)$ as a function of $v$ for several values of $\zeta$.
%%------------------------------------------------------------------------------
\begin{figure}
\centerline{\epsfxsize=5.5in\epsfbox{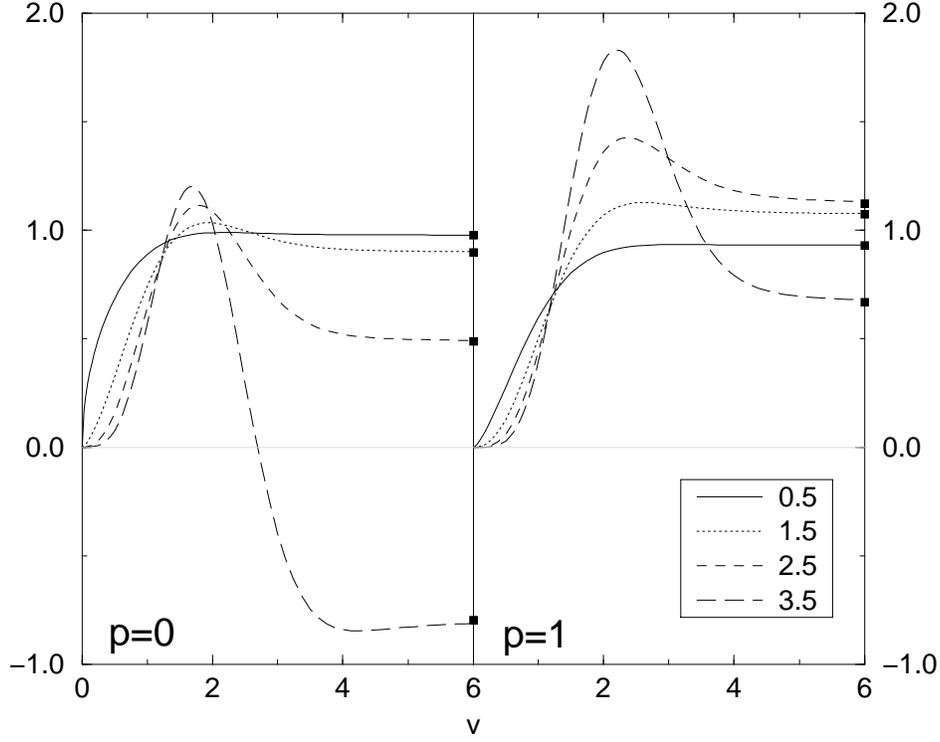}}
\caption{Scaling functions $v^{\zeta}\Omega_p(v)$ for Typ I, $N=4$, $\alpha_1=1$
and $p=0,1$. The different curves belong to values of $\zeta=N x_1$ 
as indicated and the squares are the values of $\Omega_{p,\infty}$.
\label{Abb_41}}
\end{figure}
%%------------------------------------------------------------------------------
%%------------------------------------------------------------------------------
\begin{figure}
\centerline{\epsfxsize=5.5in\epsfbox{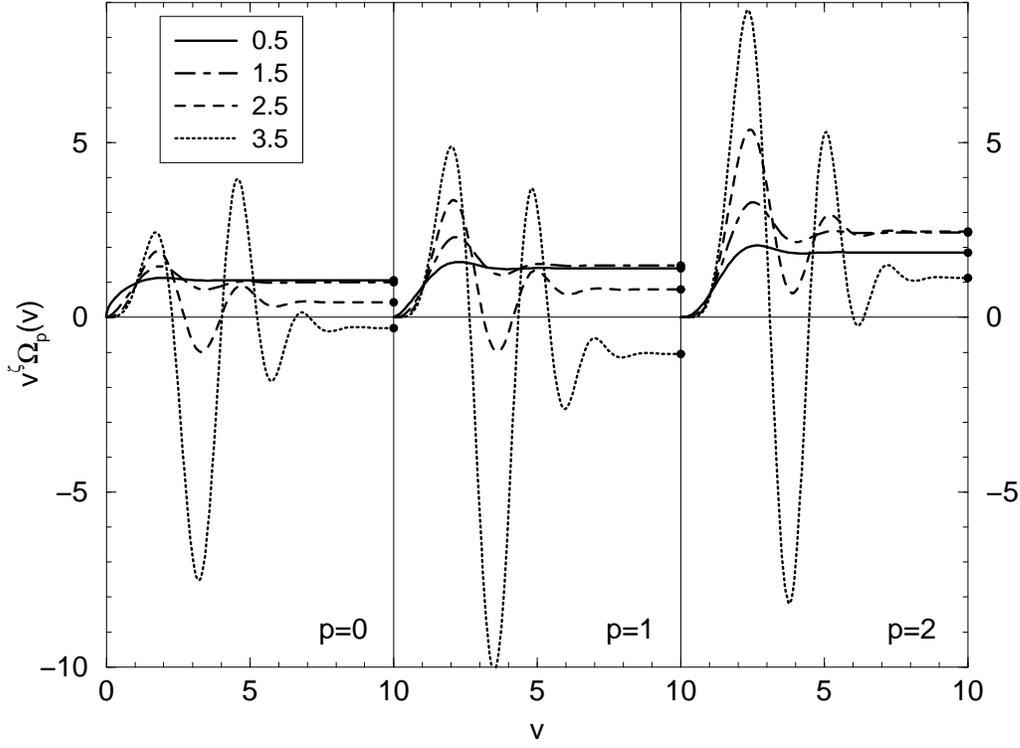}}
\caption{Scaling functions $v^{\zeta}\Omega_p(v)$ for Typ I, $N=5$, $\alpha_1=1$
and $p=0,1,2$, for different values of $\zeta=Nx_1$. The circles indicate
the values of $\Omega_{p,\infty}$.\label{Abb_42}}
\end{figure}
%%------------------------------------------------------------------------------
This is done for $N=4$ in figure~\ref{Abb_41} and for $N=5$ in 
figure~\ref{Abb_42} (we have also checked this for $N=6$). 
Besides confirming the correctness of the asymptotic
expressions (\ref{4:RandOme},\ref{4:Omeinf}) for $v\to\infty$, we also see
that for a large range of values of $\zeta$, 
the asymptotic regime is reached quite rapidly. 

Below, we shall need the explicit expressions for $\Omega_{0,1}(v)$ for $N=4$
\BEA
\Omega_{0}(v) &=& \frac{\Gamma(3/4)}{\Gamma(\zeta/4)} 
\sum_{n=0}^{\infty} \frac{\Gamma(n/2+\zeta/4)}{n! \Gamma(n/2+3/4)}
\left( - \frac{v^2}{2\sqrt{\alpha_1}}\right)^n  
\label{4:OmeN4L01}\\
\Omega_{1}(v) &=& \sqrt{\frac{\pi}{2}}\frac{v}{\Gamma((\zeta+1)/4)} 
\sum_{n=0}^{\infty} \frac{\Gamma((n+1+\zeta)/4) s(n)}
{\Gamma(n/4+1)\Gamma((n+3)/2)}
\left( - \frac{v}{\sqrt[4]{4\alpha_1}}\right)^n 
\nonumber 
\EEA
where $s(n):= \frac{1}{\sqrt{2}}\left( \cos\frac{n\pi}{4} + \sin\frac{n\pi}{4}
\right) \cos\frac{n\pi}{4}$. These expressions will be encountered again
in section 5 for the correlators of the ANNNI and ANNNS models at their 
Lifshitz points.

Next, we study what happens for $N$ not an integer. It is useful to
write the anisotropy exponent as
\BEQ
\frac{2}{\theta} = N = N_0 + \eps \;\; , \;\;
\eps = \frac{p}{q}
\EEQ
where $N_0\in\mathbb{N}$ and $p,q$ are positive coprime integers. 

For $N$ integer, we have seen that there is an unique solution which decays
as $\Omega(v)\sim v^{-\zeta}$ as $v\to\infty$. The presence of such a solution
for arbitrary $N$ may be checked by seeking solutions of the form
\BEQ
\Omega(v) = \sum_{n=0}^{\infty} a_n v^{-n/q+s} \;\; , \;\; a_0 \ne 0
\EEQ
In making this ansatz, we concentrate on those solutions of eq.~(\ref{4:DGLI})
which do {\em not\/} grow or vanish exponentially for $v$ large. 
As done before for Typ II, and under the same conditions, 
we find upon substitution 
\BEQ 
\sum_{n=p+qN_0}^{\infty} 
\frac{\alpha_1 \Gamma\left( \frac{1}{q}(-n+p+qN_0) + s+1\right)}{\Gamma\left( 
-\frac{n}{q} +2+s\right) } a_{n-p-qN_0} v^{-n/q}   
+ \sum_{n=0}^{\infty} \left( \frac{1}{q}(n-px_1-qN_0x_1)-s\right) 
a_n v^{-n/q}  = 0 
\EEQ
which must be valid for all positive values of $v$. 
Comparing the coefficients of $v$, we obtain
\BEA
s &=& - \frac{p}{q} x_1 - N_0 x_1 = - (N_0 +\eps) x_1 = - \zeta \nonumber \\
a_n &=& 0 \quad ; \quad \mbox{\rm for~~} n=1,2,\ldots,p+qN_0 -1 \\
\frac{n}{q} a_n &=& 
- \frac{\alpha_1 \Gamma\left(-\frac{1}{q}(n-p(1+x_1)-qN_0(1+x_1))+1\right)}
{\Gamma\left(-\frac{n}{q} -\frac{p}{q}x_1 -N_0 x_1 +2\right)}\,
a_{n-p-qN_0} 
\nonumber
\EEA
In principle, there might be additional $\delta$-function terms which come
from the definition (\ref{A:Ablei}) of the fractional derivative, 
see appendix A. However, if we either restrict to $v>0$ or else if $\zeta$ is
distinct from the discrete set of values $\zeta_c=2+m-n/q$, where
$n,m\in\mathbb{N}$, these terms do not occur. 

We can now let $n=(p+qN_0)\ell$ and $a_n=a_{(p+qN_0)\ell} =: b_{\ell}$ and
find the simpler recurrence
\BEQ
b_{\ell} = -\frac{\alpha_1}{N\ell}
\frac{\Gamma(1+N-(\ell+x_1)N)}{\Gamma(2-(\ell+x_1)N)}\, b_{\ell-1} 
\;\; , \;\; \ell =1,2,\ldots
\EEQ
which can be solved in a manner analogous to the one used for Typ II before,
with the result
\BEQ
b_{\ell} = b_0 \left(\frac{\alpha_1}{N^2}\right)^{\ell}
\frac{\Gamma\left(1+(\zeta-1)/N\right)}
{\ell !\, \Gamma\left(\ell+1+(\zeta-1)/N\right)}
\frac{\Gamma(1-\zeta)}{\Gamma(1-\zeta-\ell N)} 
\EEQ
In the special case $N=1$, the resulting series may be summed
straightforwardly and leads to the elementary result
$\Omega_{N=1}(v)=b_0 v^{-\zeta} e^{-\alpha_1/v}$, where $v=t r^{-2}$ and
$\zeta=x_1$. We thus recover the form (\ref{2:S2ponto}) of Schr\"odinger 
invariance for the two-point function $\langle\phi\phi^*\rangle$, as it 
should be. 

If we use the identity $\Gamma(x)\Gamma(1-x)=\pi/\sin(\pi x)$ and define
the function
\BEQ
\mathfrak{G}_{a,b}^{(N)}(z) := \Gamma(b)\Gamma(1-a) \sum_{\ell}^{\infty} 
\frac{\Gamma(\ell N+a)}{\ell !\, \Gamma(\ell+b)}\, z^{\ell}
\EEQ
the series solution with the requested behaviour 
$\Omega(v)\simeq b_0 v^{-\zeta}$ at $v\to\infty$ may be written
as follows
\BEA
\Omega(v) &=& v^{-\zeta} \sum_{\ell=0}^{\infty} b_{\ell}\, v^{-N\ell} 
\nonumber \\
&=& b_0 v^{-\zeta} \sum_{\ell=0}^{\infty} 
\frac{\Gamma\left(1+(\zeta-1)/N\right)}{\Gamma(\zeta)}
\frac{\sin\left(\pi(N\ell+\zeta)\right)}{\sin \pi\zeta}
\frac{\Gamma(\ell N+\zeta)}{\ell !\,\Gamma\left(l+1+(\zeta-1)/N\right)}
\left(\frac{\alpha_1}{N^2}\frac{1}{v^N}\right)^{\ell} ~~~~~ \label{4:Ome0Inf}\\
&=& \frac{b_0}{2\pi\II} v^{-\zeta} 
\left[ e^{\pi\II\zeta}\mathfrak{G}_{\zeta,1+\frac{\zeta-1}{N}}^{(N)}
\left(\frac{\alpha_1}{N^2}e^{\pi\II N} v^{-N}\right) 
-e^{-\pi\II\zeta}\mathfrak{G}_{\zeta,1+\frac{\zeta-1}{N}}^{(N)}
\left(\frac{\alpha_1}{N^2}e^{-\pi\II N} v^{-N}\right) \right]
\nonumber
\EEA
From these expressions, it is clear that the radius of convergence
of these series as a function of the variable $1/v$ is infinite for $N<2$ 
and zero for $N>2$. In the first case, we therefore have a convergent series 
for the scaling function, while in the second case, we have obtained an 
asymptotic expansion.

In several applications,
notably the $3D$ ANNNI model to be discussed in the next section, the
anisotropy exponent $\theta \simeq 1/2$ to a very good approximation. Therefore,
consider fractional derivatives of order $a=N_0+\eps$, where $N_0$ is an
integer and $\eps$ is small. To study perturbatively the solutions for
$\eps\ll 1$, we use the identity (\ref{A:faReihe}), set $a=N=N_0+\eps$ and 
expand to first order in $\eps$. The result is
\BEA 
\left.\partial_r^{N} f(r)\right|_{\rm reg} &=& \partial_r^{\eps}
\partial_r^{N_0} f(r) \nonumber \\
&\simeq& f^{(N_0)}(r)  + \eps {\cal L}_0 f^{(N_0)}(r) + O\left(\eps^2\right)\\
{\cal L}_0 g(r) &=&  \left[ -\left(C_E+\ln r\right)  + 
\sum_{\ell=1}^{\infty} \frac{(-1)^{\ell+1}}{\ell!\,\ell}\, r^{\ell} 
\frac{\D^{\ell}}{\D r^{\ell}} \right] g(r)
\nonumber
\EEA
where $C_E = 0.5772\ldots$ is Euler's constant. 
To find the first correction in $\eps$ with respect to the solution 
(\ref{4:OmeFinal}) when $N$ is an integer, we set again $N=N_0+\eps$ with
$N_0\in\mathbb{N}$ and $\eps>0$ and consider
\BEQ \label{4:OmegaStoe}
\Omega(v) = \Omega^{(0)}(v) + \eps \Omega^{(1)}(v) + O\left(\eps^2\right) 
\EEQ
Then $\Omega^{(0)}(v)$ solves eq.~(\ref{4:DGLI}) with $N=N_0$ 
and is consequently given by eq.~(\ref{4:OmeFinal}), whereas
$\Omega^{(1)}(v)$ satisfies the equation
\BEQ \label{4:Stoer1}
\left( \alpha_1 \partial_v^{N_0-1} - v^{2}\partial_v - \zeta v \right) 
\Omega^{(1)}(v) =  \omega(v) := - {\cal L}_{0} 
\left( v^2\partial_v+\zeta v\right) \Omega^{(0)}(v) 
\EEQ
which we now study. 

First, we consider the limiting behaviour of $\Omega^{(1)}(v)$ for $v$ either
very large or very small. If $v\gg 1$, we see from eq.~(\ref{4:Ome0Inf}) that
$\Omega^{(0)}(v) \sim v^{-\zeta}(1+{\rm O}(v^{-N_0}))$. This implies in turn
that $\omega(v)\simeq (A_{\infty}+B_{\infty}\ln v)v^{-\zeta-(N_0-1)}$, where
$A_{\infty}, B_{\infty}$ are some constants. Therefore one must have
$\Omega^{(1)}(v)\sim v^{-\zeta}(1+{\rm O}(v^{-N_0}))$ in order to reproduce 
this result for $\omega(v)$. On the other hand, if $v\ll 1$, we have 
$\Omega^{(0)}(v)\simeq {\rm cste.}$ which leads to $\omega(v)\simeq
(A_0 + B_0 \ln v)$ with some constants $A_0,B_0$. This can be reproduced 
from the limiting behaviour $\Omega^{(1)}(v) \sim {\rm O}(v, v^{N_0} \ln v)$. 
In conclusion, the first-order perturbation is compatible with the boundary
condition (\ref{3:TypIR}) for the full scaling function $\Omega(v)$.

We now work out the first correction $\Omega^{(1)}(v)$ explicitly for
$N_0=4$. This is the case we shall need in section 5. 
There are two physically acceptable solutions 
$\Omega_0^{(0)}(v)$ and $\Omega_1^{(0)}(v)$ of zeroth order in $\eps$ 
which are given in (\ref{4:OmeN4L01}). The general zeroth-order solution
is given by 
\BEQ
\Omega^{(0)}(v)=\Omega_0^{(0)}(v)+
\frac{\mathfrak{p}}{\alpha_1^{1/4}}\Omega_1^{(0)}(v)
\EEQ
where $\mathfrak{p}=b_1 \alpha_1^{1/4}/b_0$ is a universal constant. Then
all metric factors in the scaling function are absorbed into 
the argument $v/\alpha_1^{1/4}$ and the form of $\Omega(v)$
is given by the two universal parameters $\zeta$ and $\mathfrak{p}$. 

Consider the first-order correction to $\Omega^{(0)}(v)$. From the explicit 
form of ${\cal L}_0$ and (\ref{4:OmeN4L01}), we have
\BEQ
\omega(v) = \sum_{n=0}^{\infty} \left[ A_n v^{n+1}+B_n v^{n+1}\ln v\right]
\EEQ
where $A_n = - \psi(n+1) B_n$ and 
\BEQ \label{4:Bn}
B_n = \left\{ \begin{array}{ll}
\frac{\Gamma(n+\zeta/4)\,\Gamma(3/4)}{\Gamma(n+3/4)\,\Gamma(\zeta/4)}
\frac{4n+\zeta}{(2n)!} \left(\frac{1}{4\alpha_1}\right)^{n} & 
;~~ n \equiv 0 \bmod 4 \\
\frac{1}{\alpha_1^{1/4}} 
\frac{\sqrt{\pi}\,\Gamma(n+(\zeta+1)/4)}{\Gamma((\zeta+1)/4)\,\Gamma(2n+3/2)}
\frac{4n+1+\zeta}{n!\: 2}\left(\frac{1}{4\alpha_1}\right)^{n} \mathfrak{p} &
;~~ n \equiv 1 \bmod 4 \\
-\frac{1}{\sqrt{\alpha_1}}\frac{\Gamma(n+(\zeta+2)/4)}{2\,\Gamma(n+5/4)}
\frac{4n+2+\zeta}{(2n+1)!} \left(\frac{1}{4\alpha_1}\right)^{n} 
\left[\frac{\Gamma(3/4)}{\Gamma(\zeta/4)}+
\mathfrak{p}\frac{1}{\Gamma((\zeta+1)/4)}\sqrt{\frac{\pi}{2}}
\right] & 
;~~ n \equiv 2 \bmod 4 \\
0 &
;~~ n \equiv 3 \bmod 4
\end{array} \right.
\EEQ
Here $\psi(z)=\Gamma'(z)/\Gamma(z)$ is the digamma function \cite{Abra65} and 
the identity
\BEQ
\sum_{\ell=1}^{\infty} \left(\vekz{\alpha}{\ell}\right) 
\frac{(-1)^{\ell+1}}{\ell} = C_E + \psi(\alpha+1)
\EEQ
was used. The solution of the third-order differential equation 
(\ref{4:Stoer1}) is of the form
\BEQ
\Omega^{(1)}(v) = \sum_{n=0}^{\infty} \left[ a_n v^{n}+b_n v^{n}\ln v\right]
\EEQ
where in addition
\BEQ
a_0 = a_3 = 0 \;\; , \;\; b_0 = b_1 = b_2 = b_3 = 0
\EEQ
and, for all $n\geq 0$ 
\BEA
b_{n+4} &=& \frac{1}{\alpha_1\,(n+4)(n+3)(n+2)} \left[ 
B_n + b_n \left( n+\zeta\right) \right] 
\nonumber \\
a_{n+4} &=& \frac{1}{\alpha_1\,(n+4)(n+3)(n+2)} \left[ 
A_n + a_n \left( n+\zeta\right) + b_n - 
b_{n+4} \alpha_1 \left( 3n^2 + 18 n + 26\right)\right]
\label{4:abRek}
\EEA
The value of the constant $a_0$ is fixed because of the boundary condition
$\Omega(0)=1$. The first perturbative correction 
$\Omega^{(1)}(v)$ still depends on the free parameters 
$a_1=:\mathfrak{p}\,\mathfrak{a}$ and $a_2=:\mathfrak{b}$. We also observe
that because of (\ref{4:Bn}), we have $a_{4n+3}=b_{4n+3}=0$ for all 
$n\in\mathbb{N}$ and that the metric factor $\alpha_1$ merely sets the scale
in the variable $v/\alpha_1^{1/4}$ but does not otherwise affect the functional
form of $\Omega^{(1)}(v)$. 

We have seen above that for $v\gg 1$, we must recover 
$\Omega(v) \sim v^{-\zeta}$. We can therefore fix $\mathfrak{a}$ and 
$\mathfrak{b}$ such that the correction term $\Omega^{(1)}(v)$ goes to zero 
for $v$ large. Furthermore, we see from figure~\ref{Abb_41} that the asymptotic
regime is already reached for quite small values of $v$ for a large range of 
values of $\zeta$. To a good approximation, we can therefore determine 
$\mathfrak{a}$ and $\mathfrak{b}$ from the requirement that 
$\Omega^{(1)}(v_0)=0$ and $\D\Omega^{(1)}(v_0)/\D v=0$ if $v_0$ is finite,
but chosen to be sufficiently large. The recursion (\ref{4:abRek}) then gives
a system of two linear equations for $\mathfrak{a}$ and $\mathfrak{b}$. 

%%------------------------------------------------------------------------------
\begin{figure}[t]
\centerline{
\epsfxsize=3.4in\ \epsfbox{
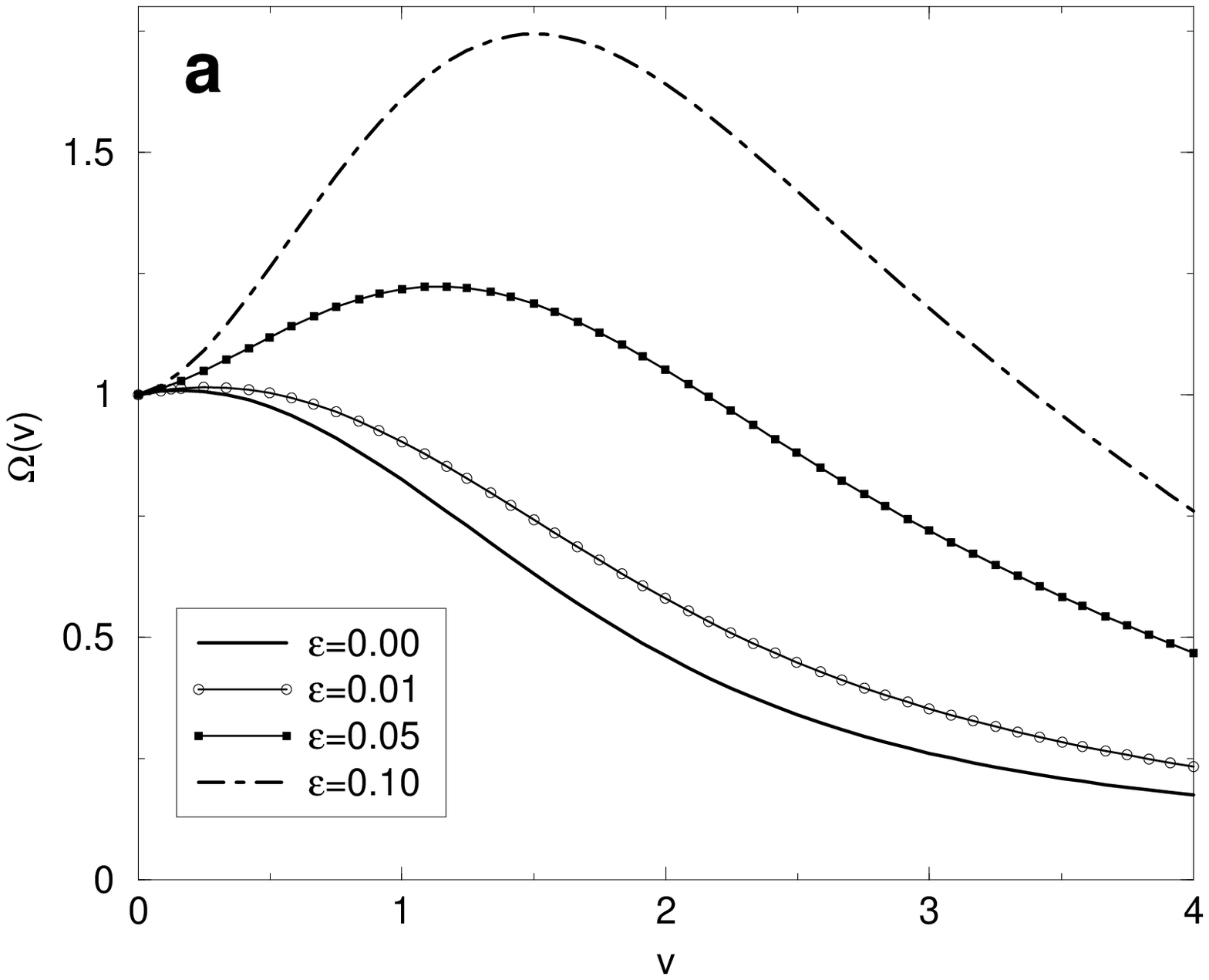}
\epsfxsize=3.4in\ \epsfbox{
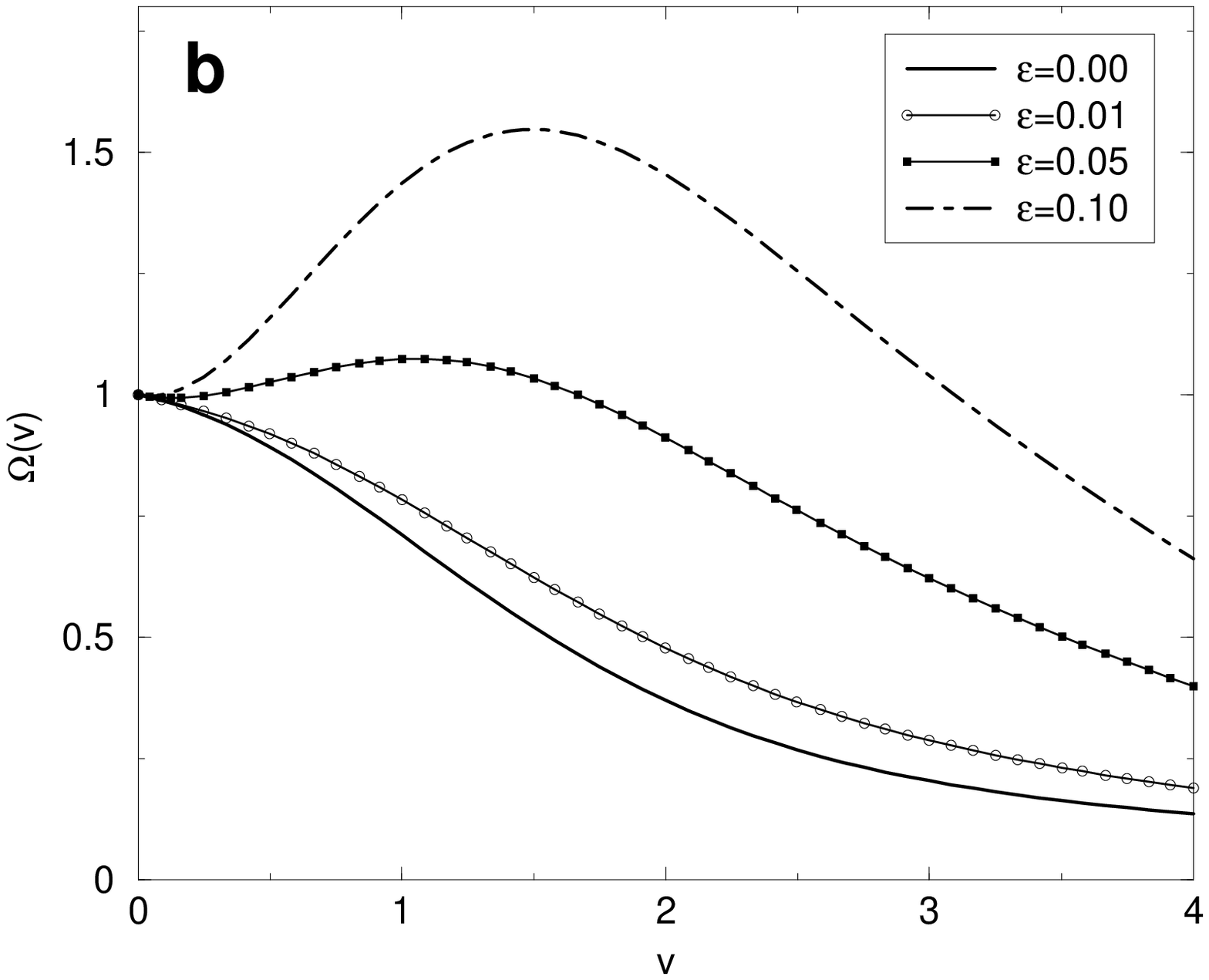}
}
\caption{Perturbative scaling functions $\Omega(v)$ 
{\protect eq.~(\ref{4:OmegaStoe})} 
for Typ I around $N_0=4$, with $\zeta=1.3$, $\alpha_1=1$ and 
(a) $\mathfrak{p}=+0.11$ and 
(b) $\mathfrak{p}=-0.11$, for several values of $\eps$. 
\label{Abb_45}}
\end{figure}
%%------------------------------------------------------------------------------
As an example, we illustrate this in figure~\ref{Abb_45} for $\zeta=1.3$. From
figure~\ref{Abb_41}, we observe that $v_0=6$ is already far in the asymptotic
regime. For the values $\mathfrak{p}=\pm 0.11$, the scaling function 
(\ref{4:OmegaStoe}), with the 
first-order correction included, is shown for several values of $\eps=N-4$. The 
first-order perturbative corrections with respect to the solution found for 
$N=4$ are quite substantial, even for small values of $\eps$. This suggests 
that a non-integer value of $N$ in the differential equation (\ref{4:DGLI}) 
should be readily detectable in numerical simulations. We shall come back to 
this in section 5 in the context of the $3D$ ANNNI model.

The full series solution leads to difficulties with the boundary condition
$\Omega(0)=1$. This is further discussed in appendix D. 

%%%%%%%%%%%%%%%%%%%%%%%%%%%%%%%%%%%%%%%%%%%%%%%%%%%%%%%%%%%%%%%%%%%%%%%%%%%%%%%%
\section{Applications}
%%%%%%%%%%%%%%%%%%%%%%%%%%%%%%%%%%%%%%%%%%%%%%%%%%%%%%%%%%%%%%%%%%%%%%%%%%%%%%%%

\subsection{Uniaxial Lifshitz points}

Uniaxial Lifshitz points \cite{Horn75} are paradigmatic examples of equilibrium
spin systems with a strongly anisotropic critical behaviour. 
They are conveniently realized
in spin systems with competing interactions. Besides the well-known
uniaxially modulated magnets, alloys and ferroelectrics 
\cite{Yeom88,Selk88,Neub98}, recently found 
new examples include ferroelectric liquid crystals, uniaxial ferroelectrics, 
block copolymers, spin-Peierls and quantum systems 
\cite{Skab00,Vyso92,Bate95,Ohwa01,Schr98}. 
For the sake of notational
simplicity, we merely consider {\em uniaxial} competing interactions, which
are described by the Hamiltonian
\BEQ \label{5:NNNHam}
{\cal H} = -J \sum_{(\vec{i},\vec{j})} s_{\vec{i}} s_{\vec{j}} 
+ \kappa_1 J \sum_{\vec{i}} s_{\vec{i}} s_{\vec{i}+2\vec{e}_{\|}}
+ \kappa_2 J \sum_{\vec{i}} s_{\vec{i}} s_{\vec{i}+3\vec{e}_{\|}}
\EEQ
where $s_{\vec{i}}$ are the spin variables at site $\vec{i}$. We shall consider
here the ANNNI model,\footnote{The axial next-nearest neighbour Ising/spherical
or ANNNI/S model is given by (\ref{5:NNNHam}) with $\kappa_2=0$. 
The case $\kappa_2\ne 0$ is sometimes referred to as the A3NNI model. For 
simplicity, we take here ANNNI/S to stand for axial {\em non}-nearest neighbour 
Ising/spherical and keep these abbreviations also for $\kappa_2\ne 0$.} 
where $s_{\vec{i}}=\pm 1$ are Ising spins and the
ANNNS model, where the $s_{\vec{i}}\in\mathbb{R}$ and satisfy the spherical
constraint $\sum_{\vec{i}} s_{\vec{i}}^2 = {\cal N}$, where $\cal N$ is the
total number of sites of the lattice. The first sum runs only over pairs
of nearest-neighbour sites of a hypercubic lattice in $d=d_{\perp}+1$ 
dimensions. In the second and third sums, additional interactions between 
second and third neighbours are added along a chosen axis ($\|$) and 
$\vec{e}_{\|}$ is the unit vector in this direction. Finally, 
$J>0$ and $\kappa_{1,2}$ are coupling constants. For reviews, see
\cite{Yeom88,Selk88,Selk92,Neub98,Dieh02a}. 

In order to understand the physics of the model, we take $\kappa_2=0$ for a
moment. If in addition $\kappa_1$ is small, the model undergoes at some 
$T_c=T_c(\kappa_1)$ 
a second-order phase transition which is in the Ising or spherical model 
universality class, respectively, for the systems considered here. However, 
if $\kappa_1$ is large and positive, the zero-temperature ground state may 
become spatially modulated and a rich phase diagram is obtained 
\cite{Yeom88,Selk88,Selk92,Neub98}. A particular
multicritical point is the meeting point of the disordered paramagnetic, the
ordered ferromagnetic and the ordered incommensurate phase. This point is 
called an uniaxial {\em Lifshitz point} (of first order) \cite{Horn75}. 
If one now lets vary $\kappa_2$, one obtains a line of Lifshitz points of 
first order. This line terminates in a {\em Lifshitz point of second order} 
\cite{Nico76,Selk77}. Lifshitz points of order $L-1$ can be defined analogously 
and exist at non-zero temperatures for $d>d_{*}$ \cite{Selk77}. 
For the ANNNS model, the lower critical dimension is 
\BEQ
d_{*} :=2+(L-1)/L
\EEQ

Close to a Lifshitz point, the scaling of the correlation functions 
$C(r_{\|},\vec{r}_{\perp})$ is strongly direction-dependent. Here $r_{\|}$
is the distance along the chosen axis with the competing interactions 
and $\vec{r}_{\perp}$ is the distance vector in the remaining $d_{\perp}$ 
directions where only nearest-neighbour interactions exist. Slightly off
criticality, correlations decay exponentially, but the scaling of the 
correlation lengths is direction-dependent
\BEQ
\xi_{\|} \sim \left( T - T_L \right)^{-\nu_{\|}} \;\; , \;\;
\xi_{\perp} \sim \left( T - T_L \right)^{-\nu_{\perp}}
\EEQ
where $T_L$ is the location of the Lifshitz point.
The anisotropy between the axial ($\|$) and the other ($\perp$) directions 
is measured in terms of the anisotropy exponent
\BEQ
\theta = \nu_{\|}/\nu_{\perp}
\EEQ  
Precisely {\em at} the Lifshitz point, one expects
\BEQ
C_{\sigma}(r_{\|},\vec{0}) \sim r_{\|}^{-2x_{\sigma}/\theta} \;\; , \;\;
C_{\sigma}(0,\vec{r}_{\perp}) \sim r_{\perp}^{-2x_{\sigma}} \;\; , \;\;
C_{\eps}(r_{\|},\vec{0}) \sim r_{\|}^{-2x_{\eps}/\theta} \;\; , \;\;
C_{\eps}(0,\vec{r}_{\perp}) \sim r_{\perp}^{-2x_{\eps}}
\EEQ
for the connected spin-spin correlator $C_{\sigma}$ and the connected
energy-energy correlator $C_{\eps}$, respectively, and where $x_{\sigma}$ and
$x_{\eps}$ are scaling dimensions. The critical exponents $\alpha,\beta,\gamma$
are defined as usual from the specific heat, the order parameter and the
susceptibility, but some of the familiar scaling relations valid for isotropic
systems (where $\theta=1$) must be replaced by
\BEQ
2 - \alpha = d_{\perp} \nu_{\perp} + \theta \nu_{\|} \;\; , \;\;
\gamma = \left(2-\eta_{\perp}\right)\nu_{\perp} 
       = \left(2/\theta-\eta_{\|}\right)\nu_{\|}
\EEQ
where the anomalous dimensions $\eta_{\|,\perp}$ are defined from the 
spin-spin correlator
\BEQ
C_{\sigma}(0,\vec{r}_{\perp}) 
\sim r_{\perp}^{-[d_{\perp}+\theta-2+\eta_{\perp}]} 
\;\; , \;\;
C_{\sigma}(r_{\|},\vec{0})
\sim r_{\|}^{-[(d_{\perp}+\theta-2)/\theta+\eta_{\|}]}
\EEQ
and are related via $\eta_{\|}=\eta_{\perp}/\theta$. Alternatively, one
often works with exponents $\nu_{\ell2}=\nu_{\perp}$, $\nu_{\ell4}=\nu_{\|}$, 
$\eta_{\ell2}=\eta_{\perp}$ and
$\eta_{\ell4}=\eta_{\|}+4-2/\theta$, see e.g. \cite{Horn75,Dieh00}.
Then $\gamma=(4-\eta_{\ell4})\nu_{\ell4}=(2-\eta_{\ell2})\nu_{\ell2}$.  

Standard renormalization group arguments lead to the following anisotropic
scaling of the correlation functions
\BEQ
C_{\sigma,\eps}(r_{\|},\vec{r}_{\perp}) = b^{-2x_{\sigma,\eps}} 
C_{\sigma,\eps}(r_{\|} b^{-\theta}, \vec{r}_{\perp} b^{-1})
= r_{\perp}^{-\zeta_{\sigma,\eps}\theta} 
\Omega_{\sigma,\eps}\left(r_{\|}/r_{\perp}^{\theta}\right)
\EEQ
for both the spin-spin and the energy-energy correlators, respectively. 
For a Lifshitz point in $(d_{\perp}+1)$ dimensions, we have
\BEQ \label{5:zetase}
\zeta_{\sigma} = \frac{2(\theta+d_{\perp})}{\theta(2+\gamma/\beta)} \;\; , \;\;
\zeta_{\eps} = \frac{2(\theta+d_{\perp})(1-\alpha)}{\theta(2-\alpha)}
\EEQ

We want to compare the form of the spin-spin correlator with the predictions
of local scale invariance. We begin with Lifshitz points of first order. 
Then, as will be discussed further below, $\theta\simeq \frac{1}{2}$ 
at least to a good approximation. In terms of the notation of sections 3 and 4,
this corresponds to $N=2/\theta=4$. For $N=4$, we recall the two-point 
function of Typ I 
\BEQ \label{5:OmegaN4}
C(r_{\|},\vec{r}_{\perp}) = r_{\perp}^{-\zeta/2}\: b_0 \: 
\left( \Omega_0(v) + \frac{\mathfrak{p}}{\alpha_1^{1/4}} \Omega_1(v) \right)
\;\; , \;\; v = t r^{-1/2} 
\EEQ
where $\Omega_{0,1}(v)$ are explicitly given in eq.~(\ref{4:OmeN4L01}).
The functional form of $\Omega(v)$ only depends on the universal parameters
$\zeta$ and $\mathfrak{p}$. The metric factor $\alpha_1$ only 
arises as a scale factor through the argument $v\,\alpha_1^{-1/4}$. 

We shall now present tests of the two-point function of Typ I of 
local scale invariance in three distinct universality classes.  

{\bf 1.} Our first example is the exactly solvable ANNNS model. 
The phase diagram is well-known and uniaxial Lifshitz points of 
first order occur along the line \cite{Nico76,Selk77,Frac93}
\BEQ \label{5:LP1ANNNS}
\kappa_2 = \frac{1}{9} \left( 1 - 4 \kappa_1\right) \;\; , \;\;
\kappa_1 < \frac{2}{5} 
\EEQ
with a known $T_L=T_L(\kappa_1,\kappa_2)$. 
The lower critical dimension $d_{*}=\frac{5}{2}$. 
We need the following exactly known critical exponents in $d$ 
dimensions \cite{Nico76,Selk77}
\BEQ \label{5:ANNNSexp}
\beta=\frac{1}{2} \;\; , \;\; \gamma = \frac{4}{2d-5} \;\; , \;\; 
\theta = \frac{1}{2} \;\; , \;\; \zeta_{\sigma}=2\left(d-\frac{5}{2}\right)
\EEQ
which means $N=4$ in our notation. 
The exact spin-spin correlator along the line (\ref{5:LP1ANNNS}) of Lifshitz 
points is \cite{Frac93}
\BEA
C_{\sigma}(r_{\|},\vec{r}_{\perp}) &=& C_0\: r_{\perp}^{-(d-d_{*})} 
\Psi\left( \frac{d-d_{*}}{2}, 
\frac{1}{4}\sqrt{\frac{3}{2-5\kappa_1}} \frac{r_{\|}^2}{r_{\perp}} \right) \\
\Psi(a,x) &=& \sum_{k=0}^{\infty} \frac{(-1)^k}{k!} 
\frac{\Gamma(k/2+a)}{\Gamma(k/2+3/4)} x^k
\EEA
where $C_0$ is a (known) normalization constant.\footnote{Properties of
the function $\Psi(a,x)$ are analysed in \cite{Frac93,Shpo01}. 
Explicit expressions are known for integer values of $\zeta=4a$ and may be
recovered as special cases of the functions listed in table~\ref{tab2}.} 
This reproduces the exponent $\zeta_{\sigma}$ from eq.~(\ref{5:ANNNSexp}). 
Comparing with the expected form (\ref{5:OmegaN4}) and the specific functions 
(\ref{4:OmeN4L01}), we see that
\BEQ
\Omega_0(v) = \frac{\Gamma(3/4)}{\Gamma(\zeta_{\sigma}/4)} 
\Psi\left(\frac{\zeta_{\sigma}}{4}, \frac{v^2}{2\sqrt{\alpha_1}}\right) 
\EEQ
With the correspondence $t \leftrightarrow r_{\|}$, 
$r\leftrightarrow r_{\perp}$ and the non-universal metric factor
$\alpha_1 = \frac{4}{3}(2-5\kappa_1)$, we therefore observe
complete agreement. In particular, we identify
the universal parameter $\mathfrak{p}=0$ \cite{Henk97}. 

{\bf 2.} Next, we consider the uniaxial Lifshitz point in the $3D$ ANNNI model. 
Two complimentary approaches have been used. First, the model may be
formulated in terms of a $n$-component field 
$\vec{\phi}(r_{\|},\vec{r}_{\perp})$ with a global O($n$)-symmetry and
spatially anisotropic interactions \cite{Horn75}. This model, which might
be called {\em ANNNO($n$) model}, reduces to the ANNNI model in the special
case $n=1$ and gives the ANNNS model in the $n\to\infty$ limit. Recently,
Diehl and Shpot \cite{Dieh00,Shpo01,Dieh01,Dieh02,Dieh02a} studied very 
thoroughly the field-theoretic renormalization group of the ANNNO($n$) model 
{\em at} the Lifshitz point at the two-loop level and derived
the critical exponents to second order in the $\eps$-expansion, where
$\eps=4.5-d$.\footnote{Another recent two-loop calculation \cite{Albu02} 
apparently used
some uncontrolled approximation in order to be able to evaluate the two-loop
integrals analytically. See \cite{Dieh01,Dieh02} for a critical discussion.} 
Second, one may resort to numerical methods, such as series expansions 
\cite{Oitm85,Mo91} or Monte Carlo simulations. While older simulational
studies \cite{Kask85} were restricted to small systems, the use of modern 
cluster algorithms \cite{Wolf89} allows to simulate considerably larger 
systems. The Wolff algorithm can be adapted to systems with competing
interactions beyond nearest neighbours such as the ANNNI model 
\cite{Plei01,Henk01a}. In addition, a recently proposed scheme \cite{Ever01}
permits the direct computation of two-point functions on an effectively
infinite lattice. That technique can be extended to ANNNI models as well
\cite{Plei01,Henk01a}. 

%%~~~~~~~~~~~~~~~~~~~~~~~~~~~~~~~~~~~~~~~~~~~~~~~~~~~~~~~~~~~~~~~~~~~~~~~~~~~~~~
\begin{table}
\caption{Estimates for the location of the uniaxial Lifshitz point of the $3D$
ANNNI model on a cubic lattice. The numbers in brackets give the uncertainty
in the last digit(s).\label{tab5}}
\begin{center}
\begin{tabular}{|ll|ll|} \hline 
\multicolumn{1}{|c}{$T_L$} & \multicolumn{1}{c|}{$\kappa_{1,L}$} & & \\ \hline 
3.73(3)    & 0.270(5) & high-temperature series & \cite{Oitm85} \\
3.77(2)    & 0.265    & Monte Carlo             & \cite{Kask85} \\ 
3.7475(50) & 0.270(4) & cluster Monte Carlo     & \cite{Plei01} \\ \hline
\end{tabular} \end{center}
\end{table}
%%~~~~~~~~~~~~~~~~~~~~~~~~~~~~~~~~~~~~~~~~~~~~~~~~~~~~~~~~~~~~~~~~~~~~~~~~~~~~~~
Before the scaling form of any correlator can be tested, the Lifshitz
point must be located precisely. In table~\ref{tab5} we show some estimates
for the coupling $\kappa_{1,L}$ and the Lifshitz point critical temperature
$T_L$. Here the ANNNI Hamiltonian (\ref{5:NNNHam}) with $\kappa_2=0$ on a
$3D$ simple cubic lattice was used. The increase in precision coming from
the new cluster algorithm is evident and we take the estimates obtained in 
\cite{Plei01} as the location of the Lifshitz point. 

%%~~~~~~~~~~~~~~~~~~~~~~~~~~~~~~~~~~~~~~~~~~~~~~~~~~~~~~~~~~~~~~~~~~~~~~~~~~~~~~
\begin{table}
\caption{Estimates for critical exponents at the uniaxial Lifshitz point of 
the $3D$ ANNNI model. The numbers in brackets give the uncertainty
in the last digit(s).\label{tab6}}
\begin{center}
\begin{tabular}{|llll|ll|} \hline 
\multicolumn{1}{|c}{$\alpha$} & \multicolumn{1}{c}{$\beta$} & 
\multicolumn{1}{c}{$\gamma$} & \multicolumn{1}{c|}{$\theta$} & & \\ \hline
0.20(15) &         & 1.62(12) &       & high-temperature series & \cite{Mo91} \\
         & 0.19(2)  & 1.40(6) &       & Monte Carlo  & \cite{Kask85} \\
0.160    & 0.220    & 1.399   & 0.487 & renormalized field theory & 
\cite{Shpo01,Dieh02a}\\
0.18(2)  & 0.238(5) & 1.36(3) &       & cluster Monte Carlo  & \cite{Plei01} \\
\hline
\end{tabular} \end{center}
\end{table}
%%~~~~~~~~~~~~~~~~~~~~~~~~~~~~~~~~~~~~~~~~~~~~~~~~~~~~~~~~~~~~~~~~~~~~~~~~~~~~~~

Next, the anisotropy exponent $\theta$ and the scaling dimension 
$\zeta_{\sigma}$ must be found. While it had been believed for a long time
that also for the ANNNI model $\theta=\frac{1}{2}$ might hold, 
it has been recently established that to the second order in the 
$\eps$-expansion $\theta= \frac{1}{2} - \mathfrak{a}\eps^2 +O(\eps^3)$, where 
$\mathfrak{a}\simeq 0.0054$ in the $3D$ ANNNI model \cite{Dieh00}. 
In table~\ref{tab6} we list two older and the most recent estimates for the
Lifshitz point critical exponents $\alpha,\beta,\gamma$ and $\theta$. 
A direct determination of $\theta$ from simulational data is not yet possible.
Since in \cite{Plei01} the exponents $\alpha,\beta,\gamma$ were determined
independently, their agreement with the scaling relation 
$\alpha+2\beta+\gamma=2$ to within $\approx 0.8\%$ allows for an {\it a 
posteriori} check on the quality of the data. For details on the simulational
methods we refer to \cite{Plei01,Henk01a}. 

If we take the exponent estimates of \cite{Plei01} and in addition set 
$\theta=\frac{1}{2}$, we find from (\ref{5:zetase}) for the $3D$ ANNNI model
($d_{\perp}=2$) 
\BEQ \label{5:zetaANNNI}
\zeta_{\sigma} = 1.30 \pm 0.05 \;\; , \;\; \zeta_{\eps} = 4.5 \pm 0.2
\EEQ
where the errors follow from the quoted uncertainties in the determination
of the exponents $\alpha,\beta,\gamma$. If we now take $\theta=0.48$ as 
suggested by two-loop results of \cite{Shpo01}, the resulting variation of
both $\zeta_{\sigma}$ and $\zeta_{\eps}$ stays within the error bars quoted
in eq.~(\ref{5:zetaANNNI}). 
In conclusion, given the precision of the available exponent estimates, any
effects of a possible deviation of $\theta$ from $\frac{1}{2}$ are not yet
notable. We shall therefore undertake the subsequent analysis of the 
correlator by making the working hypothesis $\theta=\frac{1}{2}$ \cite{Plei01}.

%%------------------------------------------------------------------------------
\begin{figure}[t]
\centerline{
\epsfxsize=4.2in\ \epsfbox{
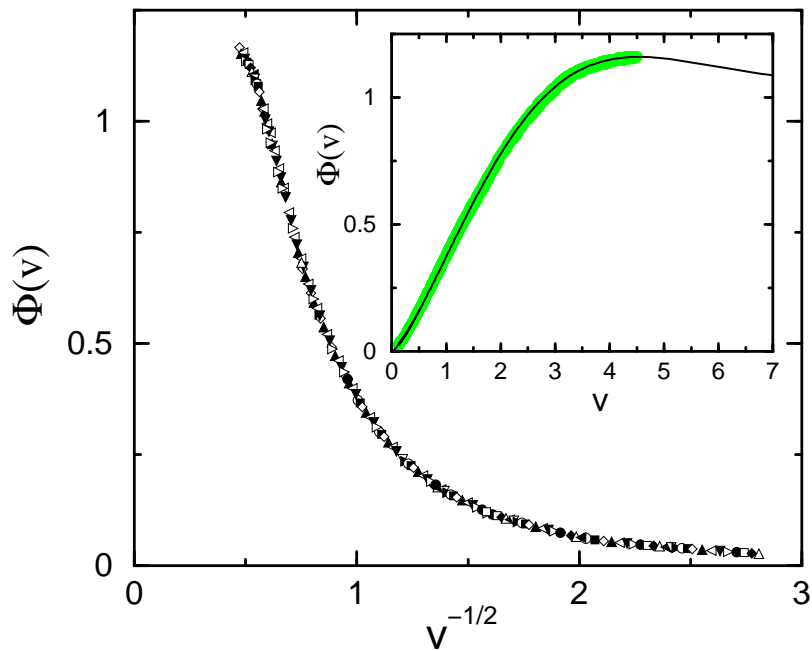}
}
\caption{Scaling function $\Phi(u)$ as defined in {\protect (\ref{5:Phi_u})} 
for the $3D$ ANNNI model, for $\kappa_1=0.270$ and
$T=3.7475$. Selected data on a $200\times 200\times 100$ lattice are shown. 
The symbols correspond to several values of $r_{\perp}$. The inset shows the
full data set (gray points) and the prediction {\protect (\ref{5:OmegaN4})}
of local scale invariance with $N=4$ and $\mathfrak{p}=-0.11$, $\alpha_1=33.2$
and $b_0=0.41$. The data are from \cite{Plei01}. 
\label{Abb_53}}
\end{figure}
%%------------------------------------------------------------------------------

In this case, we can compare with the scaling prediction 
(\ref{5:OmegaN4}) obtained for $N=2/\theta=4$. In figure~\ref{Abb_53}
we show data \cite{Plei01} for the modified scaling function of the 
spin-spin correlator
\BEQ \label{5:Phi_u} 
\Phi(v) = v^{\zeta_{\sigma}} \Omega_{\sigma}(v) 
\EEQ
The clear data collapse establishes scaling. It can be checked
that there is no perceptible change in the scaling plot for values of 
$\theta$ slightly less than $\frac{1}{2}$ \cite{Plei01}. 

For a quantitative comparison with (\ref{5:OmegaN4}), one
may consider the moments
\BEQ
M(n) = \int_{0}^{\infty} \!\D v\, v^n \Phi(v) 
\EEQ
Then it is easy to show \cite{Benz84} that the moment ratios
\BEQ
J_k\left(\left\{m_i\right\};\left\{n_j\right\}\right) = 
\left. \prod_{i=1}^{k} M(m_i) \right/ \prod_{j=1}^{k} M(n_j) \;\; , \;\;
\mbox{\rm ~~with~~} \sum_{i=1}^{k} m_i = \sum_{j=1}^{k} n_j
\EEQ
and $k\geq 2$ are independent of $b_0$ and $\alpha_1$. They only depend on the
functional form of $\Phi(u)$. This in turn is determined by $\zeta_{\sigma}$ 
and $\mathfrak{p}$. Therefore, the determination of a certain moment ratio
allows, with $\zeta_{\sigma}$ given by (\ref{5:zetaANNNI}), to find a
value for $\mathfrak{p}$. The Monte Carlo data will be consistent with
(\ref{5:OmegaN4}) if the values of $\mathfrak{p}$ found from several distinct
ratios $J_k$ coincide. In practise, these integrals cannot be calculated up to
$v=\infty$ but only to some finite value $v_0$ 
and the moments retain a dependence
on $\alpha_1$ through the upper limit of integration. Then an iteration
procedure must be used to find $\mathfrak{p}$ and $\alpha_1$ simultaneously
\cite{Plei01}. The results are collected in table~\ref{tab7}. 
  
%%~~~~~~~~~~~~~~~~~~~~~~~~~~~~~~~~~~~~~~~~~~~~~~~~~~~~~~~~~~~~~~~~~~~~~~~~~~~~~~
\begin{table}
\caption{Values of the parameters $\mathfrak{p}$ and $\alpha_1$ as determined
from different moment ratios 
$J_k\left(\left\{m_i\right\};\left\{n_j\right\}\right)$. \label{tab7}}
\begin{center}
\begin{tabular}{|lll|ll|} \hline
$k$ & \multicolumn{1}{c}{$\{m_i\}$} & \multicolumn{1}{c|}{$\{n_j\}$} & 
\multicolumn{1}{c}{$\mathfrak{p}$} & \multicolumn{1}{c|}{$\alpha_1$} \\ \hline
2 & $\{0, -0.5\}$        & $\{-0.25,-0.25\}$    & -0.102 & 32.7 \\
2 & $\{-0.25,-0.75\}$    & $\{-0.5,-0.5\}$      & -0.125 & 34.0 \\
2 & $\{0.2,-0.9\}$       & $\{0,-0.7\}$         & -0.100 & 32.8 \\
3 & $\{0.2,-0.6,-0.8\}$  & $\{-0.3,-0.4,-0.5\}$ & -0.102 & 32.8 \\
3 & $\{-0.1,-0.6,-0.7\}$ & $\{-0.4,-0.5,-0.5\}$ & -0.117 & 33.5 \\ \hline
\end{tabular}\end{center}
\end{table}
%%~~~~~~~~~~~~~~~~~~~~~~~~~~~~~~~~~~~~~~~~~~~~~~~~~~~~~~~~~~~~~~~~~~~~~~~~~~~~~~

Clearly, the two parameters can be consistently determined from different
moment ratios. The final estimate is \cite{Plei01} 
\BEQ 
\mathfrak{p} = -0.11 \pm 0.01 \;\; , \;\; 
\alpha_1 = 33.2 \pm 0.8 
\EEQ
Since we have seen above that $\mathfrak{p}=0$ for the spin-spin correlator of
the ANNNS model, it follows that the value of $\mathfrak{p}$ is characteristic
for the universality class at hand. 
 
In figure~\ref{Abb_53} the Monte Carlo data are compared with the resulting
scaling function, after fixing the overall normalization constant $b_0=0.41$. 
The agreement between the data and the prediction (\ref{5:OmegaN4}) of local
scale invariance is remarkable. 

To finish, we reconsider our working hypothesis $\theta=\frac{1}{2}$. Indeed,
in figure~\ref{Abb_45} we had shown how the form of the scaling function
$\Omega(v)$ changes when $\eps=N-4$ is increased, to first order in $\eps$.
In particular, rather pronounced non-monotonic behaviour is seen for values
of $\eps\sim 0.1$ which is the order of magnitude suggested from the results
of renormalized field theory \cite{Dieh00,Shpo01,Dieh02a}, 
see table~\ref{tab6}. 
Nothing of this is visible in the Monte Carlo data of figure~\ref{Abb_53}. 
Assuming that first-order perturbation 
theory in $\eps$ as described in section 4 is applicable here, 
we conclude from this observation that $\eps$ should be significantly smaller. 
Given the differences between the exponent estimates coming from renormalized
field theory \cite{Shpo01} and cluster Monte Carlo \cite{Plei01}, a
possible difference of $\theta$ from $\frac{1}{2}$ cannot yet be unambigously 
detected. Direct precise estimates of $\theta$ are needed. 

A similar analysis can be performed for the energy-energy correlation
function. This will be described elsewhere. 
 
In summary, having confirmed local scale invariance for the spin-spin 
correlator at the Lifshitz points in the ANNNI and the 
ANNNS models, it is plausible that the same will hold true for all 
ANNNO($n$) models with $1\leq n \leq \infty$.  

{\bf 3.} Finally, we consider the Lifshitz point of {\em second} order in the 
ANNNS model. In the ANNNS model as defined in eq.~(\ref{5:NNNHam}), 
a second-order Lifshitz point occurs at the endpoint
\BEQ
\kappa_1 = \frac{2}{5} \;\; , \;\; \kappa_{2} = - \frac{1}{15}
\EEQ
of the line (\ref{5:LP1ANNNS}) \cite{Selk77}. The lower critical dimension 
$d_{*}=\frac{8}{3}$. We need the following critical exponents \cite{Selk77}
\BEQ \label{5:ANNNSexp2}
\beta=\frac{1}{2} \;\; , \;\; \gamma = \frac{6}{3d-8} \;\; , \;\; 
\theta = \frac{1}{3} \;\; , \;\; \zeta_{\sigma}=3\left(d-\frac{8}{3}\right)
\EEQ
which in our notation corresponds to $N=6$. Therefore, the prediction of
local scale invariance is $G(t,r) = r^{-\zeta/3} \Omega(t r^{-1/3})$ with
$\Omega(v)$ given by (\ref{4:LoesFHyp},\ref{4:bpZwang}). 
At the Lifshitz point, the exact 
spin-spin correlation function is \cite{Frac93} (with $\mathfrak{c}=144/5$) 
\BEQ 
C(r_{\|}, \vec{r}_{\perp}) = C_0\: r_{\perp}^{-(d-d_{*})}\: 
\Xi\left( 3, \frac{d-d_{*}}{2}; \frac{1}{\mathfrak{c}^{1/3}} 
\left(\frac{r_{\|}}{r_{\perp}^{1/3}}\right)^2 \right)
\EEQ
where the scaling function\footnote{Properties of the function
$\Xi(3,a;x)$ are analysed in \cite{Frac93}. Explicit expressions in terms
of Airy functions are known for $a=n+\frac{1}{2},n+\frac{5}{6}$ 
with $n\in\mathbb{N}$.} 
can be written\footnote{We correct herewith a 
typographical error in eq. (4.1) in \cite{Frac93}.} in terms of generalized 
hypergeometric functions ${_1F_4}$
\BEA
\lefteqn{ \Xi(3,a;x) = \frac{\Gamma(a)}{\sqrt{\pi}\,\Gamma(5/6)}\: 
{_1F_4}\left(a;\frac{1}{3},\frac{1}{2},\frac{2}{3},\frac{5}{6};
-\frac{x^3}{27}\right)
} \\
&-& \!\frac{3\,\Gamma(a+1/3)}{\pi}\: x\, 
{_1F_4}\left(a+\frac{1}{3};\frac{2}{3},\frac{5}{6},\frac{7}{6},\frac{4}{3};
-\frac{x^3}{27}\right)
+\frac{6\,\Gamma(a+2/3)}{\sqrt{\pi}\,\Gamma(1/6)}\: x^2\, 
{_1F_4}\left(a+\frac{2}{3};\frac{7}{6},\frac{4}{3},\frac{3}{2},\frac{5}{3};
-\frac{x^3}{27}\right)
\nonumber 
\EEA
and $C_0$ is a normalization constant. This reproduces the exponent
$\zeta_{\sigma}$ from eq.~(\ref{5:ANNNSexp2}). We now compare the 
function $\Xi(3,a;x)$ with the expected form (\ref{4:LoesFHyp}) with $N=6$. 
The arguments of the functions $\Xi$ and $\Omega$ are related via
$x^3 = - \frac{1}{48}\frac{v^6}{\alpha_1}$ which implies
\BEQ
x = \frac{e^{\II \pi/3}}{(48\alpha_1)^{1/3}} v^2
\EEQ
Using this correspondence, we get
\BEQ
\Xi\left(3,\frac{\zeta}{6};x\right) = 
\frac{\Gamma(a)}{\sqrt{\pi}\,\Gamma(5/6)}\: {\cal F}_0 + 
\frac{3\,\Gamma(a+1/3)}{\pi}\frac{e^{\II \pi/3}}{(48\alpha_1)^{1/3}}\: 
v^2 {\cal F}_2 + 
\frac{6\,\Gamma(a+2/3)}{\sqrt{\pi}\,\Gamma(1/6)}\:
\frac{e^{2\II \pi/3}}{(48\alpha_1)^{2/3}}\: 
v^4 {\cal F}_4 
\EEQ
where the ${\cal F}_p$ are defined in eq.~(\ref{4:LoesFHyp}). The general
form of the scaling function for $N=6$ is $\Omega(v) = \sum_{p=0}^{4} b_p v^p
{\cal F}_p$ and we can identify the values of the free parameters $b_p$ which
apply for the spin-spin correlator at the Lifshitz point of second order in
the ANNNS model. We find $b_1=b_3=0$ and 
\BEQ
b_0 = \frac{\Gamma(a)}{\sqrt{\pi}\,\Gamma(5/6)} \;\; , \;\;
b_2 = \frac{3\,\Gamma(a+1/3)}{\pi}\frac{e^{\II \pi/3}}{(48\alpha_1)^{1/3}}
\;\; , \;\; 
b_4 = \frac{6\,\Gamma(a+2/3)}{\sqrt{\pi}\,\Gamma(1/6)}\:
\frac{e^{2\II \pi/3}}{(48\alpha_1)^{2/3}}
\EEQ
It is now straightforward to check that the constraint eq.~(\ref{4:bpZwang})
is indeed satisfied. In view of the known \cite{Frac93} power-law decay
of the function $\Xi(3,a;x)$ for $x\to\infty$ 
(and $a\ne\frac{1}{2},\frac{5}{6}$) this result might have been anticipated.  

We are not aware of any study of a second-order Lifshitz point in a  
different model.
 
\subsection{Aging in simple spin systems}

We now turn to a class of systems which display non-equilibrium dynamical
scaling. For the sake of simplicity, consider a simple ferromagnetic spin
system, e.g. an Ising model, evolving according to some dynamical rule. 
We shall exclusively consider the case of a non-conserved order parameter. 
Prepare the system in some initial state 
(an infinite-temperature initial state without any correlations is common)
and then quench it to some fixed temperature $T$ below or equal to the
equilibrium critical temperature $T_c$. Then follow the evolution of the 
system at that fixed temperature $T$. In the first case, the system
undergoes {\em phase-ordering kinetics} while in the second case one 
considers {\em non-equilibrium critical dynamics}. 
In both cases, the equilibrium state is never reached for the spatially 
infinite system. Rather, correlated domains of typical
time-dependent size $L(t)\sim t^{1/z}$ form and grow, where $z$ is the 
dynamical exponent. Consequently, the slow dynamics displays several
characteristic features which are absent from systems in thermodynamic
equilibrium. For reviews, see \cite{Bray94,Bouc98,Cate00,Godr02}. 

Main observables are the two-time correlation function $C(t,s;\vec{r})$ and the
two-time response function $R(t,s;\vec{r})$, defined as
\BEQ \label{5:DefCR}
C(t,s;\vec{r}-\vec{r}') = \langle \sigma_{\vec{r}}(t)\sigma_{\vec{r}'}(s)
\rangle \;\; , \;\;
R(t,s;\vec{r}-\vec{r}') = \left.\frac{\delta\langle\sigma_{\vec{r}}(t)\rangle}
{\delta h_{\vec{r}'}(s)} \right|_{h_{\vec{r}}=0}
\EEQ
where $\sigma_{\vec{r}}(t)$ is an (Ising) spin variable and $h_{\vec{r}}(t)$
the conjugate magnetic field at time $t$ and at
the site $\vec{r}$. It is assumed throughout that the quench occurred at time
zero and that spatial translation invariance holds. 
In particular, we shall focus here on the two-time 
autocorrelation function $C(t,s)=C(t,s;\vec{0})$ and the two-time
autoresponse function $R(t,s)=R(t,s;\vec{0})$. 
Then $s$ is the {\em waiting time} and $t$ the {\em observation time}. 
If either $T<T_c$ or $T=T_c$ the system is always out of equilibrium in the
sense that the fluctuation-dissipation ratio \cite{Cugl94,Cugl94a,Bouc98}
\BEQ
X(t,s) = T R(t,s) \left( \frac{\partial C(t,s)}{\partial s} \right)^{-1} \ne 1
\EEQ
Furthermore the two-time observables such as $C=C(t,s)$ and $R=R(t,s)$ depend
on {\em both} the waiting time $s$ and the observation time $t$ 
and not merely on their difference $\tau=t-s$. 
This breaking of time translation invariance is usally
referred to as {\em aging} \cite{Bray94,Bouc98,Cate00,Godr02} 
and will be used in this sense from now on. 
In addition it is well-established
\cite{Bray94} that the aging process is associated with dynamical scaling, 
that is in the scaling limit $s\to\infty$ and $t\to\infty$ such that
\BEQ
x = t/s > 1
\EEQ
is kept fixed, one has
\BEQ \label{5:SkalCR}
C(t,s) \sim s^{-b} f_C(t/s) \;\; , \;\;
R(t,s) \sim s^{-1-a} f_R(t/s)
\EEQ 
where $a,b$ are non-equilibrium critical exponents and $f_C$ and $f_R$ are
scaling functions. For large
arguments $x\gg 1$, these scaling functions typically behave as
\BEQ \label{5:SkalfCfR}
f_{C}(x) \sim x^{-\lambda_C/z} \;\; , \;\;
f_{R}(x) \sim x^{-\lambda_R/z}
\EEQ
where $\lambda_C, \lambda_R$ are the
autocorrelation \cite{Fish88,Huse89} and autoresponse exponents.\footnote{The
values of the exponents $\lambda_C,\lambda_R$ (and also $a,b,z$) depend on
whether $T<T_c$ or $T=T_c$, but we shall use the same notation in both cases.}
Remarkably, it can be shown that at late times the form of the growth law 
$L=L(t)$ (and thus the value of the dynamical exponent $z$) can be found for 
phase-ordering kinetics of purely dissipative systems from the scaling of the 
two-time correlation function $C(t,s;\vec{r})$ \cite{Rute95}. 

For fully disordered initial conditions, recently reviewed in
\cite{Godr02}, one has $\lambda_C=\lambda_R=\lambda$. If in addition one has
$T=T_c$, the relation $a=b=2\beta/\nu z$ holds below the upper critical
dimension, where $\beta,\nu$ are standard 
equilibrium critical exponents and the critical autocorrelation
exponent $\lambda=d-z\Theta$, where $\Theta$ is the initial-slip critical
exponent \cite{Jans89}. If on the other hand $T<T_c$, one has $b=0$,
but there does not seem to exist a general result for $a$. Indeed, in the
Glauber-Ising model $a=1/2$ in $2D$ and in $3D$, 
while in the kinetic spherical model $a=d/2-1$.  
However, these exponent identities do not necessarily hold for 
more general initial conditions \cite{Bert01,Pico02}. 
We shall need here the values of the exponents $z$, $a$ and $\lambda_R$ which 
are collected in tables~\ref{tab3} and \ref{tab4} below, for the 
Glauber-Ising model and the kinetic spherical model, respectively. 

We now derive the exact form of the scaling function $f_R(x)$ for the 
autoresponse function and then generalize towards the full spatio-temporal 
response function $R(t,s;\vec{r})$. 
Afterwards, we shall describe tests of these predictions in specific models.

We begin by assuming that the response functions transform covariantly
under local scale transformations \cite{Card85}. Recall that in the
context of Martin-Siggia-Rose theory, see \cite{Card96} and references therein,
these are given in terms of correlators 
\BEQ
R_{\phi\psi}(t_1,t_2;\vec{r}_1,\vec{r}_2) = \left.
\frac{\delta\langle\phi(t_1,\vec{r}_1)\rangle}{\delta h^{(\psi)}(t_2,\vec{r}_2)}
\right|_{h^{(\psi)}=0} 
=\langle\phi(t_1,\vec{r}_1)\wit{\psi}(t_2,\vec{r}_2)\rangle
\EEQ
of the scaling operator $\phi(t,\vec{r})$ and the 
response operator $\wit{\psi}(s,\vec{r})$ associated with the  
field $h^{(\psi)}$ canonically conjugate to the scaling operator $\psi$.
Usually, one merely considers the response function 
$R=R_{\phi\phi}=\langle\phi\wit{\phi}\rangle$ eq.~(\ref{5:DefCR}) of the
order parameter with respect to its own conjugate magnetic field.  
There is a clear analogy to the correlators 
$\langle \phi \phi^*\rangle$ which we have found in sections 3 and 4. 

However, the treatment carried out in sections 3 and 4 cannot be entirely 
taken over since in aging systems time translation invariance is broken. 
Therefore, the autoresponse functions
$R=R(t,s)$ does not transform covariantly under the entire set of
local scale transformations constructed in sections 3 and 4 but only under
those belonging to the subalgebra \cite{Henk94}
\BEQ
{\cal S} = \{ X_0, X_1, Y_m, \ldots \}
\EEQ
This contains both scale ($X_0$) and special conformal ($X_1$) transformations
as well as space translations ($Y_{-1/z}$) together with all generators which
are obtained from the commutators of these, see section 3. Since the
dynamics of aging systems may thought of as being described by some Langevin 
equation which is of first order in time, the realization of Typ II of 
local scale transformations will be adequate. 
From the explicit form of the generators 
eq.~(\ref{3:TypIIa}) we see that the line $t=0$ is kept invariant under the 
action of the $X_{0,1}$ and the $Y_m$. 

We therefore consider a two-point function $R=R(t_1,t_2;\vec{r}_1,\vec{r}_2)$ 
of two scaling operators $\phi_{1,2}$ which transform covariantly under the
action of $\cal S$. These are characterized in terms of their scaling
dimensions $x_i$ and the parameters $\beta_i,\gamma_i$, with $i=1,2$. 
The covariance of $R$ is expressed by the conditions (\ref{3:allKova})
restricted to those generators contained in $\cal S$. 
Now, spatial translation invariance can be implemented as shown in section 3 
and leads to $R=R(t_1,t_2;\vec{r})$ with $\vec{r}=\vec{r}_1-\vec{r}_2$ 
provided only that the constraints
\BEQ \label{5:betagamma}
\beta_2 + (-1)^{2-z} \beta_1 = 0 \;\; , \;\;
\gamma_2 + (-1)^{2-z} \gamma_1 = 0 
\EEQ 
hold, in complete analogy with eq.~(\ref{3:TypIIaM}). 

The autoreponse function $R=R(t,s)$ is now obtained by setting $r=|\vec{r}|=0$.
Then the conditions $Y_m R=0$ are automatically satisfied and 
the last two remaining conditions $X_0 R = X_1 R =0$ lead to the 
following differential equations
\BEA
\left( t \partial_t + s \partial_s + \frac{x_1}{z} + \frac{{x}_2}{z} 
\right) R(t,s) &=& 0 \nonumber \\
\left( t^2 \partial_t + s^2 \partial_s + \frac{2x_1}{z} t+ \frac{2{x}_2}{z}s 
\right) R(t,s) &=& 0
\EEA
with the solution
\BEQ \label{5:R1}
R(t,s) = r_0 \left( \frac{t}{s} \right)^{({x}_2-x_1)/z}
\left( t -s \right)^{-(x_1+{x}_2)/z} \, \Theta(t-s)
\EEQ
where $r_0=r_0(\beta_1,\beta_2,\gamma_1,\gamma_2)$ is a normalization constant 
which vanishes if the constraint (\ref{5:betagamma}) is not satisfied. 
We have also explicitly included the $\Theta$ function which is required 
because of causality ($\Theta(x)=1$ if $x>0$ and $\Theta(x)=0$ otherwise),
see e.g. \cite{Kreu81,Card96}. We can now compare with the
expected scaling form eq.~(\ref{5:SkalCR},\ref{5:SkalfCfR}) and then
arrive at the final result
\BEA \label{5:R}
R(t,s) &=& r_0 \left( \frac{t}{s}\right)^{1+a-\lambda_R/z} 
\left( t - s\right)^{-1-a}\, \Theta(t-s)\\
\mbox{\rm i.e.~~} f_R(x) &=& r_0\, x^{1+a-\lambda_R/z}\,(x-1)^{-1-a}\, 
\Theta(x-1)\nonumber
\EEA
Therefore, once the exponents $a$ and $\lambda_R/z$ are known, the 
functional form of the two-time autoresponse function is completely determined.

We note the following interesting consequence: from the constraint
(\ref{5:betagamma}), we see that for the response $R_{\phi\phi}$ of the scaling 
operator $\phi$ to a perturbation by its own conjugate field to be
non-vanishing  one must have $|\beta_{\phi}|=|\beta_{\wit{\phi}}|$ and
$|\gamma_{\phi}|=|\gamma_{\wit{\phi}}|$. For different scaling operators
$\phi\ne\psi$, the absolute values of $\beta_{\phi}$ and $\beta_{\psi}$ will 
{\it a priori} be different (and similarly for $|\gamma_{\phi}|$ and 
$|\gamma_{\psi}|$) and therefore we generically expect
$R_{\phi\psi}=0$. 

Next, we derive the spatio-temporal response function $R=R(t_1,t_2;\vec{r})$
from its covariance under the action of $\cal S$. 
It is convenient to write $R$ in the form
\BEQ
R = \left( \frac{t_1}{t_2}\right)^{(x_2-x_1)/z} G(t,\vec{r}) \;\; , \;\; 
t = t_1 - t_2
\EEQ
where we already took the explicit solution (\ref{5:R1}) into account. Since
the generators $Y_m$ (with $m=-1/z,-1/z+1,\ldots$) 
do not modify the temporal variables, the treatment of the conditions $Y_m G=0$
of section 3 goes through and we merely have to consider the two covariance
conditions $X_0 R = X_1 R =0$. These lead to the differential equations
for $G=G(t_1-t_2,r)$ (written here for $d=1$ spatial dimensions)
\BEA
\left( - t \partial_t - \frac{1}{z} r \partial_r - \frac{1}{z}(x_1+x_2)
\right) G &=& 0 \nonumber \\
\left( -t^2\partial_t -\frac{2}{z} t r \partial_r -\frac{2}{z}\frac{x_1+x_2}{2}t
-\left(\beta_1+\gamma_1\right) r^2 \partial_r^{2-z} 
-\gamma_1 2(2-z) r \partial_r^{1-z} \right) G & &  \\
+2 t_2 X_0 G + \frac{2}{z} r_2 Y_{-1/z+1} G &=& 0 \nonumber 
\EEA
Therefore, the function $G=G(t,r)$ does satisfy {\em exactly}
the same equations (\ref{3:TypIIaNull}) which were obtained in section 3
for a two-point function of quasiprimary operators $\phi_{1,2}$ with effective
scaling dimensions
\BEQ
x_{1,{\rm eff}} = x_{2,{\rm eff}} = \frac{1}{2} \left(x_1+x_2\right)
\EEQ 
The solution of these equations may therefore be taken from section 4 and 
our final result for the space-time reponse is
\BEQ \label{5:RR}
R(t,s;\vec{r}) = R(t,s)\, \Phi\left(\frac{|\vec{r}|}{(t-s)^{1/z}}\right) 
\EEQ
where the autoresponse function $R(t,s)$ is given in eq.~(\ref{5:R}) and the
scaling function $\Phi(u)$ can be read from eq.~(\ref{4:Phiu})
\BEQ 
\Phi(u) =  \left\{ \begin{array}{ll}  
\mathfrak{E}_{z,\Lambda}(-z^2(\beta_1+\gamma_1) u^{z}) &
\mbox{\rm ~~; ~~ $\gamma_1 \ne -\beta_1$} \\
E_{z,1}(-2z(2-z)\gamma_1 u^{z}) &
\mbox{\rm ~~; ~~ $\gamma_1 = -\beta_1$} 
\end{array} \right.
\EEQ
where $\Lambda = (z-1)/z+2(2-z)/[z(1+\beta_1/\gamma_1)]$ is a universal
constant and the functions $\mathfrak{E}_{a,b}(u)$ and 
$E_{a,b}(u)$ are defined in eq.~(\ref{4:EEFunk}). Eqs.~(\ref{5:R},\ref{5:RR})
are the main results of this section. 

For the special case $z=2$ eq.~(\ref{5:RR}) takes the simple form \cite{Henk94}
(see also eq.~(\ref{2:S2ponto-surf}))
\BEQ \label{5:RRaumZeit}
R(t,s;\vec{r}) = R(t,s)\exp\left(-\frac{\cal M}{2}\frac{\vec{r}^2}{t-s}\right)
\EEQ
where ${\cal M}=2(\beta_1+\gamma_1)$ is a dimensionful and non-universal 
scale factor. 

For simplicity of notation, the derivation has been carried out in the case
of one spatial dimension, $d=1$. If spatial rotation invariance holds, our
result (\ref{5:RR}) also holds for $d>1$. Indeed, as they stand, rotation
invariance is tacitly assumed in eqs.~(\ref{5:RR}) and
(\ref{5:RRaumZeit}). However, {\it if} rotation invariance
in space is broken (and in fact this is argued to be the case in phase-ordering 
kinetics \cite{Rute96,Rute99}), this means in our framework that the 
above calculation must be carried out
separately in each spatial direction. Consequently, eq.~(\ref{5:RR}) still 
holds phenomenologically, but the non-universal constants 
$\beta_1,\gamma_1$ should become direction-dependent.

We now turn towards tests of the conformal invariance prediction (\ref{5:R})
in specific models. Either experimentally or in simulations, it is
hard to measure $R(t,s)$ directly and one rather studies the integrated
response. For definiteness, we shall use from now on the Ising model language,
since afterwards the numerical tests will be performed in that model. Depending
on the history one obtains different forms of the integrated response. 
For example one may obtain the
{\em thermoremanent magnetization} $M_{\rm TRM}(t,s)$ by quenching the
system in a small magnetic field $h$, kept constant between the quench at time
zero and the waiting time $s$ and subsequently switched off. Alternatively,
one may quench in zero magnetic field, switch it on at the waiting time $s$
and keep it until the observation time $t$ when the {\em zero-field cooled
magnetization} $M_{\rm ZFC}(t,s)$ is measured. We then have two
integrated reponse functions
\BEA
\rho(t,s) &=& T \int_{0}^{s} \!\D u\, R(t,u) = \frac{T}{h} M_{\rm TRM}(t,s)
\nonumber \\
\chi(t,s) &=& T \int_{s}^{t} \!\D u\, R(t,u) = \frac{T}{h} M_{\rm ZFC}(t,s)
\EEA
(often, $\rho(t,s)$ is also called a {\em relaxation function} and $\chi(t,s)$ 
a {\em susceptibility function}). Other histories are possible.  
In practice, for ferromagnetic systems $h$ is a random magnetic field with 
zero mean in order to treat all phases equally \cite{Barr98}. 
The scaling of the TRM integrated response is readily found from 
eq.~(\ref{5:R1})
\BEQ
\rho(t,s) = r_0 T\: s^{1-(x_1+x_2)/z}\, x^{-2x_1/z}
{_2F_1}\left(1+\frac{x_1-x_2}{z},
\frac{x_1+x_2}{z};2+\frac{x_1-x_2}{z}; x^{-1}\right)
\EEQ
where ${_2F_1}$ is a hypergeometric function and $x=t/s$.  
In terms of the exponents $a$ and $\lambda_R/z$ this becomes
\BEQ \label{5:rho}
\rho(t,s) = r_0 T\: s^{-a}\, x^{-\lambda_R/z} 
{_2F_1}\left({1+a},\frac{\lambda_R}{z}-a;\frac{\lambda_R}{z}-a+1;x^{-1}\right)
\EEQ
Once the values of the exponents $a$ and $\lambda_R/z$ are known, the
functional form of $\rho(t,s)$ is completely fixed (the typical behaviour of 
the scaling function $s^a \rho(t,s)$ is 
exemplified in figures~\ref{Abb_51} and \ref{Abb_52} 
below). Evidently, $\chi(t,s)=\rho(t,t)-\rho(t,s)$. 

We are now ready to compare these predictions with simulational and exact 
results in specific models.

{\bf 1.} First we consider the kinetic Ising model with Glauber dynamics in
both $2D$ and $3D$ \cite{Henk01}. 
The spin Hamiltonian is ${\cal H}=-\sum_{(i,j)} \sigma_i \sigma_j$ with Ising
spins $\sigma_i=\pm 1$ at site $i$ and where the sum is over the 
nearest-neighbour pairs of a hypercubic lattice.  
Glauber or heat-bath dynamics~\cite{Glau63} is realized
through the stochastic rule $s_i(t)\to s_i(t+1)$ such that
\BEQ
s_i(t+1) = \pm 1 \mbox{\rm ~~~~~with probability 
$\frac{1}{2}[ 1\pm \tanh(h_i(t)/T)].$}
\EEQ
The local field acting on $s_i$ is $h_i(t) = h + {\sum_{j(i)}} s_j(t)$,
where $h$ is the external magnetic field, 
$j(i)$ denotes the nearest neighbours of the site $i$. The TRM integrated
response $\rho(t,s)$ was measured \cite{Henk01}, following the method of 
\cite{Barr98,Godr00b} and using a small random magnetic field, 
for systems with $300 \times 300$ spins in $2D$ and 
$50 \times 50 \times 50$ spins in $3D$, with a fully disordered initial
state. Larger systems were also simulated, 
in order to check for finite-size effects and averages over at least 
1000 different realizations of the systems were performed. For a comparison
with eq.~(\ref{5:rho}), which is equivalent to a test of (\ref{5:R}), 
the relevant exponents are collected in table~\ref{tab3},
see \cite{Fish88,Huse89,Godr02}. 
%%~~~~~~~~~~~~~~~~~~~~~~~~~~~~~~~~~~~~~~~~~~~~~~~~~~~~~~~~~~~~~~~~~~~~~~~~~~~~~~
\begin{table}[t]
\caption{Critical temperature and some non-equilibrium exponents of the $2D$ 
and $3D$ Glauber-Ising model with an infinite-temperature initial state, 
both for the phase-odrering regime ($T=0$) and for
critical dynamics ($T=T_c$). In $2D$, $T_c=2/\ln(1+\sqrt{2})$ exactly.
\label{tab3}}
\begin{center} 
\begin{tabular}{|l|l|l|l|} \hline
\multicolumn{2}{|l|}{}       & $2D$       & $3D$   \\ \hline
\multicolumn{2}{|l|}{$T_c$}  & 2.2692     & 4.5115 \\ \hline
$z$         & $T=0$          & 2          & 2      \\
            & $T=T_c$        & 2.17       & 2.04   \\ \hline
$\lambda_R$ & $T=0$          & 1.25       & 1.50   \\
            & $T=T_c$        & 1.59       & 2.78   \\ \hline
$a$         & $T=0$          & 0.5        & 0.5    \\
            & $T=T_c$        & 0.115      & 0.5064 \\ \hline
\end{tabular}\end{center}
\end{table}
%%~~~~~~~~~~~~~~~~~~~~~~~~~~~~~~~~~~~~~~~~~~~~~~~~~~~~~~~~~~~~~~~~~~~~~~~~~~~~~~

%%------------------------------------------------------------------------------
\begin{figure}[t]
\centerline{\epsfxsize=5.0in\epsfbox{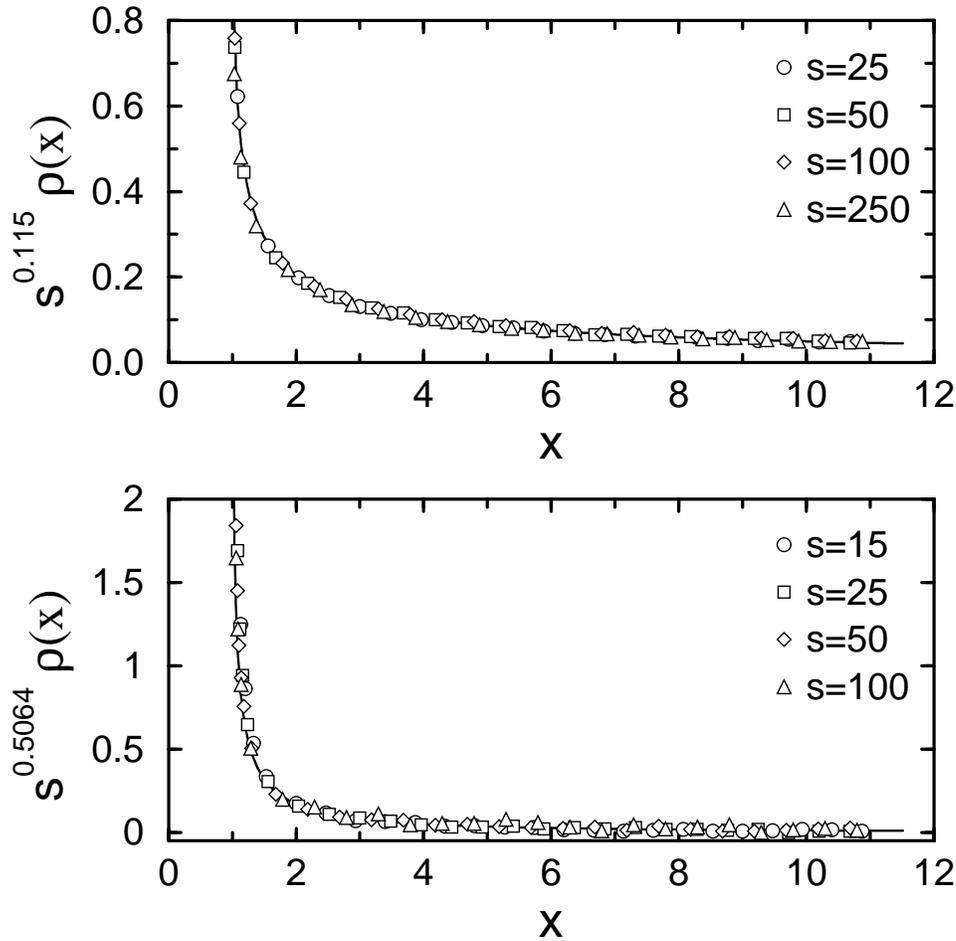}}
\caption{Scaling of the TRM integrated response function $\rho$ for the $2D$
(above) and the $3D$ (below) Glauber-Ising model at criticality ($T=T_c$).
The symbols correspond to different waiting times.
The full curve is the local scale invariance
prediction (\protect{\ref{5:rho}}) for $\rho(t,s)$. The data are from
\protect{\cite{Henk01}}.\label{Abb_51}}
\end{figure}
%%------------------------------------------------------------------------------
%%------------------------------------------------------------------------------
\begin{figure}[t]
\centerline{\epsfxsize=5.0in\epsfbox{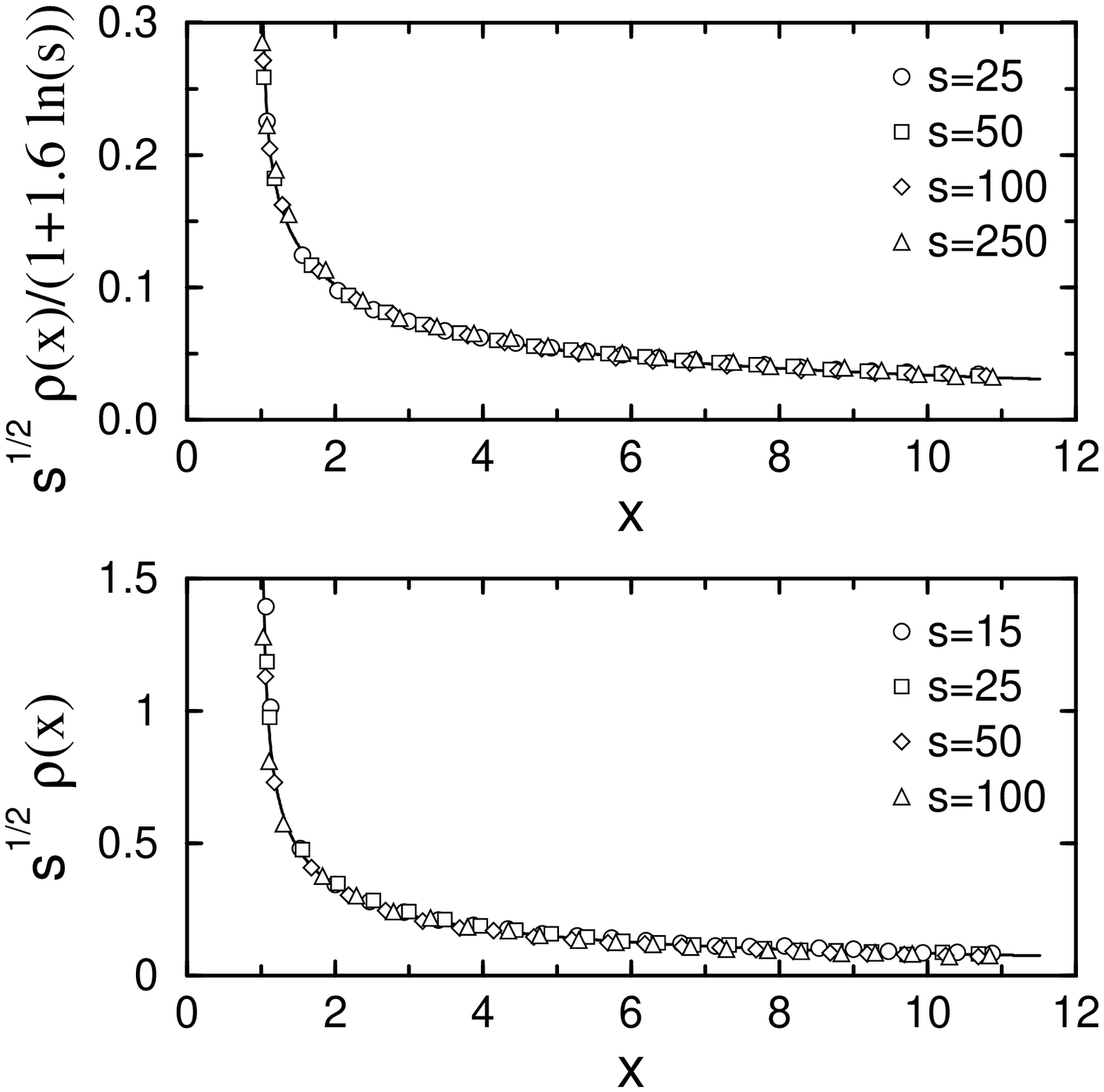}}
\caption{Scaling of the TRM integrated response function $\rho$ for the
low-temperature Glauber-Ising model
in $2D$ (above) at $T=1.5$ and in $3D$ at $T=3$ (below).
The symbols correspond to different waiting times. The full curve is
obtained from (\protect{\ref{5:rho}}). The data are from
\protect{\cite{Henk01}}.\label{Abb_52}}
\end{figure}
%%------------------------------------------------------------------------------
Clearly, from (\ref{5:rho}) we expect a data collapse if $s^a \rho(t,s)$ is 
plotted against $x=t/s$. In figure~\ref{Abb_51} Monte Carlo data at 
criticality are shown and scaling is indeed seen to hold. 
Furthermore, upon adjusting the normalization $r_0$, there is complete 
agreement between the data and eq.~(\ref{5:rho}).
Similarly, data for $T<T_c$ are shown in figure~\ref{Abb_52}. While the
expected scaling of $s^{1/2} \rho(t,s)$ as a function of $x=t/s$ works
well in $3D$, that is not the case in two dimensions. However, recall
that analytical calculations \cite{Bert99} on the scaling of the response 
function in the spirit of the OJK approximation rather suggest in two 
dimensions the presence of logarithmic corrections 
$\rho(t,s) \simeq s^{-1/2} \ln(s) f(t/s)$. Therefore,
the following ansatz for the $2D$ Glauber-Ising model \cite{Henk01}
\BEQ
\rho(t,s) = s^{-1/2} \left( r_0 + r_1 \ln s \right) E\left(\frac{t}{s}\right)
\EEQ
appears natural, where $r_{0,1}$ are non-universal constants and $E=E(x)$ is
a scaling function.\footnote{It is not impossible that the logarithm might be
explained through logarithmic Schr\"odinger invariance, which would 
have to be constructed by analogy with logarithmic conformal field
theories, see e.g. \cite{Floh01,Rahi01}. We hope to come back to this
in the future.} Indeed, a satisfactory scaling is found this way, as
can be seen from figure~\ref{Abb_52}. It has been checked that the same
kind of scaling holds in the entire low-temperature phase, where $r_0$ and
$r_1$ depend on $T$ \cite{Henk01}. Furthermore, the form of the scaling
functions thus obtained are again in perfect agreement with the prediction 
eq.~(\ref{5:rho}) of local scale invariance, for both $2D$ and $3D$. 

{\bf 2.} Second, eq.~(\ref{5:R}) has been tested extensively in the 
exactly solvable kinetic spherical model with a non-conserved order parameter,
for $d>2$ space dimensions. The kinetic spherical model may be introduced as a
spin model \cite{Cugl95,Godr00b,Zipp00,Cann01,Corb02,Pico02} 
with Hamiltonian ${\cal H} = -\sum_{i,j} J_{i,j} S_i S_j$, where
the $J_{i,j}$ are coupling constants and the $S_i$ are real variables 
subject to the spherical constraint
\BEQ \label{5:SpherZwang}
\sum_{i} S_i^2 = {\cal N}
\EEQ
where the sum runs over the entire lattice and $\cal N$ is the number of sites.
The dynamics of the model is generated through a stochastic Langevin equation
\BEQ 
\frac{\D S_{\vec{r}}}{\D t} = -\frac{\delta {\cal H}[S]}{\delta S_{\vec{r}}} -
(2d +\mathfrak{z}(t)) S_{\vec{r}} + \eta_{\vec{r}}(t)
\EEQ
where the Gaussian white noise $\eta_{\vec{r}}(t)$ has the correlation
\BEQ
\langle \eta_{\vec{r}}(t) \eta_{\vec{r}'}(t') \rangle = 
2T \delta_{\vec{r},\vec{r}'} \delta(t-t')
\EEQ
and $\mathfrak{z}(t)$ is determined by satisfying the spherical constraint. 
Alternatively, the same model may be studied as the $n\to\infty$
limit of the coarse-grained O($n$) vector model using field theory methods, 
see \cite{Bray91,Coni94,Jans89,Newm90,Cala02}. 

In many studies, the short-ranged spherical model with only nearest-neighbour
interactions $J_{i,j}$ and an infinite-temperature initial state without
any correlations was considered 
\cite{Jans89,Newm90,Cugl95,Godr00b,Zipp00,Corb02}.
Exact results for the two-time autocorrelators and autoresponse functions
were found. Writing the two-time autoreponse function in the scaling
limit as $R(t,s) = (4\pi s)^{-d/2} f_R(t/s)$, one has in the ordered phase
($T<T_c$) \cite{Newm90,Godr00b} (here and below, we always take $t>s$ or $x>1$)
\BEQ \label{5:SpherKurzR0}
f_R(x) = x^{d/4} \, (x-1)^{-d/2}
\EEQ
for all values of $d>2$.
At the critical point ($T=T_c$) one has \cite{Jans89,Godr00b}
\BEQ \label{5:SpherKurzRc}
f_R(x) = \left\{\begin{array}{ll}
x^{1-d/4} (x-1)^{-d/2} &\!\!~ ;~ \mbox{\rm if $2<d<4$}\\
(x-1)^{-d/2} &\!\!~ ;~ \mbox{\rm if $4<d$}
\end{array}\right.
\EEQ
and both results are in full agreement, upon identification of exponents, 
with eq.~(\ref{5:R}). The second expression of (\ref{5:SpherKurzRc}) 
corresponds to the mean-field case and coincides with the result found for
a free Gaussian field \cite{Cugl94}, as expected. We point out that
the form of the correlation scaling functions depends strongly on $d$ and on 
whether $T<T_c$ or $T=T_c$ since the behaviour of the fluctuation-dissipation
ratio $X(t,s)$ is different \cite{Godr00b} 
in each of the three cases considered so far. 

More recently, these results have been generalized in two directions. First, 
the two-time autocorrelation and autoresponse functions were
calculated exactly in the kinetic spherical model with spatially
long-range interactions, of the form \cite{Cann01}
\BEQ
J_{i,j}= J(\vec{r}_{ij}) =  J_0 |\vec{r}_{ij}|^{-d-\sigma} 
\left( {\sum_{j}}'|\vec{r}_{ij}|^{-d-\sigma}\right)^{-1} 
\EEQ
where ${\sum_{j}}'$ runs over all sites $j\ne i$, $\vec{r}_{ij}$ is the
distance between sites $i$ and $j$, $J_0$ is a constant and $\sigma$ is a free 
parameter. For $d>2$ and $\sigma>2$,
one recovers the short-ranged spherical model discussed above. On the
other hand, if either (i) $d>2$ and $0<\sigma<2$ or else (ii) $d\leq 2$ and
$0<\sigma<d$ the model has an equilibrium  phase transition at a non-vanishing 
$T_c$ between an ordered and a paramagnetic phase \cite{Cann01}. 
In this case and below 
criticality ($T<T_c$), the dynamical exponent $z=\sigma$ and the response 
function is in the aging scaling limit \cite{Cann01}
\BEQ
R(t,s) = r_0 \, (t/s)^{d/(2\sigma)} (t-s)^{-d/\sigma},
\EEQ
which again fully confirms eq.~(\ref{5:R}). One identifies the exponents
$a=d/\sigma-1$ and $\lambda_R=d/2$ in the ordered phase. 
We are not aware of any published results on $R$ in this model at criticality. 

Second, the case of nearest-neighbour interactions but with long-ranged 
correlations characterized by the form 
\BEQ \label{5:SpherCIni}
C_{\rm ini}(\vec{r})\sim |\vec{r}|^{-d-\alpha}
\EEQ 
of the spin-spin correlator in the initial state has been
studied \cite{Newm90,Bray91,Coni94,Pico02}. 
The uncorrelated initial state considered
above is recovered as the special case $\alpha=0$. Indeed, the effect of 
initial correlations is only notable in the long-time behaviour if $\alpha<0$. 
In the ordered phase
($T<T_c$), the exact autoresponse function is in the scaling limit 
$R(t,s)=(4\pi s)^{-d/2} f_R(t/s)$, where \cite{Newm90,Pico02}
\BEQ
f_R(x) = x^{(d+\alpha)/4}\, (x-1)^{-d/2}
\EEQ
in complete agreement with (\ref{5:R}). For $\alpha=0$, 
eq.~(\ref{5:SpherKurzR0}) is reproduced. 
At criticality ($T=T_c$), {\em five} distinct regimes of non-equilibrium
critical dynamics exist \cite{Pico02}. These are distinguished by the values of
$d$ and $\alpha$ as listed in table~\ref{tab4} which also gives the
values of the required non-equilibrium exponents. 
%%~~~~~~~~~~~~~~~~~~~~~~~~~~~~~~~~~~~~~~~~~~~~~~~~~~~~~~~~~~~~~~~~~~~~~~~~~~~~~~
\begin{table}
\caption{Some non-equilibrium exponents of the kinetic spherical model with
short-ranged interactions and correlated initial conditions of the form
(\protect{\ref{5:SpherCIni}}), for the five distinct regimes I,\ldots,V at
criticality ($T=T_c$) and also in the ordered phase ($T<T_c$). Here 
$D=d+\alpha+2$.\label{tab4}}
\begin{center} \begin{tabular}{|c|ccc|c|ccc|} \hline
Regime & \multicolumn{3}{c|}{conditions} & $\digamma$ & $a$ & $z$ & $\lambda_R$ 
\\ \hline
I   & $2<d<4$ & $2<D<4$ & & $\alpha/4-1/2$   & $d/2-1$ & 2 & $d-\alpha/2-1$ \\
II  & $4<d$ & $2<D<4$ & & $(d+\alpha)/4-1/2$ & $d/2-1$ & 2 & $(d-\alpha)/2+1$\\
III & $2<d<4$ & $4<D$   & & $1-d/4$          & $d/2-1$ & 2 & $3d/2-2$ \\
IV  & $4<d$   & $4<D$   & $\alpha>-2$ & $0$  & $d/2-1$ & 2 & $d$ \\ 
V   & $4<d$   & $4<D$   & $\alpha<-2$ & $0$  & $d/2-1$ & 2 & $d$ \\ \hline
$T<T_c$ & $2<d$ &       & & $(d+\alpha)/4$   & $d/2-1$ & 2 & $(d-\alpha)/2$ \\ 
\hline
\end{tabular}\end{center}
\end{table} 
%%~~~~~~~~~~~~~~~~~~~~~~~~~~~~~~~~~~~~~~~~~~~~~~~~~~~~~~~~~~~~~~~~~~~~~~~~~~~~~~
The autoresponse function is $R(t,s)=(4\pi)^{-d/2} s^{-1-a} f_R(t/s)$ in all
five regimes where \cite{Pico02}
\BEQ
f_{R}(x) = x^{\digamma}\, (x-1)^{-d/2}
\EEQ
and the values of the exponent $\digamma$ can be read off from 
table~\ref{tab4}. 
Once more we find complete agreement with eq.~(\ref{5:R}). The results
quoted in eq.~(\ref{5:SpherKurzRc}) are reproduced in the critical 
regimes III and IV. Again, the behaviour of the autocorrelations differs
widely between the five regimes as discussed in detail in \cite{Pico02}.  

{\bf 3.} Recently, the off-equilibrium response and correlation functions
of the O($n$)-symmetric vector model with a non-conserved order parameter
(model A in the terminology of \cite{Halp77}) and a fully disordered initial 
state were calculated in one-loop order
in $d=4-\eps$ dimensions \cite{Cala02}. From earlier calculations of 
correlators and responses with the initial state at criticality \cite{Jans89}, 
the critical exponents
\BEQ
z = 2 + {\rm O}\left(\eps^2\right) \;\; , \;\; 
a = \frac{d}{2} -1 + {\rm O}\left(\eps^2\right) \;\; , \;\;
\lambda_R = d - \frac{\eps}{2}\frac{n+2}{n+8} + {\rm O}\left(\eps^2\right) 
\EEQ
were already known. At criticality, the autoresponse 
function is in the scaling limit \cite{Cala02}
\BEQ
R(t,s) = r_0 \left(\frac{t}{s}\right)^{\frac{\eps}{4}\frac{n+2}{n+8}}
\, \left( t-s\right)^{-d/2} + {\rm O}\left(\eps^2\right)
\EEQ
which again agrees, to first order in $\eps$, with the prediction
(\ref{5:R}). It would be extremely interesting to see whether this
agreement can be extended to higher orders in the $\eps$-expansion. 
We are not aware of any results on $R(t,s)$ in the low-temperature phase. 

{\bf 4.} A particularly simple textbook system which reproduces eq.~(\ref{5:R}) 
is the free random walk, see \cite{Cugl94}. It is described in the continuum 
through the following Langevin equation for $y(t)$ 
and the correlator of the Gaussian 
white noise $\eta$
\BEQ
\frac{\D y(t)}{\D t} = \eta(t) \;\; , \;\; 
\langle \eta(t) \eta(s) \rangle = 2T \delta(t-s)
\EEQ
The autocorrelation function and the autoresponse function are readily found
\cite{Cugl94}
\BEA
C(t,s) &=& \langle y(t) y(s) \rangle \hspace{3.1truemm}= 2T\min(t,s)\nonumber\\
R(t,s) &=& \left.\frac{\delta \langle y(t)\rangle}{\delta h(s)}\right|_{h=0} 
= \frac{1}{2T} \langle y(t) \eta(s) \rangle = \Theta(t-s)
\EEA
and do satisfy the scaling fom (\ref{5:SkalfCfR}). The system is out of
equilibrium since the fluctuation-dissipation ratio $X(t,s)=1/2$. 
The exact expression for $R(t,s)$ matches eq.~(\ref{5:R}) with the exponents 
$a=-1$ and $\lambda_R/z=0$. 

{\bf 5.} Having studied so far simple ferromagnets, we now consider the 
spherical model spin glass \cite{Cugl95} as a very simple example of a
disordered system. The Hamiltonian ${\cal H}= - \sum_{(i,j)} J_{i,j} S_i S_j$ 
describes nearest-neighbour interactions between the spherical spins subject
to the constraint (\ref{5:SpherZwang}). 
The couplings $J_{i,j}$ are independently
distributed quenched random variables with zero mean and variance inversely 
proportional to the number of sites $\cal N$. For uniform initial conditions,
the mean-field two-time autoresponse function is in the scaling 
limit \cite{Cugl95}
\BEQ \label{5:SpherGlass}
R(t,s) = r_0 \left(\frac{t}{s}\right)^{3/4}\, (t-s)^{-3/2}
\EEQ
which again agrees with (\ref{5:R}). Could this be a hint that the form
(\ref{5:R}) for $R(t,s)$ might also hold for glassy systems ? 
In fact, the result (\ref{5:SpherGlass}) coincides 
with eq.~(\ref{5:SpherKurzRc}) 
for $d=3$. This is a consequence of the known \cite{Zipp00} similarity between 
the $3D$ spherical ferromagnet and the mean-field spherical spin glass. 
Therefore, although the result (\ref{5:SpherGlass}) may appear suggestive, 
it is not yet clear at all whether or not the predictions 
(\ref{5:R},\ref{5:rho}) of local scale invariance may be reproducible in 
physically interesting glassy systems. 

{\bf 6.} The tests described so far all concerned the autoresponse function
$R(t,s)$. Spatio-temporal responses $R(t,s;\vec{r})$ 
have so far only been tested in the exactly 
solvable spherical model. For short-ranged interactions, the dynamical 
exponent $z=2$ and from local scale invariance (or 
Schr\"odinger invariance in this case) we expect (\ref{5:RRaumZeit})
to hold. Indeed, this had been checked long ago for a disordered initial state
both at and below $T_c$ \cite{Jans89,Newm90,Henk94} and recently for 
an initial state with the long-range correlations of the form 
(\ref{5:SpherCIni}), again both at and below $T_c$ \cite{Pico02}. 
To leading order in the $\eps$-expansion, eq. (\ref{5:RRaumZeit}) also holds
in the O($n$)-model \cite{Cala02}. 
Tests of the spatio-temporal response (\ref{5:RR}) in the Glauber-Ising model
both below and at criticality are currently being performed and will be
reported elsewhere. 

Summarising, we have seen that the form eq.~(\ref{5:R}) of the two-time
autoreponse function has been reproduced in a considerable variety of 
non-equilibrium ferromagnetic spin systems with a non-conserved order 
parameter. The examples for which (\ref{5:R}) has so far be seen to hold 
suggest its validity 
{\em independently} of the following characteristics of specific models,
namely (see also note added)
\begin{enumerate}
\item the value of the dynamical exponent $z$.
\item the value of the space dimension, provided $d>1$.
\item the number of components of the order parameter and the global symmetry
group. 
\item the spatial range of the spin interactions.
\item the presence of spatial long-range correlations in the initial state.
\item the value of $T$, in particular whether $T<T_c$ or $T=T_c$, provided 
$T_c>0$.
\end{enumerate}
The response functions of different universality classes are only distinguished
by the values of the exponents $a$ and $\lambda_R/z$. 

Clearly, further tests in different models will be most useful to either 
confirm further or else invalidate this conjecture. On the other hand, as yet 
there are only few tests of the full spatio-temporal response (\ref{5:RR}). 
At present, there is no prediction from local scale invariance for 
the two-time {\it correlation} functions in non-equilibrium dynamical scaling. 

Finally, we comment on some examples where eq.~(\ref{5:R}) does not hold. 
From the exact solution of the $1D$ Glauber-Ising model at $T=0$ one has in the 
aging regime $R(t,s)=[2\pi^2 s (t-s)]^{-1/2}$ \cite{Godr00a,Lipp00}. For the
$2D$ XY model in the critical low-temperature phase, one has in the spin-wave
approximation $R(t,s)\sim s^{-1-\eta/2} f_R(t/s)$ with 
$f_R(x)=[(x+1)^2/x]^{\eta/4}(x-1)^{-1-\eta/2}$ \cite{Bert01}. 
Here $\eta=\eta(T)$ is the usual equilibrium temperature-dependent exponent, 
see \cite{Bert01,Drou88,Zinn89,Card96} and references therein. It is
well-established, however, that in these models already the growth law
$L=L(t)$ is unusual: the generic description which assumes 
that energy dissipation is dominated by the motion of single defect structures 
at the scale $L=L(t)$ no longer applies here \cite{Rute95,Bray00}. From 
that perspective, it is not too surprising that also the form of the scaling 
functions should be non-generic in these models. We shall come back elsewhere 
to the question whether these systems may or may not be treated by some 
form of local scale invariance \cite{Pico02a}. 

\subsection{Equilibrium critical dynamics}

Having discussed the local scaling of the response function out of equilibrium,
we now briefly turn towards the case of {\em equilibrium} critical dynamics. 
It will be of interest to compare the predictions of local scale invariance
as introduced here with those obtained from dynamical conformal invariance
\cite{Card85}. 

Consider a spin system {\em at} its equilibrium critical point. Two-time
correlation functions and response functions can be defined as before, see
eq.~(\ref{5:DefCR}). However,
and in contrast to the non-equilibrium case, time translation invariance is
expected to hold and we should have
\BEQ
C = C(t-s;\vec{r}) \;\; , \;\; R = R(t-s;\vec{r})
\EEQ
The response functions should transform covariantly under local scale
transformations, as originally proposed in \cite{Card85}. In consequence, 
carrying over the treatment of section 4, $R$ is again given by 
eq.~(\ref{5:RR}), viz. 
\BEQ
R(\tau;\vec{r}) = \langle \phi(\tau,\vec{r}) \wit{\phi}(0;\vec{0})\rangle
= \tau^{-2x_1/z} \Phi\left(r \tau^{-1/z}\right)
\EEQ
but with the additional constraint $x_1=x_2$ coming from time translation 
invariance (alternatively, this constraint may be written as $\lambda_R/z=1+a$).
The scaling function $\Phi(u)$ is given as before by eq.~(\ref{4:Phiu}). 
The constraints discussed above in section 5.2 for the non-equilibrium response
also apply.  

Since at equilibrium the fluctuation-dissipation theorem
\BEQ
T_c R(\tau;\vec{r}) = - \frac{\partial}{\partial \tau} C(\tau;\vec{r})
\EEQ
holds \cite{Card96,Kreu81}, where $\tau=t-s$, the scaling form of the two-time 
correlator may be given as well. Integrating, we have
\BEQ \label{5:CTc}
C(t;\vec{r}) = C(0;\vec{r}) - T_c z\: r^{z-2x_1} 
\int_{r t^{-1/z}}^{\infty} \!\D u\, u^{2x_1-z-1} \Phi(u)
\EEQ
Here $C(0;\vec{r})=\langle \phi(t,\vec{r})\phi(t,\vec{0})\rangle$ is the
equal-time correlator, which is $t$-independent, because time translation 
invariance holds at the equilibrium critical point.  

We point out that the exponent $x_1$ which enters (\ref{5:CTc}) is {\em not} 
a static critical exponent. Rather, it is related to the static 
scaling dimension $x_1^{\rm (g)}$ by
\BEQ
2 x_1^{\rm (g)} = 2 x_1 -z 
\EEQ
For example, for a $2D$ critical point, the $x_1^{\rm (g)}$ are the scaling 
dimensions which can be obtained from the representations of the Virasoro
algebra \cite{Bela84}. 
In terms of these, we have, with $u=r t^{-1/z}$
\BEA
R(t;\vec{r}) &=& t^{-1-2x_1^{\rm (g)}/z} \Phi\left(r t^{-1/z}\right) 
\label{5:RGg} \\
C(t;\vec{r}) &=& C(0;\vec{r}) -T_c z\: r^{-2x_1^{\rm (g)}} 
\int_{u}^{\infty} \!\D w\, w^{2x_1^{\rm (g)}-1} \Phi(w)
\label{5:CGg} 
\EEA
Because of the known behaviour of $\Phi(w)$ at the $w\to 0$ and $w\to\infty$ 
boundaries, it is easy to see that the integral in (\ref{5:CGg}) is convergent 
if $x_1^{\rm (g)}>0$ and $z>0$. 
Since we are at criticality, we expect $C(0;\vec{r})\sim r^{-2x_1^{\rm (g)}}$ 
and finally obtain
\BEQ \label{5:CGgFinal}
C(t;\vec{r}) = r^{-2x_1^{\rm (g)}} \left( C_0 + C_1 
\int_{0}^{u} \!\D w\, w^{2x_1^{\rm (g)}-1} \Phi(w) \right) \;\; , \;\; 
u = r t^{-1/z}
\EEQ
where $C_{0,1}$ are non-universal normalization constants.  

We now compare the scaling form (\ref{5:RGg}) with the prediction of dynamical 
conformal invariance for a non-conserved order parameter 
$R(t,r)\sim t^{-1-2x/z} \exp(-r^z/t)$ \cite{Card85}, up to suppressed 
non-universal constants. In our theory, we have found a simple exponential 
scaling function in two cases: (i) $z=2$ with $\beta_1+\gamma_1={\cal M}/2\ne 0$
and (ii) $z=1$ with $\beta_1+\gamma_1=0$. In these special cases, the two-time
correlator takes the following form: in the case $z=2$, ${\cal M}\ne 0$ 
\BEQ
C(t,\vec{r}) = r^{-2x_1^{\rm (g)}} \left[ 
C_0 + C_1\, \gamma\left( x_1^{\rm (g)},\frac{\cal M}{2} 
\frac{r^2}{t}\right)\right]
\EEQ
and in the case $z=1$, $\beta_1+\gamma_1=0$
\BEQ
C(t,\vec{r}) = r^{-2x_1^{\rm (g)}} \left[ 
C_0 + C_1\, \gamma\left( 2 x_1^{\rm (g)}, 2 \gamma_1 \frac{r}{t}\right)\right]
\EEQ
respectively. Here $\gamma(a,z)$ is
an incomplete gamma function, see eqs. (6.5.4), (6.5.29) in \cite{Abra65}
and $C_{0,1}$ are normalization constants. In all other cases, the scaling
functions will take a form quite distinct from these two examples, see
figures~\ref{Abb_43} and \ref{Abb_44}. We recall that the fundamental
hypothesis of dynamical conformal invariance was conformal invariance in
{\em space} \cite{Card85} and not in time, as we have assumed throughout 
this paper.  

Tests of the consequences (\ref{5:RGg},\ref{5:CGgFinal}) of the hypothesis of 
local scale invariance in equilibrium critical
dynamics in specific models would be most welcome.

%%%%%%%%%%%%%%%%%%%%%%%%%%%%%%%%%%%%%%%%%%%%%%%%%%%%%%%%%%%%%%%%%%%%%%%%%%%%%%%%
\section{Conclusions}
%%%%%%%%%%%%%%%%%%%%%%%%%%%%%%%%%%%%%%%%%%%%%%%%%%%%%%%%%%%%%%%%%%%%%%%%%%%%%%%%

Our starting point has been the heuristic idea that it might be possible to 
extend strongly anisotropic or dynamical scaling from mere dilatation 
invariance with a given anisotropy exponent $\theta$ (or dynamical
exponent $z$) to a larger dynamical symmetry involving local scale 
transformations with a space-time-dependent dilatation factor. We have studied
this idea by seeking confirmation on the purely phenomenological level
of being able to reproduce certain two-point function in the context of
specific models. The agreement found provides evidence in favour of, but does
not prove, this hypothesis of local scale invariance. 

In attempting to construct local scale transformations for an arbitrary
anisotropy exponent $\theta$, we have tried to follow
the two known cases of conformal and Schr\"odinger invariance as closely
as possible. The common feature of these groups is the presence of conformal
transformations in {\em time} and we have made this the central assumption
in our formulation of local scale invariance. 

In carrying out the construction of infinitesimal local scale transformations,
we have seen that there exist two types, the first one (Typ I) being related to 
strongly anisotropic {\em equilibrium} systems, while the second one (Typ II)
describes systems with {\em time-dependent} dynamical scaling. The main
properties of local scale invariance when applied to quasiprimary operators
are collected in table~\ref{tab1}. Conformal invariance and 
Schr\"odinger invariance are recovered as special cases, for $\theta=1$
and $\theta=2$, respectively. 
On the other hand, if $\theta\ne 1,2$, the generators only close
into a Lie algebra on certain states only and thus form a {\em weak} 
Lie algebra. 

Local scale transformations form a (weak) dynamic symmetry group of the
equation of motion ${\cal S}\psi=0$ of certain, in general non-local, 
free-field theories (see table~\ref{tab1} for the precise form of $\cal S$). 

In addition, two-point functions formed from quasiprimary operators satisfy
certain linear (fractional) differential equations, from which the form of
these two-point functions can be determined. 

Indeed, these explicit predictions (see eqs.~(\ref{4:LoesFHyp},\ref{4:bpZwang})
for Typ I with $N=2/\theta\in\mathbb{N}$ and eq.~(\ref{4:Phiu}) for Typ II) 
allow for a test of the applicability of our notion of local scale 
invariance in concrete models. These tests have been performed in two
different settings:
\begin{enumerate}
\item Uniaxial Lifshitz points are a classic example of a strongly 
anisotropic equilibrium system. The form of the spin-spin correlator
at the first-order Lifshitz points in both the ANNNS model and the $3D$ ANNNI 
model and also at the second-order Lifshitz point in the ANNNS model agrees
fully with local scale invariance, for $\theta\simeq\frac{1}{2}$ and
$\theta=\frac{1}{3}$, respectively \cite{Henk97,Plei01}. 
\item Dynamical scaling occurs in the aging behaviour of simple ferromagnets
which undegoes phase-ordering kinetics or non-equilibrium critical dynamics. 
We found a particularly simple scaling form of the two-time autoresponse 
function in the aging regime
\BEA \label{6:RForm}
R(t,s) &=& 
\left.\frac{\delta\langle \sigma(t)\rangle}{\delta h(s)}\right|_{h=0}
= s^{-1-a} f_R(t/s)  \nonumber \\
f_R(x) &=& r_0\, x^{A} (x-1)^{B}
\EEA
where the values of the exponents $A$ and $B$ can be matched to known
dynamical exponents, see section 5.2. This
form can be reproduced in a large variety of models, with the dynamical 
exponent $z$ taking values both below and above 2, notably in the $2D$ and 
$3D$ Glauber-Ising models, the O($n$)-symmetric kinetic model A to first
order in the $\eps$-expansion, several variants of the kinetic
spherical model with non-conserved order parameter and brownian motion
\cite{Henk01,Cala02,Godr00b,Cann01,Pico02,Cugl94}. 

The form (\ref{6:RForm}) for the scaling function $f_R(x)$ is likely
to be very robust and of broad validity, see section 5.2. 
\end{enumerate}
We stress that the $3D$ ANNNI model and the $2D$ and $3D$ Glauber-Ising models
{\em cannot} be expressed as free-field theories. This renders the confirmation
of local scale invariance in these models non-trivial. 

Technically, our results depend on the construction of {\em commuting}
fractional derivatives. That property was essential in the derivation of the
fractionall differential equations satisfied by the scaling function of the
two-point function. However, these differential equations themselves do 
{\em not} depend on any other property of the fractional derivatives. On the 
other hand, a specific choice must be made for the fractional derivative if an 
explicit solution is requested. At present, there is not yet any test available
which would inform us whether the specific form used in this work or else
any of the known alternatives is realized in the context of local scale 
invariance. 

The possibility of local scale invariance has recently been discussed 
(under the name of re\-pa\-ra\-me\-tri\-za\-tion 
invariance) in the aging of certain glassy systems, where the covariance of 
correlators and response functions was studied \cite{Kenn01,Cham01}. In
the models studied there, the scaling dimensions $x_{\phi}=0$ of the 
order parameter $\phi$ and $x_{\wit{\phi}}=1$ of the 
response operator $\wit{\phi}$ take rather
simple values, however. The covariance of non-equilibrium two-time correlators
under time reparametrizations also plays a role in a recent study on 
general constraints on the scaling of these \cite{Kurc01}. Finally,
a recently introduced model for a fluctuating interface with $z=1$ 
can be described in the thermodynamic limit in terms of the characters of
a conformal field theory \cite{Gier02}.

All in all, we have proposed a new type of dynamical symmetry which might 
generalize usual anisotropic or dynamical scaling. The existing 
phenomenological confirmations provide evidence for local scale invariance to 
be realized, at least to a very good approximation, in certain statistical 
systems. Of course, the theory must be built further and in
particular, one would like to be able to show that the field theories
underlying these scale-invariant statistical systems indeed satisfy local
scale invariance. Work along several of these lines is in progress. 

\zeile{3}
\noindent {\large\bf Note added in proof:}\\

\noindent 
For the critical weakly disordered kinetic Ising model with a non-conserved
order parameter, Ca\-la\-bre\-se and Gambassi \cite{Cala02a} find, to one-loop
order, the response function at vanishing external momentum $\vec{q}$ to be
in full agreement with the prediction of local scale invariance. 

\zeile{3}
\noindent {\large\bf Acknowledgements}\\

\noindent
It is a pleasure to thank 
L. Berthier, C. Godr\`eche, H. Hinrichsen, D. Karevski, 
J.-M. Luck, A. Picone, J. Unterberger 
and especially M. Pleimling 
for useful discussions and/or correspondence. 
Parts of this work were done at the Centro de F\'{\i}sica da Mat\'eria 
Condensada (CFMC) of the 
Universidade de Lisboa and at the Fachbereich Physik of the Universit\"at
Saarbr\"ucken, whom I thank for warm hospitality. Support of the CINES
Montpellier (projet pmn2095) is gratefully acknowledged.  

\newpage 

%%%%%%%%%%%%%%%%%%%%%%%%%%%%%%%%%%%%%%%%%%%%%%%%%%%%%%%%%%%%%%%%%%%%%%%%%%%%%%%%
\appsection{A}{On fractional derivatives}
%%%%%%%%%%%%%%%%%%%%%%%%%%%%%%%%%%%%%%%%%%%%%%%%%%%%%%%%%%%%%%%%%%%%%%%%%%%%%%%%

Fractional derivation and integration is a well-established topic, see
\cite{Samk93,Mill93,Podl99,Hilf00} for reviews. However, for a derivative
operator $\partial^a$ of real order $a$ the commutativity
$\partial^a \partial^b = \partial^{a+b} = \partial^b \partial^a$ is not 
trivial. The well-known Riemann-Liouville and Gr\"unwald-Letnikov fractional
derivatives do not satisfy it. Since our construction of local scaling
operators is built around this property, we shall present here fractional
derivatives in a self-contained manner such that commutativity is guaranteed, 
along with some other simple calculational rules. 
In particular, the definition used here allows for the solving of fractional 
differential equations of rational order by series expansion methods. 
For simplicity, we merely consider functions $f(r)$ of a single variable $r$. 

Consider a set $E$ of numbers $e$ such that any $e\in E$ is not a
negative integer, $e\ne -(n+1)$ with $n\in \mathbb{N}$. We call such a set an 
{\it E-set}. Let $I$ be some (possibly infinite) positive real interval. 
We define the {\it $\cal M$-space} of generalized functions 
associated with the E-set $E$
\BEQ \label{A:MRaum}
{\cal M} := {\cal M}_{E}(I,\mathbb{R}) = 
\left\{ f : I \subset \mathbb{R} \to \mathbb{R} 
\left| f(r) = \sum_{e\in E} f_e r^e + 
\sum_{n=0}^{\infty} F_n \delta^{(n)}(r) \; ; 
f_e \in \mathbb{R}\; , F_n \in\mathbb{R} \right. \right\}
\EEQ
where $\delta^{(n)}(r)$ is the $n^{\rm th}$
derivative of the Dirac delta function. The part of $f\in{\cal M}$ parametrized
by the constants $f_e$ is called the {\em regular part} of $f$ 
and the part parametrized by the $F_n$ is call the {\em singular part} of $f$. 
A well-known theorem \cite[p. 81]{Gelf64} states that any generalized function 
$f(r)$ concentrated at $r=0$ is indeed given by a {\em finite} sum 
$\sum_{n} F_n \delta^{(n)}(r)$. It is understood throughout that only finitely
many of the coefficients $F_n$ in (\ref{A:MRaum}) are non-vanishing. 
Furthermore, we assume that 
the regular part in (\ref{A:MRaum}) converges `well enough' that all 
`reasonable' operations can be carried out (see Lemma 2 below for a sufficient
condition). For example, we could choose $E=\mathbb{N}$ 
and take $\cal M$ to be the set of analytic functions on $I$. More generally,
if we take $E=\mu\mathbb{N}+\lambda$ with $\mu>0$ and 
$\lambda\ne -(\mu (n+1)+m+1)$ 
with $n,m\in\mathbb{N}$, ${\cal M}_E$ is the space of functions of the form 
$r^{\lambda} f\left(r^{\mu}\right)$ with $f(r)$ analytic. Although
everything here is specified in terms of real numbers, the formal extension 
to complex-valued functions is immediate.   

\noindent {\bf Definition:} Let $a\in\mathbb{R}$, $E$ be an E-set and let 
$E':=\{ e'| e'=e-a; e\in E\}$. Analogously to (\ref{A:MRaum}) one 
has the space ${\cal M}'={\cal M}_{E'}$. An operator 
$\partial^a : {\cal M}\to{\cal M}'$ is called a {\it derivative of order $a$}, 
iff it satisfies the properties:
\BEA
\mbox{\rm i)} & & \partial^a \left( \lambda f(r) + \mu g(r) \right) =
\lambda \partial^a f(r) + \mu \partial^a g(r) \;\; 
\forall \lambda,\mu \in\mathbb{R}\; \mbox{\rm and all } f,g\in {\cal M} 
\nonumber \\
\mbox{\rm ii)} & & \partial^a r^{e} =\frac{\Gamma(e+1)}{\Gamma(e-a+1)} r^{e-a}
+ \sum_{n=0}^{\infty} \delta_{a,e+n+1} \Gamma(e+1) \delta^{(n)}(r)
\label{A:Ablei} \\
\mbox{\rm iii)} & & \partial^a\delta^{(n)}(r)=\frac{r^{-1-n-a}}{\Gamma(-a-n)} 
+ \sum_{m=0}^{\infty} \delta_{a,m} \delta^{(n+m)}(r)
\nonumber 
\EEA
where $\Gamma(x)$ is the Gamma function. In particular, it follows from
(\ref{A:Ablei}) that the prefactor for any monomial $r^{-n-1}$ with 
$n\in\mathbb{N}$ indeed vanishes. For our purposes, we may consider the set 
$E'$ therefore also as an E-set and ${\cal M}'$ as an ${\cal M}$-space. 

In particular, it is implied that $\partial^a$ can be applied term-by-term
to any function $f\in{\cal M}$. Often, we
shall also write $\partial^a = \partial_r^a$ if we want to specify explicitly
the variable $r$ on which $\partial^a$ is supposed to act. We point out that
$\partial^a$ is not defined on negative integer powers $r^{-n-1}$ with $n\geq0$.
%%If $a\in\mathbb{N}$, we recover properties of the usual derivative.

In order to show that this definition is not empty, recall the definition of
fractional derivatives as given by Gelfand and Shilov \cite[p. 115]{Gelf64}. 
For the real line, they consider the generalized function
\BEQ
r_+^{\alpha} := \left\{ \begin{array}{ll} 
r^{\alpha} & ;~~ r > 0 \\ 0 & ;~~ r \leq 0  \end{array} \right.
\EEQ
and for generalized functions concentrated on the half-line $r\geq 0$, 
they define 
\BEQ \label{A:DefGS}
\partial^a f(r) := f(r) * \frac{r_+^{-a-1}}{\Gamma(-a)} = 
\frac{1}{\Gamma(-a)} \int_{0}^{r} \!\D\rho\: f(\rho)\left( r-\rho\right)^{-a-1}
\EEQ
It is understood that the integral must be regularized \cite{Gelf64}. 
The conditions (\ref{A:Ablei}) are immediately verified \cite{Gelf64}, using
$\int_0^1\!\D t\, t^{a-1} (1-t)^{b-1} = \Gamma(a) \Gamma(b)/\Gamma(a+b)$ and
analytic continuation. For comparison with the 
Riemann-Liouville/Gr\"unwald-Letnikov/Marchaud fractional derivatives
$D^a = {}_0D_r^a$, we recall that $D^a r^e = 
(\Gamma(e+1)/\Gamma(e-a+1))\,r^{e-a}$ \cite{Samk93,Mill93,Podl99,Hilf00}.
The singular terms which arise in the definition (\ref{A:Ablei}) are absent. 

The practical interest of the definition (\ref{A:Ablei}) 
comes from the following simple rules for calculation. 

\noindent{\bf Lemma 1:} {\it If $E$ is an E-set, $\cal M$ the associated 
$\cal M$-space, $f\in{\cal M}$ and $\partial^a$ the derivative 
of order $a$ such that all $\partial^a f$ considered below exist, one has 
on ${\cal M}$}
\BEA
\partial^{a+b} f(r) &=& \partial^{a} \partial^{b} f(r) = 
\partial^b \partial^a f(r)  \label{A:Lem11}\\
\left[ \partial^a , r \right] f(r) &=& a \partial^{a-1} f(r) \label{A:Lem12}\\
\partial_{r}^{a} f(\lambda r) &=& 
\lambda^a \partial_{\lambda r}^{a} f(\lambda r) \label{A:Lem13} \\
\partial_{\lambda r}^a f(r) &=& \lambda^{-a} \partial_r^a f(r)\label{A:Lem14}
\EEA
{\it where $\lambda> 0$ is a real constant. 
If $f$ is analytic without singular terms and $g\in {\cal M}$, one has}
\BEQ
\partial^{a}\left( f(r) g(r) \right) = \sum_{\ell=0}^{\infty} 
\left( \begin{array}{c} a \\ \ell \end{array}\right) 
\frac{\D^{\ell} f(r)}{\D r^{\ell}}\, 
\partial^{a-\ell} g(r) \label{A:Lem15}
\EEQ
{\it where $\D^{\ell}/\D r^{\ell}$ are ordinary derivatives of 
integer order $\ell$.}
These rules are the natural generalizations of the familiar properties of the
usual derivative. The commutator $[.,.]$ is defined as usual. 
Well-known fractional derivatives such as the Riemann-Liouville,
Gr\"undwald-Letnikov or Marchaud fractional derivatives satisfy the
commutativity relation (\ref{A:Lem11}) only if further 
conditions are imposed on $a$ and $b$ or on the function $f(r)$ 
\cite{Samk93,Mill93,Podl99} (see also the example (\ref{A:Beispiel}) below). 
On the other hand, 
(\ref{A:Lem11}) does hold for the Gelfand-Shilov definition (\ref{A:DefGS}) 
\cite{Gelf64,Podl99}, the Weyl fractional derivative \cite{Mill93} and for
the recent complex multivalued definition presented in \cite{Zava98}.
The generalized Leibniz rule (\ref{A:Lem15}) is usually proven for both $f,g$ 
analytic \cite{Samk93} or infinitely often differentiable \cite{Podl99}. 
In \cite{Mill93}, $f$ is analytic and $g\in{\cal M}_{M}$, with the E-set
$M=\mathbb{N}+\mu$, $\mu>-1$ (logarithms are also admitted). 
Generalized (and simpler) Leibniz rules based on the convolution 
product were discussed in \cite{Klim01}.

So far, the definition of $\partial^a$ was performed for variables $r>0$
(and we should have written $r_+^e$ instead of $r^e$ everywhere). 
We may use (\ref{A:Lem13}) with $\lambda$ negative to formally extend the 
definition of $\partial^a$ to any value of $r\ne 0$. 
Then (\ref{A:Lem11},\ref{A:Lem12},\ref{A:Lem14}) remain valid. 

\noindent {\it Proof:} We show that the rules of the lemma can be reduced to 
the basic properties (\ref{A:Ablei}). In order to prove (\ref{A:Lem11}), 
we consider separately two cases. Let
\BEQ
{\rm ind}_{b\not\in\mathbb{N}} := \left\{ \begin{array}{ll}
1 & \mbox{\rm ~~;~  if~~ $b\not\in\mathbb{N}$} \\
0 & \mbox{\rm ~~;~  if~~ $b\in\mathbb{N}$} 
\end{array}\right.
\EEQ
First, consider
\BEA
\lefteqn{ \partial^a \partial^b \delta(r) \; = \;
\partial^a \left( \frac{r^{-1-b}}{\Gamma(-b)} + \sum_{m\in\mathbb{N}}
\delta_{b,m} \delta^{(m)}(r) \right) } \nonumber \\
&=& \frac{{\rm ind}_{b\not\in\mathbb{N}}}{\Gamma(-b)} 
\left( \frac{\Gamma(-b)r^{-1-b-a}}{\Gamma(-b-a)} 
+\sum_{n\in\mathbb{N}} \delta_{a,-b+n} \Gamma(-b) \delta^{(n)}(r) \right)
%%\nonumber \\
%%& & 
+ \sum_{m\in\mathbb{N}} \delta_{b,m} \left( 
\frac{r^{-1-m-a}}{\Gamma(-a-m)}+\sum_{n\in\mathbb{N}}
\delta_{a+b,n}\delta^{(n)}(r)  \right) 
\nonumber \\
&=& \frac{1}{\Gamma(-b-a)} r^{-1-b-a} + \sum_{n\in\mathbb{N}} \delta_{a+b,n}
\delta^{(n)}(r) \; = \; \partial^{a+b} \delta(r)
\nonumber
\EEA
and this also implies commutativity on $\delta^{(n)}(r)=\partial^n\delta(r)$. 
Second, we have
\BEA
\partial^{a}\partial^{b} r^{e} &=&
\partial^a \left(   
\frac{\Gamma(e+1)}{\Gamma(e-b+1)} r^{e-b} 
+\sum_{n\in\mathbb{N}} \delta_{b,e+n+1} \Gamma(e+1) \delta^{(n)}(r)\right)
\nonumber \\ 
&=& {\rm ind}_{b-e-1\not\in\mathbb{N}}\frac{\Gamma(e+1)}{\Gamma(e-b+1)} 
\left(\frac{\Gamma(e-b+1)}{\Gamma(e-a-b+1)} r^{e-b-a} 
+ \sum_{m\in\mathbb{N}}\delta_{a,e-b+m+1} \Gamma(e-b+1)\delta^{(m)}(r)\right) 
\nonumber \\
& &+ \sum_{n\in\mathbb{N}} \delta_{b,e+n+1} \left( 
\frac{\Gamma(e+1)}{\Gamma(-a-n)} r^{-1-n-a} + \sum_{m\in\mathbb{N}}
\delta_{a,m} \Gamma(e+1) \delta^{(n+m)}(r) \right) 
\nonumber \\
&=& \frac{\Gamma(e+1)}{\Gamma(e-b-a+1)} r^{e-b-a} 
+\sum_{m\in\mathbb{N}} \delta_{a+b,e+m+1} \Gamma(e+1) \delta^{(m)}(r)
\nonumber \\
&=& \partial^{a+b} r^e
\nonumber
\EEA 
Having thus checked the claim on the `basis set' spanned by $r^e$ and
$\delta^{(n)}$ it holds on $\cal M$ by linear superposition. Clearly 
$\partial_r^a \partial_r^b f = \partial_r^b \partial_r^a f$. In the sequel
we shall need the identities 
\BEQ \label{A:DeltaId}
r \delta^{(n)}(r) = - n \delta^{(n-1)}(r) \;\; , \;\;
\delta^{(n)}(\lambda r) = \lambda^{-n-1} \delta^{(n)}(r)
\EEQ
To prove (\ref{A:Lem12}), consider
\BEA
\lefteqn{ \left[\partial_r^a, r\right] r^e 
= \partial_r^a r^{e+1} - r \partial_r^a r^e }
\nonumber \\
&=& \left(\frac{\Gamma(e+2)}{\Gamma(e-a+2)}-\frac{\Gamma(e+1)}{\Gamma(e-a+1)}
\right) r^{e-a+1}
\nonumber \\
& & +\sum_{n=0}^{\infty} \delta_{a,e+n+2} \Gamma(e+2) \delta^{(n)}(r) 
-\sum_{n=0}^{\infty} \delta_{a,e+n+1} \Gamma(e+1) r\delta^{(n)}(r) 
\nonumber \\
%&=& \left( \frac{e+1}{e-a+1} -1 \right) 
%\frac{\Gamma(e+1)}{\Gamma(e-a+1)} r^{e-(a-1)}
%+\sum_{n=0}^{\infty} \delta_{a-1,e+n+1} (a-1-n) \Gamma(e+1) \delta^{(n)}(r) 
%+\sum_{n=1}^{\infty} \delta_{a,e+n+1} n\Gamma(e+1) \delta^{(n-1)}(r) 
%\nonumber \\
&=& \frac{a\Gamma(e+1)}{\Gamma(e+1-(a-1))} r^{e-(a-1)}
+ a\sum_{n=0}^{\infty} \delta_{a-1,e+n+1} \Gamma(e+1) \delta^{(n)}(r)
\nonumber \\
&=& a \partial_r^{a-1} r^e \nonumber
\EEA
where in the third line the first identity (\ref{A:DeltaId}) was used. Next, 
\BEA
\left[\partial^a, r\right] \delta^{(p)}(r) &=& 
-p  \partial^{a} \delta^{(p-1)}(r) - r \partial^{a} \delta^{(p)}(r)
\nonumber \\
&=& -\left( \frac{p}{\Gamma(-a-p+1)} + \frac{1}{\Gamma(-a-p)}\right) 
r^{-p-a} -\sum_{m=0}^{\infty} \delta_{a,m} \left( p\delta^{(p-1+m)}(r) 
+r \delta^{(p+m)}(r) \right) 
\nonumber \\
&=& \frac{a}{\Gamma(-(a-1)-p)}r^{-1-p-(a-1)} + a \sum_{m=0}^{\infty}
\delta_{a,m} \delta^{(p+a-1)}(r) 
\; = \; a \partial_r^{a-1}\delta^{(p)}(r)
\nonumber
\EEA
where the first identity (\ref{A:DeltaId}) was used again. Having checked
(\ref{A:Lem12}) on $r^e$ and $\delta^{(n)}$, it holds on $\cal M$ by linear 
superposition. For (\ref{A:Lem13}), we set $s=\lambda r$ and use the
second identity (\ref{A:DeltaId}). Then 
\BEA
\partial_r^a f(\lambda r) &=& 
\partial_r^a\left( \sum_{e\in E} f_e\:(\lambda r)^e 
+ \sum_{n=0}^{\infty} F_n \delta^{(n)}(\lambda r) \right)
= \partial_r^a\left( \sum_{e\in E} \lambda^e f_e\:r^e 
+ \sum_{n=0}^{\infty} \lambda^{-n-1} F_n \delta^{(n)}(r) \right) 
\nonumber \\
&=& \sum_{e\in E} \lambda^e f_e \Gamma(e+1)\left( 
\frac{r^{e-a}}{\Gamma(e-a+1)} + 
\sum_{m=0}^{\infty} \delta_{a,e+m+1} \delta^{(m)}(r)  \right)
\nonumber \\  
& & + \sum_{n=0}^{\infty} F_n \lambda^{-n-1} \left( 
\frac{r^{-1-n-a}}{\Gamma(-a-n)} + 
\sum_{m=0}^{\infty} \delta_{a,m} \delta^{(n+m)}(r) \right) 
\nonumber \\
&=& \lambda^a\left( \sum_{e\in E} f_e\partial_{\lambda r}^a(\lambda r)^e 
+ \sum_{n=0}^{\infty} F_n \partial_{\lambda r}^a\delta^{(n)}(\lambda r) \right) 
\; =\;  \lambda^a \partial_s^a f(s)
\nonumber
\EEA
For the forth relation (\ref{A:Lem14}), let again $s=\lambda r$ and, using
(\ref{A:Lem13})
\begin{displaymath}
\partial_{\lambda r}^a f(r) = \partial_s^a f(\lambda^{-1} s) \:
=\: \lambda^{-a} \partial_{\lambda^{-1}s}^a f(\lambda^{-1} s) \: = \: 
\lambda^{-a} \partial_r^a f(r) 
\end{displaymath}
The fifth rule (\ref{A:Lem15}) may be obtained in the same manner, but we 
shall prove below it with the help of the commutator identity (\ref{A:Lem31})
which permits a shorter proof. \hfill q.e.d.

In particular, one can set $g(r)=1$ in eq.~(\ref{A:Lem15}). Then, 
for $f$ analytic and non-singular one has the regular part, 
see also \cite{Samk93,Mill93,Podl99,Klim01} 
\BEQ \label{A:faReihe}
\left.\partial_r^a f(r)\right|_{\rm reg} 
= r^{-a} \sum_{\ell=0}^{\infty} \left( \vekz{a}{\ell}\right) 
\frac{1}{\Gamma(\ell-a+1)} \,r^{\ell}\,\frac{\D^{\ell} f(r)}{\D r^{\ell}}
= r^{-a} \sum_{\ell=0}^{\infty} \frac{\Gamma(a+1)}{\ell !}
\frac{\sin(\pi(a-\ell))}{\pi(a-\ell)}\,r^{\ell}\,\frac{\D^{\ell} f(r)}
{\D r^{\ell}}
\EEQ
which expresses the regular part of $\partial^a f$ in terms of ordinary 
derivatives. Clearly, if $a\to k\in\mathbb{N}$, one recovers from the
second form (\ref{A:faReihe}) via
$\left.\partial_r^a f(r)\right|_{\rm reg} \to \D^k f(r)/\D r^k$ 
the ordinary derivative of integer order $k\geq 0$. 
On the other hand, if $a=k+\alpha$ with
$k\in\mathbb{N}$ and $0<\alpha<1$, we have
\BEQ
\partial_r^{k+\alpha} \exp(r) = \sum_{n=0}^{\infty} 
\frac{r^{n-k-\alpha}}{\Gamma(n+1-k-\alpha)}
=\sum_{n=-k}^{\infty} 
\frac{r^{n-\alpha}}{\Gamma(n+1-\alpha)} = r^{-k-\alpha} E_{1,1-k-\alpha}(r)
\EEQ
where $E_{\alpha,\beta}(r)=\sum_{k=0}^{\infty}z^k/\Gamma(\alpha k+\beta)$ is
a Mittag-Leffler type function. If $\alpha=0$, we recover 
$\left.\partial_r^k \exp(r)\right|_{\rm reg} = \exp(r)$, but that
property does not hold anymore if $\alpha>0$ (for the Weyl fractional 
derivative one has ${\partial^a}_W e^r = e^r$ \cite{Mill93}). 

Finally, it may be useful to 
illustrate the commutativity of $\partial^a$ on some example. 
Following Miller and Ross \cite[p. 210]{Mill93}, take a positive integer $q$
and let 
\BEQ \label{A:Beispiel}
\mathfrak{e}(t) := \sum_{k=0}^{q-1} \alpha^{q-1-k} t^{-k/q} 
E_{1,1-k/q}(\alpha^q t)
\EEQ
For the Riemann-Liouville derivative $D^a={}_0D_t^a$, one has indeed
$D^{1/q}\mathfrak{e}(t)=\alpha\mathfrak{e}(t)$, but $D^{2/q}$ and 
$D^{1/q}D^{1/q}$ are clearly different, since
\begin{displaymath}
D^{2/q} \mathfrak{e}(t) = \alpha^2 \mathfrak{e}(t) + 
\frac{t^{-1-1/q}}{\Gamma(-1/q)} \ne \alpha^2 \mathfrak{e}(t) = 
D^{1/q} \left( D^{1/q} \mathfrak{e}(t) \right)
\end{displaymath}
and this fact had led Miller and Ross to the definition of {\em sequential}
fractional derivatives \cite{Mill93}. 
On the other hand, using (\ref{A:Ablei}), we find
\BEA
\partial_t^{1/q} \mathfrak{e}(t) &=& \alpha \mathfrak{e}(t) + \delta(t) 
\nonumber \\
\partial_t^{2/q} \mathfrak{e}(t) &=& \alpha^2 \mathfrak{e}(t) + 
\frac{t^{-1-1/q}}{\Gamma(-1/q)}  + \alpha \delta(t) \; = \; 
\partial_t^{1/q}\left( \partial_t^{1/q} \mathfrak{e}(t) \right)
\nonumber
\EEA
and the singular terms are seen to be essential for the commutativity property
(\ref{A:Lem11}). 

Consider an E-set $E$ which is countable and ordered. Therefore, the elements
$e\in E$ can be labelled by an $n\in\mathbb{N}$, viz. $e=e_n$. Let
$\nu_n := e_{n+1}-e_n > 0$. Call such an E-set $E$ {\em well-separated 
with separation constant $\eps$}, 
if there is an $\eps>0$ such that $\nu_n\geq \eps$.
For the regular part of a function $f\in{\cal M}_E$, we have
\begin{displaymath}
\left.f(r)\right|_{\rm reg} = \sum_{e\in E} f_e r^e = 
\sum_{n=0}^{\infty} f_n r^{e_n} \;\; ; \;\;
f_n := f_{e_n}
\end{displaymath}
Questions of existence of $\partial_r^a f(r)$ and its relation to ordinary
derivatives are dealt with in the following

\noindent {\bf Lemma 2:} {\it 
Let $E$ be a well-separated E-set with separation constant $\eps$,
$f\in{\cal M}_E$, $e_n >0$, $e_n-a>0$ and
\BEQ
\rho^{-1} := \limsup_{n\to\infty} \, |f_n|^{1/e_n} \geq 0
\EEQ
Then the following holds.\\
(i) $f(r)$ converges absolutely for $|r|<\rho$. \\
(ii) If $\nu_n/e_n < B$ for some constant $B$, 
$\partial_r^a f(r)$ converges absolutely for $|r|<\rho 
\min\left(1,(1+B)^{-a/\eps}\right)$. \\
(iii) If $f:I\to\mathbb{R}$ is analytic with 
a radius of convergence $\rho>0$ around $r=0$, then the series 
(\ref{A:faReihe}) for $\partial_r^a f(r)$ converges absolutely for 
$|r|< \rho/2$.} 
Property (i) is well-known for psi-series and still holds if only $e_n/\ln n
\to \infty$ as $n\to\infty$ \cite{Hill76}. The conditions imposed here are
sufficiently wide to include functions of the form 
$r^{\lambda}f\left(r^{\mu}\right)$ with $f(r)$ analytic, $\mu>0$ and 
$\lambda\ne-\mu m - n$, $n,m\in\mathbb{N}$, which is enough for the 
applications we have in mind. In this case, we have effectively $B=0$, since
$\nu_n/e_n = \mu/(\mu n +\lambda)\to 0$ as $n\to\infty$. 

\noindent {\it Proof:} The conditions $e_n>0$ and $e_n-a>0$ 
are only needed in order to make the regular parts of $f(0)$ and 
$\partial^a f(0)$ well-defined.
Since the singular parts of $f$ and $\partial^a f$ are finite sums, 
they do not affect the convergence and can be suppressed here. 

\noindent (i) In order to show the convergence of $f(r)$, 
consider $N\in\mathbb{N}$ sufficiently
large. We then have the remainder stimate
\begin{displaymath}
R_N := \sum_{n=N}^{\infty} \left| f_n r^{e_n} \right| =
\sum_{n=N}^{\infty} \left| f_n^{1/e_n} r \right|^{e_n}
\leq \sum_{n=N}^{\infty} \left| \frac{r}{\rho} \right|^{e_n}
\end{displaymath}
which holds for $N$ sufficiently large.\footnote{More precisely, let $\delta>0$
and consider $|r|<\rho-\delta$. $\delta$ can be made arbitrarily small if $N$
is large enough.} 
Since $E$ is well-separated,
we have $e_n - e_N \geq (n-N)\eps$ and $e_N\geq e_0 + N\eps$. 
Therefore, for $|r|<\rho$
\begin{displaymath}
R_N \leq  \left|\frac{r}{\rho}\right|^{e_N}
\sum_{n=N}^{\infty}\left|\frac{r}{\rho}\right|^{e_n-e_N}
\leq \left|\frac{r}{\rho}\right|^{e_N}
\sum_{n=N}^{\infty}\left|\frac{r}{\rho}\right|^{\eps(n-N)}
\leq  \left|\frac{r}{\rho}\right|^{\eps N}
\frac{|r/\rho|^{e_0}}{1-|r/\rho|^{\eps}}
\end{displaymath}
and $R_N\to 0$ as $N\to\infty$.

\noindent (ii) We need the asymptotic identity (6.1.47) 
in \cite{Abra65} for $z\to\infty$ and constants $a,b$
\BEQ \label{A:asympG}
\frac{\Gamma(z+a)}{\Gamma(z+b)} \simeq z^{a-b} \left( 1+ O(z^{-1}) \right) 
\EEQ
For the convergence of
\begin{displaymath}
\left.\partial^a f(r)\right|_{\rm reg} 
= \sum_{n=0}^{\infty} f_n \frac{\Gamma(e_n+1)}{\Gamma(e_n-a+1)} r^{e_n-a}
\end{displaymath}
we consider the remainder
\begin{displaymath}
Q_N :=\sum_{n=N}^{\infty}\left| f_n\frac{\Gamma(e_n+1)}
{\Gamma(e_n-a+1)} r^{e_n-a}\right|
= |r^{-a}|
\sum_{n=N}^{\infty}\left|
\frac{\Gamma(e_n+1)}{\Gamma(e_n-a+1)}\right|\,\left|f_n^{1/e_n} r\right|^{e_n}
\leq |r^{-a}| \sum_{n=N}^{\infty} e_n^a \left|
\frac{r}{\rho} \right|^{e_n} 
\end{displaymath}
where eq.~(\ref{A:asympG}) was used and $N$ was taken to be sufficiently large.
Now, we use the following known fact
\cite{Cour65}: if $y_n>0$ and $\sum_{n=0}^{\infty} y_n < \infty$ and 
furthermore $|x_{n+1}/x_n|< y_{n+1}/y_n$, then the series 
$\sum_{n=0}^{\infty} x_n$ is absolutely convergent. We apply this to the
sequence $x_n := e_n^a |r/\rho|^{e_n}$ and have the estimate
\begin{displaymath}
\frac{x_{n+1}}{x_n} = \left(\frac{e_{n+1}}{e_n}\right)^a 
\left|\frac{r}{\rho}\right|^{e_{n+1}-e_n} =
\left(1+\frac{\nu_n}{e_n}\right)^a \left|\frac{r}{\rho}\right|^{e_{n+1}-e_n} 
< \left( 1+B\right)^a \left|\frac{r}{\rho}\right|^{e_{n+1}-e_n}
\end{displaymath}
Therefore, if we take $y_n = (1+B)^{an} |r/\rho|^{e_n}$, it is only left to
prove that $\sum_{n=0}^{\infty} y_n$ is convergent. But this is obvious, since
for $|r|<\rho$, one has
\begin{displaymath}
Q_N' := \sum_{n=N}^{\infty} y_n = \sum_{n=N}^{\infty} (1+B)^{an} \left|
\frac{r}{\rho}\right|^{e_n} 
\leq \left| \frac{r}{\rho}\right|^{e_0} \sum_{n=N}^{\infty} \left(
(1+B)^a \left|\frac{r}{\rho}\right|^{\eps}\right)^n
\end{displaymath}
which indeed tends to zero for $N\to\infty$, 
if $(1+B)^a (|r|/\rho)^{\eps} < 1$.

\noindent 
(iii) The series $f(2r) = \sum_{k=0}^{\infty} (k!)^{-1} r^k
f^{(k)}(r)$, where $f^{(k)}(r) := \D^k f(r)/\D r^k$, converges absolutely for
$|2r|<\rho$, thus
\begin{displaymath}
\left| \frac{r}{n+1} \frac{f^{(n+1)}(r)}{f^{(n)}(r)}\right|\;
\mathop{<}_{n\to\infty} 1 \;\; \mbox{\rm ~;~ if $|r|<\rho/2$}
\end{displaymath}
If $a\in\mathbb{N}$, there is nothing to show. If
$a\not\in\mathbb{N}$, let $b=-a$
and recall that for $a\in\mathbb{R}$
\begin{displaymath}
\left(\vekz{a}{\ell}\right) = \frac{a(a-1)\ldots (a-\ell+1)}{\ell !}
= (-1)^{\ell} \frac{b(b+1)\ldots(b+\ell-1)}{\ell !}
\end{displaymath}
We have from the first form of (\ref{A:faReihe}) the estimate
\begin{displaymath}
\left| \left.\partial_r^{a} f(r)\right|_{\rm reg} \right|
= \left| \left. \partial_r^{-b} f(r)\right|_{\rm reg} \right| \leq 
\left| r^b \right| \sum_{\ell=0}^{\infty}
\left| \frac{b(b+1)\ldots(b+\ell-1)}{\ell ! \, \Gamma(\ell+b)}
r^{\ell}\, f^{(\ell)}(r) \right|
=: \sum_{\ell=0}^{\infty} \mu_{\ell} 
\end{displaymath}
We call the $\ell^{\rm th}$ coefficient in this series  $\mu_{\ell}$ and have
(since $b$ is not a negative integer) 
\begin{displaymath}
\frac{\mu_{n+1}}{\mu_n} = 
\left| \frac{(b+n)r}{(n+1)(n+b)} \frac{f^{(n+1)}(r)}{f^{(n)}(r)}\right|\;
\mathop{<}_{n\to\infty} 1 \;\; \mbox{\rm ~;~ if $|r|<\rho/2$}
\end{displaymath}
which implies absolute convergence. \hfill q.e.d.

Practical calculations with $\partial^a$ are simplified by the following

\noindent {\bf Lemma 3:} {\it (i) If $\partial^a$ is the derivative of real 
order $a$ and $n\in\mathbb{N}$, one has}
\BEQ \label{A:Lem31}
\left[ \partial_t^a, t^{n} \right] = \sum_{k=1}^{n}
\left( \vekz{a}{k}\right)\left(\vekz{n}{k}\right)\, 
k!\, t^{n-k}\, \partial_t^{a-k}
\EEQ
{\it (ii) If $f$ is analytic without any singular terms and
$g\in{\cal M}$ the relations (\ref{A:Lem12},\ref{A:Lem15},\ref{A:Lem31}) 
are equivalent. \\
(iii) If $f\in{\cal M}$ but without singular terms and $n\in\mathbb{N}$, 
one has with the ordinary derivative 
$f^{(\ell)}(x) = \D^{\ell} f(x)/\D x^{\ell}$} of integer order $\ell$ 
\BEQ \label{A:Lem32}
\left[ f(\partial_r), r^n\right] = \sum_{\ell=1}^{n} 
\left(\vekz{n}{\ell}\right) r^{n-\ell} f^{(\ell)}(\partial_r)
\EEQ
{\it Proof:} (i) We proceed by induction over $n$. The case $n=1$ is the
identity (\ref{A:Lem12}). For the induction step, let $g\in{\cal M}$ and
consider, using (\ref{A:Lem12}) again twice
\BEA 
\lefteqn{ 
\left[ \partial_t^a, t^{n+1} \right] g(t) = 
\left[ \partial_t^a, t^n \right] tg(t) 
+ t^n \left[ \partial_t^a, t \right] g(t) }
\nonumber \\
&=& \left( \sum_{k=1}^{n} \left( \vekz{a}{k}\right)\left(\vekz{n}{k}\right)\, 
k!\, t^{n-k}\,  
\left( t \partial_t^{a-k} + \left[ \partial_t^{a-k},t\right]\right) 
+ a t^n \partial_t^{a-1} \right) g(t) 
\nonumber \\
&=& \left( \sum_{k=1}^{n} \left( \vekz{a}{k}\right)\left(\vekz{n}{k}\right)\, 
k!\, t^{n+1-k}\, \partial_t^{a-k} \right. 
\nonumber \\
& & \left. 
+ \sum_{k=1}^{n} \left( \vekz{a}{k+1}\right)\left(\vekz{n}{k}\right)\, 
(k+1)!\, t^{n+1-(k+1)}\, \partial_t^{a-(k+1)} 
+ a t^n \partial_t^{a-1} \right) g(t) 
\nonumber \\
&=& \left( a (n+1) t^n \partial_t^{a-1}  + 
\sum_{k=2}^{n} \left( \vekz{a}{k}\right)\left(\vekz{n+1}{k}\right)\, 
k!\, t^{n+1-k}\, \partial_t^{a-k} 
+ \left(\vekz{a}{n+1}\right)\, 
(n+1)!\,  \partial_t^{a-n-1} \right) g(t)
\nonumber \\
&=& \sum_{k=1}^{n+1} \left( \vekz{a}{k}\right)\left(\vekz{n+1}{k}\right)\, 
k!\, t^{n+1-k}\, \partial_t^{a-k} g(t)
\nonumber 
\EEA
and the assertion follows. \\
(ii) For an analytic functions without singular terms, one has
$f(r)=\sum_{n=0}^{\infty} f_n r^n$ and for all $k\in\mathbb{N}$ the
ordinary derivative of $t^n$ is
\begin{displaymath}
\frac{\D^k t^{n}}{\D t^k} = 
\left(\vekz{n}{k}\right) \, k!\, t^{n-k}
\end{displaymath}
From (\ref{A:Lem31}) we have
\begin{displaymath}
\partial_t^a \left( t^n g(t) \right) = 
\sum_{k=0}^{n} \left( \vekz{a}{k}\right)\left(\vekz{n}{k}\right)\, 
k!\, t^{n-k}\, \partial_t^{a-k} g(t)
\end{displaymath}
and (\ref{A:Lem15}) indeed follows, under the stated assumptions on $f$. 
Conversely, starting from the generalized Leibniz rule (\ref{A:Lem15}) and 
setting $f(t)=t^{n}$ with $n\in\mathbb{N}$ we have 
\begin{displaymath}
\left[ \partial_t^a, t^{n} \right]g(t) = 
\partial_t^a\left( t^{n} g(t)\right) - t^{n}\, \partial_t^a g(t) =
\sum_{k=1}^{\infty}
\left(\vekz{a}{k}\right) 
\frac{\D^k t^{n}}{\D t^k}\, \partial_t^{a-k}g(t)
\end{displaymath}
and we recover indeed (\ref{A:Lem31}). The special case $n=1$ then reproduces
(\ref{A:Lem12}). 
\\
(iii) For $f\in{\cal M}$ non-singular one has $f(r)=\sum_e f_e r^e$. Therefore
\begin{displaymath}
\left[ f(\partial_r), r\right] = \left[ \sum_e f_e \partial_r^e, r \right] 
= \sum_e f_e e \partial_r^{e-1} = f^{(1)}(\partial_r)
\end{displaymath}
where in the second step (\ref{A:Lem12}) was used. The assertion now follows 
immediately by induction over $n$. \hfill q.e.d.

%%%%%%%%%%%%%%%%%%%%%%%%%%%%%%%%%%%%%%%%%%%%%%%%%%%%%%%%%%%%%%%%%%%%%%%%%%%%%%%%
\appsection{B}{Generators of the Schr\"odinger algebra for $d>1$}
%%%%%%%%%%%%%%%%%%%%%%%%%%%%%%%%%%%%%%%%%%%%%%%%%%%%%%%%%%%%%%%%%%%%%%%%%%%%%%%%

We list the generators of the infinite-dimensional Lie algebra of the 
Schr\"odinger group in $d=2$ spatial dimensions. They read
\BEA
X_n &=& -t^{n+1}\partial_t - \frac{n+1}{2} t^n (r_1 \partial_1 + r_2\partial_2)
-\frac{n(n+1)}{4} {\cal M} t^{n-1} (r_1^2 + r_2^2) 
- \frac{x}{2} (n+1) t^n \nonumber \\
Y_m^{(1)} &=& -t^{m+1/2} \partial_1 -\left(m+\frac{1}{2}\right) 
{\cal M} t^{m-1/2} r_1 \nonumber \\
Y_m^{(2)} &=& -t^{m+1/2} \partial_2 -\left(m+\frac{1}{2}\right)
{\cal M} t^{m-1/2} r_2 \\
M_n &=& - t^n {\cal M} \nonumber \\
R &=& r_1 \partial_2 - r_2 \partial_1 \nonumber 
\EEA
where $\partial_j = \partial/\partial r_j$ with $j=1,2$ and where $\cal M$ 
is the mass. Here $r_j\in\mathbb{R}$ are the two spatial coordinates.
The indices $n\in\mathbb{Z}$ and $m \in\mathbb{Z}+\frac{1}{2}$. With respect to
the $d=1$ case treated in the text, there are now two sets of generators for 
generalized Galilei transformations and the new generator $R$ of spatial 
rotations.

A straightforward calculation gives the commutators ($i,j=1,2$):
\BEA
[ X_n , X_{n'} ] &=& (n-m) X_{n+n'} \nonumber \\ {}
[ X_n , Y_m^{(j)} ] &=& \left( \frac{n}{2} -m \right) Y_{n+m}^{(j)} 
\nonumber \\ {}
[ X_n , M_{n'} ] &=& -n' M_{n+n'} \nonumber \\ {}
[ Y_m^{(i)} , Y_{m'}^{(j)} ] &=& \delta_{i,j} \, (m-m') M_{m+m'} \\ {}
[ Y_m^{(j)} , M_n ] &=& [ M_n , M_{n'} ] = 0 \nonumber \\ {}
[ X_n , R ] &=& [ M_n , R ] = 0 \nonumber \\ {}
[ Y_m^{(1)} , R ] &=& Y_m^{(2)} \nonumber \\ {}
[ Y_m^{(2)} , R ] &=& - Y_m^{(1)} \nonumber 
\EEA
which closes into an infinite-dimensional Lie algebra. Similarly, the extension
coming from including a parameter $B_{20}$ in the generators 
(see (\ref{3:XIIgen},\ref{3:YIIgen}) for the $1D$ case) can be written down 
straightforwardly. 

The special case of the finite-dimensional Lie subalgebra, corresponding
to (\ref{2:Schr}) for $d=2$, is given by the set 
\BEQ
\{ X_{-1,0,1}, Y_{-1/2,+1/2}^{(1)}, Y_{-1/2,+1/2}^{(2)}, M_0, R\}.
\EEQ 
%
%%On voit donc qu'on a bien un alg\`ebre de Lie de dimension infinie. 
%%Par analogie avec l'alg\`ebre conforme, si on utilise $X_n$ pour m\'esurer 
%%les dimensions des champs quantiques $X(t) = \sum_n t^n X_n$, 
%%$Y(t) =\sum_m t^m Y_m$ et $M(t) = \sum_n t^n M_n$, on lit \`a partir des 
%%commutateurs que
%%\BEQ
%%\dim X =2 \;\; , \;\; \dim Y = \frac{3}{2} \;\; , \;\; \dim M = 1
%%\EEQ
%
The generalization to spatial dimensions $d>2$ proceeds along the same lines.  

%%%%%%%%%%%%%%%%%%%%%%%%%%%%%%%%%%%%%%%%%%%%%%%%%%%%%%%%%%%%%%%%%%%%%%%%%%%%%%%%
\appsection{C}{Infinitesimal generators for generalized Galilei transformations}
%%%%%%%%%%%%%%%%%%%%%%%%%%%%%%%%%%%%%%%%%%%%%%%%%%%%%%%%%%%%%%%%%%%%%%%%%%%%%%%%

As an alternative to the construction of local scale transformation
given in the text, we present here a different construction of the 
infinitesimal local scaling generators and proceed to the 
calculation of scaling functions. 
This had been the first case where generators of a local scale invariance
could be explicitly constructed and scaling functions could be found. 

Starting from the generators
\BEQ \label{C:XXY}
X_{-1} = -\partial_t \;\; , \;\;
X_{0} = -t \partial_t - \frac{1}{\theta} r \partial_r \;\; , \;\;
Y_{-1/2} = - \partial_r
\EEQ
which satisfy the commutation relations
\BEQ
\left[ X_0, X_{-1} \right] = X_{-1} \;\; , \;\;
\left[ X_{-1}, Y_{-1/2} \right] =0 \;\; , \;\;
\left[ X_0, Y_{-1/2} \right] = \frac{1}{\theta} Y_{-1/2}
\EEQ
we want to find, for $\theta$ as general as possible, a generalized Galilei
transformation $Y_{1/2}$ such that a closed Lie algebra results. 
In particular, we want to {\it construct this algebra $\mathbb{A}$ such that
$X_0$ acts as a counting operator, that is
for any generator $A\in\mathbb{A}$, we have $[X_0,A] =a A$}. Motivated
from the Schr\"odinger case $\theta=2$, 
where $Y_{1/2} = -t\partial_r - {\cal M} r$ (and after having tried out many
different forms), we make the ansatz
\BEQ
Y_{1/2} = - t \partial_r - {\cal M}(\partial_r) r
\EEQ
In what follows, we use the formal properties of the derivative $\partial_r^a$
as defined in appendix~A. 

First, we formally calculate the commutator, using (\ref{A:Lem32})
\BEQ
\left[ X_0, Y_{1/2} \right] =
\frac{\theta-1}{\theta} t \partial_r - \frac{1}{\theta} {\cal M}'(\partial_r)
\partial_r r + \frac{1}{\theta} {\cal M}(\partial_r) r
\stackrel{!}{=} - \frac{\theta-1}{\theta} Y_{1/2}
\EEQ
where the last equation is motivated from the first term in the generator
$Y_{1/2}$ and our construction principle. This leads to
\BEQ
(\theta-1) {\cal M}(\partial_r) r = - {\cal M}'(\partial_r) \partial_r r +
{\cal M}(\partial_r) r
\EEQ
If we let $x := \partial_r$, we find $(\theta-2){\cal M}(x) = - {\cal M}'(x) x$
which has the solution
\BEQ
{\cal M}(x) = {\cal M}_0 x^{2-\theta}
\EEQ
where ${\cal M}_0$ is a constant. Next, we find the commutators
\BEQ
\left[ X_{-1} , Y_{1/2} \right] = - Y_{-1/2} \;\; , \;\;
\left[ Y_{1/2}, Y_{-1/2} \right] = - {\cal M}_0 \partial_r^{2-\theta} =: M_0
\EEQ
where $M_0 = - {\cal M}_0 \partial_r^{2-\theta}$ is a new generator. 
For $\theta=2$, we recover the Galilei algebra, where $M_0$ is central. 
Otherwise, we find through a formal calculation
\BEQ
\left[ X_{-1}, M_0 \right] = \left[ Y_{-1/2}, M_0 \right] = 0 \;\; , \;\;
\left[ X_0, M_0 \right] = - \frac{2-\theta}{\theta} M_0 \;\; , \;\; 
\left[ Y_{1/2}, M_0 \right] = -(2-\theta) {\cal M}_0^2 \partial_r^{3-2\theta}
\EEQ
and we see that we must define a new generator $N$. In the special case
$\theta=3/2$, we have $N := - \frac{1}{2} {\cal M}_0^2$. Then 
\BEQ
\left[ Y_{1/2}, M_0 \right] = N
\EEQ
and $N$ is central. Therefore, {\it if $\theta=3/2$, the set} 
\BEQ
\mathbb{A} := \{ X_{-1}, X_0, Y_{-1/2}, Y_{1/2}, M_0, N \}
\EEQ
{\it closes as a Lie algebra and satisfies the condition that $X_0$ acts as a
counting operator.} 

We now derive the form of the two-point function covariant under the 
transformations generated by the set $\mathbb{A}$. Scaling operators will be
characterized by their scaling dimension $x$ and their `mass'. For the 
Schr\"odinger invariant case $\theta=2$, scaling operators are doubletts
$(\phi,\phi^*)$, with `masses' $({\cal M},-{\cal M})$, where $\cal M$ is a
non-negative constant \cite{Henk94}. Here, for $\theta=3/2$, it turns out
that scaling operators are quadrupletts $\phi^{(\alpha)}$, with 
$\alpha=0,1,2,3$ and `masses' $\mathfrak{M} := \II^{\alpha}{\cal M}$, 
where again ${\cal M}$ is a positive constant. We consider two-point functions 
of the form
\BEQ
F = F^{(\alpha,\beta)}(t_1,t_2;r_1,r_2) = 
\langle \phi_1^{(\alpha)}(t_1,r_1)\phi_2^{(\beta)}(t_2,r_2)\rangle
\EEQ
and the covariance conditions are
\BEQ \label{C:Kov}
X_0 F = \frac{2x}{3} F \;\; , \;\; 
X_{-1} F = Y_{\pm 1/2} F = M_0 F = N F = 0
\EEQ
where $x=x_1+x_2$. Here, $X_{-1,0}$ and $Y_{-1/2}$ are given in (\ref{C:XXY}),
while the other generators read
\BEQ
Y_{1/2} = -t \partial_r - \mathfrak{M}\, \partial_r^{1/2}\, r \;\; , \;\;
M_0 = - \mathfrak{M}\, \partial_r^{1/2} \;\; , \;\;
N = -\frac{1}{2} \mathfrak{M}^2
\EEQ

Spatio-temporal translation invariance yields $F=F(t,r)$, where $t=t_1-t_2$
and $r=r_1-r_2$. Invariance under the action of $N$ leads to the condition
${\cal M}_2^2 = (-1)^{\beta-\alpha+1}{\cal M}_1^2$ or alternatively
\BEQ \label{C:NInv}
{\cal M}_2 = - \II^{\beta-\alpha+1}{\cal M}_1 \;\; , \;\;
\mbox{\rm where $\beta=\alpha+1 \mbox{\rm ~mod } 2$}
\EEQ
Then the action of $M_0$ on $F$ is, using translation invariance and 
(\ref{A:Lem13})
\BEA 
M_0 F(t,r) &=& - \left( {\cal M}_1 \II^{\alpha} \partial_r^{1/2} 
+ {\cal M}_2 \II^{\beta+1} \partial_r^{1/2} \right) F(t,r) \nonumber \\
&=& -\II^{\alpha} \left( {\cal M}_1  
+ {\cal M}_2 \II^{\beta-\alpha+1} \right) \partial_r^{1/2} F(t,r) \nonumber \\
&=& -\II^{\alpha} {\cal M}_1 \left( 1 - (-1)^{\beta-\alpha+1} \right)
\partial_r^{1/2} F(t,r) = 0 
\EEA
because of (\ref{C:NInv}) and thus $F$ is always invariant 
under the action of $M_0$.
To calculate the action of of $Y_{1/2}$, we recall from the Leibniz rule
(\ref{A:Lem15}) that for $f\in{\cal M}_E$
\BEQ
\partial_r^{1/2} \left( r f(r) \right) = r \partial_r^{1/2} f(r) + \frac{1}{2}
\partial_r^{-1/2} f(r) 
\EEQ
and we find, using again (\ref{A:Lem13})
\BEA
Y_{1/2} F(t,r) &=& \left[ -t\partial_r -
\mathfrak{M}_1 \left( r_1 \partial_{r_1}^{1/2} 
+\frac{1}{2} \partial_{r_1}^{-1/2}\right) 
- \mathfrak{M}_2 \left( r_2 \partial_{r_2}^{1/2} 
+\frac{1}{2} \partial_{r_2}^{-1/2}\right) \right] F(t,r) \nonumber \\
&=& \left[ -t\partial_r -
{\cal M}_1 \II^{\alpha} \left( r_1 -r_2 \II^{2(\beta-\alpha+1)}\right)
\partial_r^{1/2} -
\frac{1}{2}{\cal M}_1 \II^{\alpha} \left( 1 + \II^{2(\beta-\alpha+1)}\right)
\partial_r^{-1/2} \right] F(t,r) \nonumber \\
&=& -\left( t\partial_r + \mathfrak{M} 
\left( r \partial_r^{1/2} + \partial_r^{-1/2}\right) \right) F(t,r)
\label{C:Y12Gen}
\EEA
where we have set $\mathfrak{M} := {\cal M}_1 \II^{\alpha}$. Finally,
\BEQ
X_0 F(t,r) = - \left( t\partial_t + \frac{2}{3} r \partial_r \right) F(t,r)
\EEQ
Now, the scaling ansatz
\BEQ
F(t,r) = t^{-2x/3} \Psi(u) \;\; , \;\; u = r t^{-2/3}
\EEQ
solves the first of the covariance conditions (\ref{C:Kov}), while the last
remaining condition (\ref{C:Kov}) leads, via (\ref{C:Y12Gen}) and 
(\ref{A:Lem13}), to a fractional 
differential equation for the scaling function $\Psi(u)$
\BEQ
\left( \partial_u + \mathfrak{M}\, u \partial_u^{1/2} + 
\mathfrak{M}\, \partial_u^{-1/2} \right) \Psi(u) = 0 
\EEQ
which coincides with eq.~(\ref{3:TypIIaDGL}) found for Typ IIa with $\beta=0$
and $\gamma=\frac{2}{3}\mathfrak{M}$. 
Inspection of the terms present in this equation 
leads to the following series ansatz
\BEQ
\Psi(u) = \sum_{n=0}^{\infty} \Psi_n u^{s+3n/2} \;\; , \;\; \Psi_0 \ne 0
\EEQ
which promptly gives $s=0$ and the recursion relation, valid for all $n\geq 1$
\BEQ
\Psi_n = -\mathfrak{M}\, \Gamma\left( \frac{3n-1}{2}\right) 
\Gamma\left( \frac{3n}{2}\right)^{-1} \Psi_{n-1}
\EEQ 
This is best solved by rewriting it in the form 
$\Psi_{n+2}=\Psi_n (2\mathfrak{M}^2/3)(n+1)\left( (n+2/3)(n+4/3) \right)^{-1}$ 
and we finally have for desired scaling function
\BEA
\Psi(u) &=& \Psi_0 \sqrt{\frac{4\pi}{3}} \sum_{n=0}^{\infty} 
\frac{\Gamma\left(\frac{n}{2}+\frac{1}{2}\right)}
{\Gamma\left(\frac{n}{2}+\frac{1}{3}\right)
\Gamma\left(\frac{n}{2}+\frac{2}{3}\right)}
\left( -\frac{\mathfrak{M}}{\sqrt{3}} u^{3/2} \right)^n 
\\
&=& \Psi_0 \left[ {_2F_2}\left( 1,\frac{1}{2};\frac{1}{3},\frac{2}{3};
\frac{\mathfrak{M}^2}{3} u^3 \right) 
- \sqrt{\frac{4}{\pi} \mathfrak{M}^2\, u^{3}\:}\: 
{_2F_2}\left( 1,1;\frac{5}{6},\frac{7}{6};
\frac{\mathfrak{M}^2}{3} u^3 \right) \right] \nonumber
\EEA
where $_2F_2$ is a generalized hypergeometric function and 
$\Psi_0=\Psi(0)$ is the initial value.

\begin{figure}[htb]
\centerline{\epsfxsize=3.25in\epsfbox{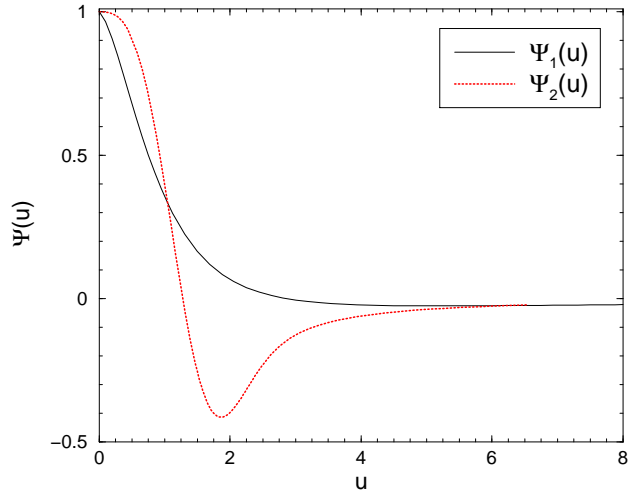}}
\caption{Scaling functions $\Psi_{1,2}(u)$ for the two-point functions
$F_i(t,r) = t^{-2x/3} \Psi_i(u)$, for
${\cal M}=1$ and $\Psi_0=1$. 
\label{Abb_C1}}
\end{figure}

For a physical interpretation, recall that the four scaling operators
$\phi^{(\alpha)}$ had the `masses' $\mathfrak{M} = \II^{\alpha} {\cal M}_1$,
where ${\cal M}_1>0$ and $\alpha=0,1,2,3$. In addition, considering global
scale invariance with either $t=0$ or $r=0$, it follows that for $u\to\infty$,
the boundary condition $\Psi(u)\to 0$ must be satisfied. That is possible
in two cases: (i) when $\mathfrak{M}>0$ and (ii) when $\mathfrak{M}$ is 
imaginary and the real part of the two-point function is retained. 
In these cases, the two-point functions read
\BEA
F_1 &=& F^{(0,1)}(t,r) = t^{-2x/3} \Psi(u) \;\; ; \;\; 
\mbox{\rm~~~~ where $\mathfrak{M}={\cal M}_1 = {\cal M}_2 >0$} \nonumber \\
F_2 &=& \frac{1}{2} \left( F^{(1,2)}(t,r) + F^{(3,0)}(t,r)\right) 
= t^{-2x/3} \Psi_0\; {_2F_2}\left( 1,\frac{1}{2};\frac{1}{3},\frac{2}{3};
-\frac{{\cal M}_1^2}{3} u^3 \right) \\
& & \hspace{4.5truecm}; \hspace{0.25truecm} 
\mbox{\rm~~~~ where $\mathfrak{M}^2=-{\cal M}_1^2$, 
${\cal M}_1 = {\cal M}_2 >0$} \nonumber 
\EEA
The scaling functions $\Psi_{1,2}(u)$ obtained form these are shown 
in figure~\ref{Abb_C1}. 
%%----------------------------------------------------------------------------%%
%%----------------------------------------------------------------------------%%

%%%%%%%%%%%%%%%%%%%%%%%%%%%%%%%%%%%%%%%%%%%%%%%%%%%%%%%%%%%%%%%%%%%%%%%%%%%%%%%%
\appsection{D}{ }
%%%%%%%%%%%%%%%%%%%%%%%%%%%%%%%%%%%%%%%%%%%%%%%%%%%%%%%%%%%%%%%%%%%%%%%%%%%%%%%%

We discuss the solution of the fractional-order differential equation 
(\ref{4:DGLI}) in terms of series expansion, following standard lines
\cite{Mill93,Podl99}. As in section 4, we have $N=N_0+p/q$ with 
$N_0\in\mathbb{N}$. We make the ansatz
\BEQ
\Omega(v) = \sum_{n=0}^{\infty} a_n v^{n/q+s} \;\; , \;\; a_0 \ne 0
\EEQ
and from substitution into eq.~(\ref{4:DGLI}), we find for $N_0 \geq 2$,
\BEQ 
\sum_{n=-p-qN_0}^{\infty} 
\frac{\alpha_1 \Gamma\left( \frac{1}{q}(n+p+qN_0) + s+1\right)}{\Gamma\left( 
\frac{n}{q} +2+s\right) } a_{n+p+qN_0} v^{n/q}   
- \sum_{n=0}^{\infty} \left( \frac{1}{q}(n+px_1+qN_0x_1)+s\right) 
a_n v^{n/q} = 0 
\EEQ
where the definition (\ref{A:Ablei}) has been used. 
The singular terms can be dropped if $v$ is positive. 
Since this equation holds for all $v>0$, we obtain the following conditions. 
First, since $a_0\ne 0$, we must have
\BEQ \label{D:3}
\frac{\alpha_1 \Gamma\left(s+1\right)}{\Gamma\left( -\frac{p}{q} -N_0 
+2 +s\right) } = 0
\EEQ
Second, we have $a_{\ell}=0$ for $\ell=1,\ldots,p+qN_0-1$. Finally, we get
the recurrence
\BEQ
\frac{\alpha_1 \Gamma\left( \frac{1}{q}(n+p+qN_0) + s+1\right)}{\Gamma\left( 
\frac{n}{q} +2+s\right)} a_{n+p+qN_0} 
= \left( \frac{1}{q}(n+px_1+qN_0x_1)+s\right) a_n
\EEQ

From the first condition eq.~(\ref{D:3}), we find the possible values of $s$,
namely
\BEQ
s = s_m := \frac{p}{q} + m \;\; , \;\; m = 0,1,\ldots, N_0 -2
\EEQ
Negative values of $s$ would lead to a singularity as $v\to 0$ and are 
therefore excluded. The second and third conditions are solved 
by writing $n=(p+qN_0)\ell$ and by letting
$\omega_{\ell} := a_{(p+qN_0)\ell} = a_n$. We then find
\BEQ
\omega_{\ell+1} = \frac{N^2}{\alpha_1} 
\left( \ell+\frac{s_m+\zeta}{N}\right) 
\left( \ell+\frac{1+\zeta}{N}\right) 
\frac{\Gamma(N\ell +s_m+1)}{\Gamma(N(\ell+1)+s_m+1)}
\, \omega_{\ell}
\EEQ
The $N_0-1$ linearly independent solutions are, with $\eps=N-N_0$
\BEA 
\Omega_{(m)}(v) &=& \sum_{\ell=0}^{\infty} \omega_{\ell}\, v^{N\ell+\eps+m} 
\nonumber \\
&=& v^{\eps+m} \omega_{0,m} \sum_{\ell=0}^{\infty} 
\frac{\Gamma(l+(s_m+\zeta)/N)\Gamma(l+(s_m+1)/N)}
{\Gamma((s_m+\zeta)/N)\Gamma((s_m+1)/N)}
\frac{\Gamma(s_m+1)}{\Gamma(N\ell+s_m+1)} 
\left( \frac{N^2}{\alpha_1} v^N\right)^{\ell} ~~~~~
\label{4:OmegamFrac}
\EEA
where $m = 0,1,\ldots, N_0-2$ and $\omega_{0,m}$ are arbitrary constants. 
It is easy to see that these series have an infinite radius of 
convergence, provided $N>2$. Although this solution was
only derived for rational values of $N=N_0+\eps$, we can make the analytical
continuation to all real values of $N$. 

We have seen in section 4 that there are solutions of eq.~(\ref{4:DGLI}) such
that $\Omega(v)\sim v^{-\zeta}$ for $v$ large. These may be constructed from 
the series (\ref{4:OmegamFrac}) 
in the same way as done for $N$ integer in section 4. On the
other hand, for $v$ small, one has $\Omega_{(0)}(v)\sim v^{\eps}$.
Therefore, the boundary condition $\Omega(0)=1$ cannot be satisfied for 
$\eps>0$, that is any non-integer value of $N$. 

This difficulty with the $v\to 0$ boundary condition is a specific
property of the fractional derivative (\ref{A:Ablei}) which it has in
common with the Riemann-Liouville derivative. It is known that
initial conditions for ordinary fractional differential equations are 
specified in terms of {\em fractional} derivatives $\partial^a f(0)$ 
\cite{Mill93,Samk93,Podl99,Hilf00} and {\em not}
in terms of ordinary derivatives $f^{(N)}(0)$ of integer order $N$.
Indeed, it has been suggested to avoid this problem by using the
fractional Caputo derivative instead \cite{Podl99,Hilf00}.  
However, the Caputo derivative does not commute
and it appears to be an open mathematical problem if the Caputo
definition can be modified such as to obtain a commutative operator.

%\newpage

%%


{\small
\begin{thebibliography}{999}
\bibitem{Abra65} M.A. Abramowitz and I.A. Stegun, {\it Handbook of Mathematical
Functions}, Dover (New York 1965)
\bibitem{Ahar76} A. Aharanoy, in C. Domb and M.S. Green (eds),
{\it Phase Transitions and Critical Phenomena}, Vol. 6, Academic
(London 1976), p. 358
\bibitem{Albu02} L.C. de Albuquerque and M.M. Leite, J. Phys. {\bf A35}, 
1807 (2002); {\bf A34}, L327 (2001).  
\bibitem{Barg54} V. Bargmann, Ann. of Math. {\bf 59}, 1 (1954).
\bibitem{Barr98} A. Barrat, Phys. Rev. {\bf E57}, 3629 (1998).
\bibitem{Baru73} A.O. Barut, Helv. Phys. Acta {\bf 46}, 496 (1973). 
\bibitem{Bate95} F.S. Bates, W. Maurer, T.P. Lodge, M.F. Schulz, 
M.W. Matsen, K.\ Almdal, and K.\ Mortensen, 
Phys. Rev.\ Lett. {\bf 75}, 4429 (1995).
\bibitem{Bela84} A.A. Belavin, A.M. Polyakov and A.B. 
Zamolodchikov, Nucl. Phys. {\bf B241}, 333 (1984).
\bibitem{Benz84} J. Benzoni, J. Phys. {\bf A17}, 2651 (1984).
\bibitem{Bert99} L. Berthier, J.L. Barrat and J. Kurchan, 
Eur. Phys. J. {\bf B11}, 635 (1999).
\bibitem{Bert01} L. Berthier, P.C.W. Holdsworth, and M. Sellitto,
J. Phys. {\bf A34}, 1805 (2001).
\bibitem{Bray91} A.J. Bray, K. Humayun and T.J. Newman, Phys. Rev. {\bf B43}, 
3699 (1991).
\bibitem{Bray94} A.J. Bray, Adv. Phys. {\bf 43}, 357 (1994).
\bibitem{Bray00} A.J. Bray in \cite{Cate00}, sect 5.4. 
\bibitem{Bouc98} J.P. Bouchaud, L.F. Cugliandolo, J. Kurchan and M. M\'ezard,
in A.P. Young (ed.) {\it Spin Glasses and Random Fields},
World Scientific (Singapore 1998); ({\tt cond-mat/9702070}).
\bibitem{Cala02} P. Calabrese and A. Gambassi, Phys. Rev. {\bf E65}, 
066120 (2002).
\bibitem{Cann01} S.A. Cannas, D.A. Stariolo and F.A. Tamarit, 
Physica {\bf A294}, 362 (2001).
\bibitem{Card85} J.L. Cardy, J. Phys. {\bf A18}, 2771 (1985).
\bibitem{Card96} J.L. Cardy, {\it Scaling and Renormalization in 
Statistical Mechanics} Cambridge University Press, (Cambridge, 1996).
\bibitem{Cate00} M.E. Cates and M.R. Evans (eds), {\it Soft and Fragile Matter},
Proc. 53$^{\rm rd}$ Scottish University Summer Schools in Physics, 
(Bristol 2000).
\bibitem{Cham01} C. Chamon, M.P. Kennett, H. Castillo and L.F. Cugliandolo,
{\tt cond-mat/0109150}.  
\bibitem{Chop98} B. Chopard and M. Droz, {\it Cellular Automata Modelling
of Physical Systems}, Cambridge University Press (Cambridge 1998).
\bibitem{Coni94} A. Coniglio, P. Ruggiero and M. Zanetti, Phys. Rev. {\bf E50},
1046 (1994). 
\bibitem{Corb02} F. Corberi, E. Lippiello and M. Zanetti, 
Phys. Rev. {\bf E65}, 046136 (2002). 
\bibitem{Cour65} R. Courant and F. John, {\it Introduction to Calculus and
Analysis}, Vol. I, Wiley (New York 1965), p. 566
\bibitem{Cugl94} L.F. Cugliandolo, J. Kurchan, and G. Parisi, J. Physique 
{\bf I4}, 1641 (1994).
\bibitem{Cugl94a} L.F. Cugliandolo and J. Kurchan, J. Phys. {\bf A27}, 
5749 (1994). 
\bibitem{Cugl95} L.F. Cugliandolo and D.S. Dean, J. Phys. {\bf A28}, 
4213 (1995).
\bibitem{Datt93} S. Dattagupta, Physica {\bf A194}, 137 (1993). 
\bibitem{Dieh00} H.W. Diehl and M. Shpot, Phys. Rev. {\bf B62}, 12338 (2000).
\bibitem{Dieh01} H.W. Diehl and M. Shpot, J. Phys. {\bf A34}, 9101 (2001).
\bibitem{Dieh02} H.W. Diehl and M. Shpot, J. Phys.{\bf A35}, 6249 (2002).
\bibitem{Dieh02a} H.W. Diehl, Acta physica slovaka {\bf 52}, 271 (2002). 
\bibitem{Drou88} J.M. Drouffe et C. Itzykson, {\it Th\'eorie statistique des 
champs}, 2 Vols., Editions CNRS (Paris 1988) 
\bibitem{Ever96} H.G. Evertz and D.P. Landau, Phys. Rev. {\bf B54}, 
12302 (1996).
\bibitem{Ever01} H.G. Evertz and W.v.d. Linden, Phys. Rev. Lett. {\bf 86},
5164 (2001). 
\bibitem{Fish83} M.E. Fisher in F.J.W. Hahne (ed),  
{\it Critical Phenomena}, Springer Lecture Notes in Physics {\bf 186}, 
Springer (Heidelberg 1983), p. 1
\bibitem{Fish88} D.S. Fisher and D.A. Huse, Phys. Rev. {\bf B38}, 373 (1988).
\bibitem{Floh01} M. Flohr, {\tt hep-th/0111228}.
\bibitem{Frac93} L. Frachebourg and M. Henkel, Physica {\bf A195}, 577 (1993).
\bibitem{Fran97} P. di Francesco, P. Mathieu and D. S\'en\'echal, 
{\it Conformal Field Theory}, Springer (Heidelberg 1997).
\bibitem{Gelf64} I.M. Gelfand and G.E. Shilov, {\it Generalized Functions},
Vol. 1, Academic Press (New York 1964). 
\bibitem{Glau63} R.J. Glauber, J. Math. Phys. {\bf 4}, 294 (1963).
\bibitem{Gier02} J. de Gier, B. Nienhuis, P.A. Pearce and V. Rittenberg,
{\tt cond-mat/0205467}. 
\bibitem{Giul96} D. Giulini, Ann. of Phys. {\bf 249}, 222 (1996). 
\bibitem{Godr00a} C. Godr\`eche and J.M. Luck, J. Phys. {\bf A33}, 1151 (2000).
\bibitem{Godr00b} C. Godr\`eche and J.M. Luck, J. Phys. {\bf A33}, 9141 (2000).
\bibitem{Godr02} C. Godr\`eche and J.M. Luck, J. Phys. Cond. Matt. {\bf 14}, 
1589 (2002).
\bibitem{Gryn94} M.D. Grynberg, T.J. Newman and R.B. Stinchcombe, 
Phys. Rev. {\bf E50}, 957 (1994). 
\bibitem{Hage72} C.R. Hagen, Phys. Rev. {\bf D5}, 377 (1972).
\bibitem{Halp77} B.I. Halperin and P.C. Hohenberg, Rev. Mod. Phys. 
{\bf 49}, 435 (1977). 
\bibitem{Henk92} M. Henkel, Int. J. Mod. Phys. {\bf C3}, 1011 (1992).
\bibitem{Henk94} M. Henkel, J. Stat. Phys. {\bf 75}, 1023 (1994). 
\bibitem{Henk94a} M. Henkel and G.M. Sch\"utz, Int. J. Mod. Phys. {\bf B8},
3487 (1994). 
\bibitem{Henk97} M. Henkel, Phys. Rev. Lett. {\bf 78}, 1940 (1997).
\bibitem{Henk98} M. Henkel and D. Karevski, J. Phys. {\bf A31}, 2503 (1998).
\bibitem{Henk99} M. Henkel, {\it Phase Transitions and Conformal Invariance},
Springer (Heidelberg 1999).
\bibitem{Henk01} M. Henkel, M. Pleimling, C. Godr\`eche and J.-M. Luck,
Phys. Rev. Lett. {\bf 87}, 265701 (2001). 
\bibitem{Henk01a} M. Henkel and M. Pleimling, {\tt cond-mat/0108454}, 
Comm. Comp. Phys. in press 
\bibitem{Hilf00} R. Hilfer (ed), {\it Applications of Fractional Calculus in 
Physics}, World Scientific (Singapore 2000). 
\bibitem{Hert76} J.A. Hertz, Phys. Rev. {\bf B14}, 1165 (1976).
\bibitem{Hill76} E. Hille, {\it Ordinary Differential Equations in the 
Complex Domain}, Wiley (New York 1976). 
\bibitem{Hinr00} H. Hinrichsen, Adv. Phys. {\bf 49}, 1 (2000).
\bibitem{Horn75} R.M. Hornreich, M. Luban and S. Shtrikman, Phys. Rev.
Lett.\ {\bf 35}, 1678 (1975).
\bibitem{Huse89} D.A. Huse, Phys. Rev. {\bf B40}, 304 (1989).
\bibitem{Jans89} H.K. Janssen, B. Schaub and B. Schmittmann, 
Z. Phys. {\bf B73}, 539 (1989).
\bibitem{Kand90} D. Kandel, E. Domany and B. Nienhuis, J. Phys. {\bf A23}, L755
(1990).
\bibitem{Kask85} K. Kaski and W. Selke, Phys. Rev. {\bf B31}, 3128 (1985). 
\bibitem{Kenn01} M.P. Kennett and C. Chamon, Phys. Rev. Lett. {\bf 86}, 
1622 (2001). 
\bibitem{Klim01} M. Klimek, J. Phys. {\bf A34}, 6167 (2001).
\bibitem{Kreu81} H.J. Kreuzer, {\it Nonequilibrium Thermodynamics and its 
Statistical Foundation}, Clarendon Press (Oxford 1981). 
\bibitem{Krug97} J. Krug, Adv. Phys. {\bf 46}, 139 (1997). 
\bibitem{Kurc01} J. Kurchan, {\tt cond-mat/0110628}. 
\bibitem{Lahn98} V.I. Lahno, J. Phys. {\bf A31}, 8511 (1998).
\bibitem{Levy67} J.-M. Levy-Leblond, Comm. Math. Phys. {\bf 4}, 157 (1967);
{\bf 6}, 286 (1967)
\bibitem{Lipp00} E. Lippiello and M. Zanetti, Phys. Rev. {\bf E61}, 3369 (2000).
\bibitem{Marr99} J. Marro and R. Dickman, {\it Nonequilibrium Phase Transitions
in Lattice Models}, Cambridge University Press (Cambridge 1999).
\bibitem{Mehe00} T. Mehen, I.W. Stewart and M.B. Wise, Phys. Lett. 
{\bf B474}, 145 (2000).
\bibitem{Mill93} K.S. Miller and B. Ross, {\it An introduction to the 
fractional calculus and fractional differential equations}, Wiley 
(New York 1993).
\bibitem{Mo91} Z. Mo and M. Ferer, Phys. Rev. {\bf B43}, 10890 (1991). 
\bibitem{Neub98} B. Neubert, M. Pleimling and R. Siems, 
Ferroelectrics {\bf 208-209}, 141 (1998). 
\bibitem{Newm90} T.J. Newman and A.J. Bray, J. Phys. {\bf A23}, 4491 (1990).
\bibitem{Nico76} J.F. Nicoll, G.F. Tuthill, T.S. Chang and H.E. Stanley,
Phys. Lett. {\bf A58}, 1 (1976).
\bibitem{Nied72} U. Niederer, Helv. Phys. Acta {\bf 45}, 802 (1972);
{\bf 46}, 191 (1973); {\bf 47}, 119 and 167 (1974); {\bf 51}, 220 (1978). 
\bibitem{Ohwa01} K. Ohwada, Y. Fujii, N. Takesue, M. Isobe, Y. Ueda, N. Nakao,
Y. Wakabayashi, Y. Murakami, K. Ito, Y. Amemiya, Y. Fujihisa, K. Aoki, T. Shobu,
Y. Noda and N. Ikeda, Phys. Rev. Lett. {\bf 87}, 086402 (2001). 
\bibitem{Oitm85} J. Oitmaa, J. Phys. {\bf A18}, 365 (1985).
\bibitem{Perr77} M. Perroud, Helv. Phys. Acta {\bf 50}, 233 (1977).
\bibitem{Pico02} A. Picone and M. Henkel, J. Phys. {\bf A35}, 5575 (2002).
\bibitem{Pico02a} A. Picone and M. Henkel, to be published. 
\bibitem{Plei01} M. Pleimling and M. Henkel, Phys. Rev. Lett. {\bf 87}, 125702
(2001). 
\bibitem{Podl99} I. Podlubny, {\it Fractional differential equations},
Academic Press (New York 1999). 
\bibitem{Poly70} A.M. Polyakov, Sov. Phys. JETP {\bf 12}, 381 (1970).
\bibitem{Priv96} V. Privman (Ed.) {\it Nonequilibrium Statistical Mechanics in
One Dimension}, Cambridge University Press (Cambridge 1996).
\bibitem{Rahi01} M.R. Rahimi Tabar, {\tt hep-th/0111327}.
\bibitem{Rute95} A.D. Rutenberg and A.J. Bray, Phys. Rev. {\bf E51}, 
5499 (1995). 
\bibitem{Rute96} A.D. Rutenberg, Phys. Rev. {\bf E54}, R2181 (1996).
\bibitem{Rute99} A.D. Rutenberg and B.P. Vollmayr-Lee, Phys. Rev. Lett.
{\bf 83}, 3772 (1999). 
\bibitem{Sach00} S. Sachdev, {\it Quantum Phase Transitions}, Cambridge
University Press (Cambridge 2000). 
\bibitem{Samk93} S.G. Samko, A.A. Kilbas and O.I. Marichev, 
{\it Fractional Integrals and derivatives}, Gordon and Breach (Amsterdam 1993).
\bibitem{Scha75} L. Sch\"afer, J. Phys. {\bf A9}, 377 (1975).
\bibitem{Schu94} G.M. Sch\"utz and S. Sandow, Phys. Rev. {\bf E49}, 2726 (1994).
\bibitem{Schu00} G.M. Sch\"utz, in C. Domb and 
J.L. Lebowitz (eds), {\it Phase Transitions and Critical Phenomena}, Vol. 19,
Academic (New York 2000).
\bibitem{Schm95} B. Schmittmann and R.K.P. Zia, in C. Domb and 
J.L. Lebowitz (eds), {\it Phase Transitions and Critical Phenomena}, Vol. 17,
Academic (New York 1995).
\bibitem{Schr98} 
A. Schr\"oder, G. Aeppli, E. Bucher, R. Ramazashvili, and P. Coleman, 
Phys. Rev. Lett. {\bf 80}, 5623 (1998).
\bibitem{Selk77} W. Selke, Z. Phys. {\bf B27}, 81 (1977) and Phys. Lett.
{\bf A61}, 443 (1977). 
\bibitem{Selk88} W. Selke, Phys. Rep. {\bf 170C}, 213 (1988).
\bibitem{Selk92} W. Selke, in C. Domb and J.L. Lebowitz (eds) 
{\it Phase Transitions and Critical Phenomena}, Vol.15, Academic Press 
(New York, 1992).
\bibitem{Shpo01} M. Shpot and H.W. Diehl, Nucl. Phys. {\bf B612}, 340 (2001).
\bibitem{Skab00}
M. \u{S}karabot, R. Blinc, I. Mu\u{s}evi\u{c}, A. Rastegar, and
Th. Rasing, Phys. Rev. {\bf E61}, 3961 (2000).
\bibitem{Turb02} L. Turban and F. Igl\'oi, Phys. Rev. {\bf B66}, 014440 (2002). 
\bibitem{Vyso92} Y.\ M.\ Vysochanskii and V.\ U.\ Slivka, Usp.\ Fiz.\
Nauk {\bf 162}, 139 (1992).
\bibitem{Wolf89} U. Wolff, Phys. Rev. Lett. {\bf 62}, 361 (1989). 
\bibitem{Wrig35} E.M. Wright, Proc. London Math. Soc. {\bf 46}, 389 (1940); 
J. London Math. Soc. {\bf 27}, 256 (1952) erratum. 
\bibitem{Yeom88} J.M. Yeomans, Solid State Physics {\bf 41}, 151 (1988).
\bibitem{Zava98} P. Z\'avada, Comm. Math. Phys. {\bf 192}, 261 (1998). 
\bibitem{Zinn89} J. Zinn-Justin, {\it Quantum Field Theory 
and Critical Phenomena}, Clarendon Press (Oxford 1989, 1996$^3$)
\bibitem{Zipp00} W. Zippold, R. K\"uhn and H. Horner, Eur. Phys. J. {\bf B13},
531 (2000). 
\bibitem{Cala02a} P. Calabrese and A. Gambassi, {\tt cond-mat/0207487}. 
%%
%%\bibitem{Meno97} G. Menon, M. Barma, and D. Dhar, 
%%                 J. Stat. Phys. {\bf 86}, 1237 (1997).

\end{thebibliography}
}
\end{document}